\documentclass{aa}

\usepackage{graphicx}
\usepackage{subcaption}
\usepackage{multirow}
\usepackage{float} 

\usepackage{txfonts}
\usepackage{xcolor}
\usepackage{hyperref}
 \usepackage{natbib} 
\bibpunct{(}{)}{;}{a}{}{,} 
\usepackage{array}      
\usepackage{ragged2e}  

\begin{document}

   \title{Quenching of galaxies at cosmic noon}

   \subtitle{Understanding the effect of environment}

     \author{Akriti Singh,
           \inst{1,2,15}
           Lucia Guaita,\inst{2,15}
           Pascale Hibon,\inst{1}
           Boris H\"au\ss ler,\inst{1}
           Kyoung-Soo Lee\inst{3}, Vandana Ramakrishnan\inst{3}, Ankit Kumar\inst{2}, 
            Nelson Padilla\inst{4}, Nicole M. Firestone\inst{5}, Hyunmi Song\inst{6}, Maria Celeste Artale\inst{2}, Ho Seong Hwang \inst{7,8}, Paulina Troncoso Iribarren \inst{9}, Caryl Gronwall\inst{10,11}, Eric Gawiser\inst{5}, Julie Nantais\inst{2}, Francisco Valdes\inst{12}, Changbom Park \inst{13}, Yujin Yang \inst{14}
  }

    \institute{European Southern Observatory, Alonso de Córdova 3107, Vitacura, Santiago de Chile, Chile\\
               \email{akriti.ramnee@gmail.com}
          \and
              Universidad Andres Bello, Facultad de Ciencias Exactas, Departamento de Fisica, Instituto de Astrofisica, Fernandez Concha 700, Las Condes, Santiago RM, Chile
              \and Department of Physics and Astronomy, Purdue University, 525 Northwestern Avenue, West Lafayette, IN 47907, USA \and Instituto de Astronom\'{\i}a Te\'orica y Experimental (IATE), CONICET-UNC, Laprida 854, X500BGR, C\'ordoba, Argentina
 \and Department of Physics and Astronomy, Rutgers, the State University of New Jersey, Piscataway, NJ 08854, USA \and Department of Astronomy and Space Science, Chungnam National University, 99 Daehak-ro, Yuseong-gu, Daejeon, 34134, Republic of Korea \and Department of Physics and Astronomy, Seoul National University, 1 Gwanak-ro, Gwanak-gu, Seoul 08826, Republic of Korea \and SNU Astronomy Research Center, Seoul National University, 1 Gwanak-ro, Gwanak-gu, Seoul 08826, Republic of Korea \and Universidad Central de Chile, Avenida Francisco de Aguirre 0405, 171-0614 La Serena, Coquimbo, Chile \and  Department of Astronomy and Astrophysics, The Pennsylvania State University, University Park, PA 16802, USA \and Institute for Gravitation and the Cosmos, The Pennsylvania State University, University Park, PA 16802, USA \and NSFs National Optical-Infrared Astronomy Research Laboratory, 950 N. Cherry Ave., Tucson, AZ 85719, USA \and Korea Institute for Advanced Study, 85 Hoegi-ro, Dongdaemun-gu, Seoul 02455, Republic of Korea \and Korea Astronomy and Space Science Institute, 776 Daedeokdae-ro, Yuseong-gu, Daejeon 34055, Republic of Korea \and  Millennium Nucleus for Galaxies (MINGAL)}

   \date{Received XX XX, 2024; accepted XX XX, 2024}

 \abstract 
 {}
 {We aim to identify and analyse quiescent galaxies at $z \approx 3.1$ within the 2-deg$^2$ COSMOS field and explore how galaxy environment influences quenching processes. By examining how the quenched fraction and physical properties of these galaxies vary across different environmental contexts—including local densities, protoclusters, and cosmic filaments—we investigate the connection between environmental factors and galaxy quenching at cosmic noon.} 
 {We select massive quiescent galaxies at $z \sim 3.1$ using deep photometric data from the COSMOS2020 catalogue combined with narrowband-selected Lyman-$\alpha$ emitters (LAEs) from the One-hundred-square-degree DECam Imaging in Narrowbands (ODIN) survey. We perform spectral energy distribution (SED) fitting using the BAGPIPES code to derive star-formation histories (SFHs) and quenching timescales. We construct Voronoi tessellation density maps using
LAEs and, independently,  photometrically-selected galaxies to characterize galaxy environments.}
 {We identify 24 massive quiescent galaxies (MQGs) at $z \approx 3.1$, each having stellar masses above $10^{10.6} M_{\odot}$. These MQGs share remarkably uniform star-formation histories, with intense starburst phases followed by rapid quenching within short timescales ($\leq 400$ Myr). The consistency of these quenching timescales suggests a universal and highly efficient quenching mechanism at this epoch. We find no significant correlation between environmental density—either local or large-scale—and galaxy quenching parameters such as quenching duration, quenched fraction, or timing. MQGs show no preferential distribution with respect to protoclusters or filaments compared to massive star-forming galaxies. Some MQGs reside close to gas-rich filaments but show no evidence of rejuvenated star formation, implying gas-heating mechanisms instead of gas exhaustion. These results indicate that quenching processes at $z \approx 3.1$ likely depend little on the immediate galaxy environment.}
 {Our findings suggest that environmental processes alone, such as galaxy mergers, interactions, or gas stripping, cannot fully explain galaxy quenching at $z\approx3.1$. Instead, internal mechanisms such as AGN feedback, stellar feedback, virial shock heating, or morphological quenching play an important role in quenching. Future spectroscopic observations must confirm the quiescent nature and precise redshifts of these galaxies. Additionally, observational studies of gas dynamics, gas temperature, and ionization conditions within and around MQGs will clarify the physical mechanisms driving galaxy quenching during this critical epoch of galaxy evolution.}

\titlerunning{Quenching as a function of the environment}
\authorrunning{Singh et al.}
   
\maketitle

\section{Introduction}

Observations and simulations have shown a clear bi-modality in various properties of galaxies such as their colours, morphologies, and star formation rates (SFRs) \citep{1980ApJ...236..351D}. Quiescent galaxies are characterised by redder colours and typically a spheroidal morphology, whereas star-forming galaxies exhibit bluer colours and typically show disk-like morphology. However, there is a gap in our understanding of the mechanisms that cause the star formation in bluer galaxies to stop, quenching them into the redder quiescent galaxies.

Quenching mechanisms are generally categorized as mass quenching or environmental quenching \citep{peng2010environmental}. Mass quenching refers to internal processes that correlate with the stellar mass of a galaxy, such as stellar feedback and AGN feedback\citep[e.g.,][]{croton2006many,ceverino2009role,fabian2012observational,cicone2014massive}. Environmental quenching refers to external processes such as ram pressure stripping \citep[e.g.,][]{gunn1972infall}, strangulation \citep{larson1990galaxy} , and galaxy harassment \citep{moore1996galaxy,2009ApJ...699.1595P} which depend on the environment in which the galaxy resides. Although both mass and environmental quenching affect the cessation of star formation, their relative importance depends on the mass of the galaxy and the cosmic epoch. In the local universe, there is a well-established relation between SFR and environment density according to which quenched galaxies preferentially occur in dense environments such as clusters. In contrast, star-forming galaxies predominantly exist in relatively low-density environments.  This is because in local virialized clusters, environmental processes can remove or heat cold gas from galaxies \citep{2006PASP..118..517B}, therefore 
accelerating quenching phenomena and increasing the fraction of quenched galaxies.  

In recent years, the existence of quiescent galaxies (QGs) at redshifts $z$ $\sim$ 3–4 \citep[e.g.,][]{spitler2014exploring,straatman2014substantial,2019A&A...632A..80G,Carnall_2018,carnall2023surprising} has been corroborated by spectroscopic confirmations \citep[e.g.,][]{glazebrook2017massive,glazebrook2022revelation,schreiber2018near,d2020typical,nanayakkara2024population}. 
While at lower redshift ($z < 1$), environmental quenching typically occurs on timescales of several billion years \citep{mao2022revealing}, at z = 3 the Universe is only about 2 billion years old and the quenching processes must act faster.  
This raises the question of what processes quenched galaxies at $z$$>3$. 
Some studies suggest that high-density environments promote star formation and are dominated by starburst galaxies\citep{casey2016ubiquity,calvi2023bright}, whereas other found quiescent galaxies in compact  groups  \citep{ito2023cosmos2020,tanaka2024protocluster}.
In this study, we conduct a statistical analysis  quiescent galaxies and their environments at z=3.1. The aim of this work is to determine whether there is a systematic relationship between environment and quenching.

To trace the environmental density, we use two tracers, all the galaxies listed in the COSMOS2020 catalogue \citep{2021AAS...23721506W}, selected based on their photometric redshift (photo-z), and Lyman-$\alpha$ emitting galaxies (LAEs) from the One-hundred-square-degree DECam Imaging in Narrowbands (ODIN) survey \citep{2024ApJ...962...36L}. 
The first galaxy sample is meant not to be biased in terms of physical properties, while LAEs are characterised by a strong Lyman-$\alpha$ emission line. The Ly$\alpha$ emission line ($\lambda_{rest-frame}\simeq1216 \AA$) is the strongest recombination line of neutral Hydrogen and it has been used to trace star formation, AGN activity, and the gravitational collapse of dark matter halos at redshift $\geq$ 2 \citep{ouchi2020observations}. Given the wavelength of Ly$\alpha$ emission, LAEs can be detected up to $z$$\sim$ 6 using ground-based photometric surveys, making them an ideal tool for tracing the underlying density distribution and the large-scale structure over wide areas \citep[e.g.,][]{hu1996detection, ouchi2003subaru, gawiser2007lyalpha, kovavc2007clustering}. With the photo-z selected galaxies and LAEs we create two density maps that we used to study the environment of MQGs.

Some studies based on simulations have suggested that AGN feedback is the primary quenching mechanism for high-redshift MQGs \citep{kurinchi2024origin, hartley2023first}. \cite{kurinchi2024origin} found in the IllustrisTNG simulation that the number of merger events in a galaxy's history is not sufficient to distinguish between quiescent and star-forming galaxies, and that AGN feedback is required to quench the galaxy. By using the \textit{Magneticum}  simulation,  \cite{kimmig2025blowing} showed that, while AGN-driven gas removal quenches galaxies, quenching is also influenced by the environment. They found that for a galaxy to be quenched, it must reside in a significant underdensity, which prevents replenishment of gas. 
The aim of our study here is to observationally assess the extent of the environmental impact on galaxy quenching at $z=3.1$.

To look for quiescent galaxies we make use of archival photometric data in the optical, near- and mid-IR bands. The data used to select the galaxies and make density maps criteria are given in Sections \ref{odin} and \ref{cosmos}. The SED fitting analysis needed to determine the passive nature of MQGs is described in Section \ref{sec2}. We describe the selection of quiescent galaxies in Section \ref{sec3}. 
In section \ref{denscalc} we describe the density maps constructed using LAEs and photo-z selected galaxies.
A summary of the results is presented in Section \ref{results}. A discussion of our results and a comparison with previous observations and current simulations is presented in Section \ref{discussion}. Throughout this paper, a standard cold dark matter  cosmology is adopted with the Hubble constant $H_0 = 70$ km s$^{-1}$ Mpc$^{-1}$, the total matter density $\Omega_M = 0.3$ and the dark energy density $\Omega_\Lambda = 0.7$. All magnitudes are expressed in the AB system and $\log$ is the base 10 logarithm, if not otherwise specified.

\section{Data $\&$ methods}
 In this section, we describe the archival and the new observed data we employ to select galaxies at $z$ $\sim$ 3.1. These galaxies are used to make density maps to trace the large-scale environment as described in Sections~\ref{laemap}, \ref{vorlae} and \ref{denscalc}.
We also describe here the SED fitting method used to identify quiescent galaxies(QG) from the selected galaxies.

\subsection{The One-hundred-deg$^2$ DECam Imaging in Narrowbands Survey}\label{odin}
The One-hundred-square-degree DECam
Imaging in Narrowbands (ODIN) survey uses Lyman-alpha (Ly$\alpha$) emitting galaxies to trace the cosmic structures and overdensities at the following redshift slices: $z$$\sim$4.5, 3.1, 2.4 (t$_{age}$ =1.3, 2.0, 2.7 Gyr since the Big Bang, respectively).  The adopted \textit{N419}, \textit{N501}, and \textit{N673} filters have central wavelengths of 419 nm, 501 nm and 673 nm, respectively, which cover the Lyman-alpha line at the redshifts mentioned above and have full width at half maxima (FWHM) of 7.5 nm, 7.6 nm and 10.0 nm, respectively. 
LAEs are selected as narrow-band excess compared to the existing broad-band data from Hyper-Suprime-Cam Subaru.
ODIN aims to select more than 100,000 LAEs at the three cosmic epochs \citep{2024ApJ...962...36L}.

The ODIN survey reaches a 5$\sigma$ sensitivity of 25.7 mag (\textit{N501} filter) at $z \approx$ 3.1, with a redshift precision of $\Delta z$ = 0.062. The corresponding Lyman-$\alpha$ luminosity is $
L_{\text{Ly}\alpha} \approx 1.43 \times 10^{42} \text{ erg s}^{-1}$.  LAEs detected at $z\sim 3.1$ have median $log(SFR / M_{\odot} yr^ {-1}) = 0.77 \pm 0.2$ and $log(M / M.\odot) = 8.8 \pm 0.2$, and a median dust attenuation of $A_V = 0.5 \pm 0.1$. The LAE number density at this redshift is 0.20 arcmin$^ {-2}$.
The survey design and LAE selection in the ODIN survey are described in detail in \cite{2024ApJ...962...36L} and \cite{firestone2024odin}, respectively. In this paper, we use the LAEs selected in the
Cosmic Evolution Survey (COSMOS) field \citep{scoville2007cosmic} as described in \cite{firestone2024odin}.
We concentrate on redshift $\sim$ 3.1 in the COSMOS field because the ODIN observations at this redshift ($N501$ filter) have been fully acquired and reduced and the LAE density map has been published \citep{Ramakrishnan_2023}.

\subsection{Publicly available data }\label{cosmos}
The COSMOS field is a 2-square-degree field with the deepest available near-infrared wide-field coverage \citep{scoville2007cosmic,mccracken2012ultravista,2021AAS...23721506W,dunlop2023ultravista}.The COSMOS2020 photometric catalogue \citep{2021AAS...23721506W},  includes ultra-deep optical data from the Subaru Hyper-Suprime-Cam \citep{aihara2019second}, ultra-deep U-band CLAUDS data \citep{2019MNRAS.489.5202S},  ultra-deep near-infrared UltraVISTA data \citep{dunlop2023ultravista}, and \textit{Spitzer}IRAC data \citep{2021AAS...23721506W}.
The deep Ultra-VISTA data and Spitzer data are well suited to look for optically faint galaxies like QGs. We use the COSMOS2020 catalogue, which is produced using a profile fitting photometry method (known as the COSMOS ``FARMER'' catalogue).  Galaxies in COSMOS2020 are selected from a near-infrared
izY JHKs CHI-MEAN co-added detection image. Among these, the i-band is the most sensitive, reaching a 3$\sigma$ depth of approximately 27 mag, while the Ks-band is the shallowest at around 25 mag 3$\sigma$ depth. The Ks band plays a key role in selecting mass-complete samples up to z $\leq$ 4.5. Its increased depth-0.5 to 0.8 mag deeper than the COSMOS2015 catalogue—improves completeness, particularly for low-mass galaxies. Combined with IRAC data, these observation also enhance the detection of old, red, and dust-obscured sources.
For the brightest galaxies ($i <$ 22.5), the catalogue achieves a high redshift accuracy, with an outlier rate below 1$\%$. Even for fainter sources in the 25 $< i <$ 27 range, photometric redshift estimates maintain $\sim 4\%$ precision, though with a higher outlier rate of $\sim 20\%$.
Stellar mass completeness thresholds—defined as the redshift-dependent limits where samples remain $\sim 70\%$ complete—are established as follows:
    \begin{itemize}
    \item All galaxies: $M_{\star}$=$-3.23 \times 10^7 (1+z) + 7.83 \times 10^7 (1+z)^2$
    \item Star-forming galaxies:  $M_{\star}$= $-5.77 \times 10^7 (1+z) + 8.66 \times 10^7 (1+z)^2$
    \item Quiescent galaxies: $M_{\star}$= $-3.79 \times 10^8 (1+z) + 2.98 \times 10^8 (1+z)^2$
\end{itemize}
For further details of the photometry, SED fitting, mass function and luminosity function, we refer the reader to \cite{2021AAS...23721506W} and \cite{weaver2023cosmos2020}.  We also use photometric redshifts  estimated from LePhare \citep{arnouts1999}, provided in the catalogue.

\subsection{Selection of galaxies at redshift  $z$$\sim$3.1
}\label{sec1}
For the COSMOS2020 catalogue, \cite{2021AAS...23721506W} discuss the accuracy of photometric redshifts obtained using the LePHARE code \citep{arnouts1999} by comparing them to spectroscopic redshifts. They report that the LePHARE photometric redshifts show a high degree of consistency with the spectroscopic redshifts in the COSMOS field, achieving subpercent photometric redshift accuracy. The precision of photometric redshifts is approximately 0.01$\times$(1+$z$)
for galaxies with \textit{i-band} magnitude < 22.5. For fainter magnitudes, the precision decreases but remains better than 0.025$\times$(1+$z$) for \textit{i}<25. Given this good quality of photo-z in the COSMOS2020 catalogue, we use the LePHARE photo-z to select the sample of galaxies at $z \sim 3.1$.

We do not apply an additional mass cut in selecting galaxies from the COSMOS 2020 catalogue. From the  relations in Section \ref{cosmos}, at $z = 3.1$, the stellar mass completeness thresholds are:

\begin{itemize}
    \item All galaxies: $M_{\text{lim}} = 1.18 \times 10^9 \, M_{\odot}$
    \item Star-forming galaxies: $M_{\text{lim}} = 1.22 \times 10^9 \, M_{\odot}$
    \item Quiescent galaxies: $M_{\text{lim}} = 3.46 \times 10^9 \, M_{\odot}$
\end{itemize}

To ensure a balance between sample completeness and purity, we adopt a redshift range of $\Delta z=0.12$ centred at $z=3.1$. We select galaxies such that 68$\%$ of the redshift probability distribution function of our selected galaxies in confined to $3.004 \leq z_{phot}  \leq 3.16$.
With this criterion, we select 6431 galaxies. This selection ensures a clean sample with low uncertainty in the redshift range. 
To identify the quiescent galaxies from this sample, we perform the SED fitting using the photometric data available in the COSMOS2020 catalogue.
\subsection{SED fitting using BAGPIPES} \label{sec2}
The Bayesian Analysis of Galaxies for Physical Inference and Parameter EStimation \citep[BAGPIPES,][]{Carnall_2018,carnall2023surprising} is a state-of-the-art Python code to model galaxy spectra and fit spectroscopic and photometric observations simultaneously. In recent years, BAGPIPES has been used successfully in the literature to study 
MQGs \citep[see, e.g.,][]{jin2024cosmic}.
We use the BAGPIPES code to fit the COSMOS2020 SED of all 6431 $z\simeq 3.1$ galaxies and estimate their specific star formation rate (sSFR).

For our study, we perform spectral energy distribution (SED) fitting utilizing all 29 photometric bands available from the COSMOS2020 catalogue. We adopt a double power-law star formation history (SFH) model, which effectively captures the rising and declining phases of the SFH with two distinct power-law slopes. The functional form of the SFH is given by:
\begin{equation}   
{\rm SFR(t)} \propto \left[ \left( \frac{t}{\tau} \right)^\alpha + \left( \frac{t}{\tau} \right)^{-\beta} \right]^{-1} 
\end{equation}
where $\alpha$ and $\beta$ are the falling and rising slopes respectively, and $\tau$ is related to the time at which star formation peaks. \cite{Carnall_2018} conducted a comparative study of various star-formation history (SFH) parametrizations for quenched galaxies generated using the MUFASA simulation \citep{dave2016mufasa}.
They found that for quenched galaxies, the classical exponentially declining SFH model tends to overestimate stellar masses by an average of 0.06 dex ($\approx$ 15 $\%$), with around 80 $\%$ of the objects having overestimated stellar masses. Additionally, this model tends to underestimate the timing of formation and quenching events by approximately 0.4 Gyr on average. In contrast, the double power-law model showed significantly better agreement with observed data, yielding more accurate estimates for stellar mass, formation redshift, and quenching timescale. Specifically, the stellar-mass estimates are offset by only 0.02 dex, and the bias in the median quenching timescale is reduced to 100 Myr. The better agreement achieved when using the double power law model led us to adopt this model  for fitting our COSMOS2020 sample.

We follow the method described in \cite{carnall2023surprising} for the SED fitting of our galaxies using BAGPIPES.
We assume the 2016 updated version of the \cite{bruzual2003stellar} stellar population models \citep{chevallard2016modelling} using the MILES stellar spectral library \citep{sanchez2006medium,falcon2011updated} and the updated stellar evolutionary tracks of \cite{bressan2012parsec} and \cite{marigo2017new}. Nebular line and continuum emissions are implemented using an approach based on the CLOUDY photoionisation code. We fix the ionization parameter to U $=10^{-3}$. Dust attenuation is taken into account using the model proposed by \cite{salim2020dust}. This model has a variable slope that is parametrized with a power-law deviation, $\delta$, from the \cite{calzetti2000dust} model. The \cite{salim2020dust} model  has an additional free parameter to model the 2175 $\AA$ bump strength, B. We set a uniform prior on B from 0 to 5, where the Milky Way law has B = 3\citep{calzetti2000dust}. Given the good agreement between the LePhare photometric and spectroscopic redshifts, we fix the redshift prior in the fit to lie between the lower and upper limits that encompass 68$\%$ of the redshift distribution around the central value. Table \ref{sedparams} shows the priors and ranges for all the free parameters of the SED fitting. 
\begin{table*}[]
\caption{The priors of the free parameters of the BAGPIPES model we use to fit photometric data.}
\label{table:1}
\centering
\begin{tabular}{lccc}
\hline\hline
Component & Parameter & Range & Prior \\ \hline

General & Total stellar mass formed ($M_\ast / M_\odot$) & $(1, 10^{13})$ & Logarithmic \\ 
General & Stellar and gas-phase metallicity ($Z / Z_\odot$) & (0.2, 2.5) & Logarithmic \\ \hline

Star-formation history & Double-power-law falling slope ($\alpha$) & $(0.01, 1000)$ & Logarithmic \\ 
Star-formation history & Double-power-law rising slope ($\beta$) & $(0.01, 1000)$ & Logarithmic \\ 
Star-formation history & Double power-law turnover time ($\tau / \text{Gyr}$) & $(0.1, t_{\text{obs}})$ & Uniform \\ \hline

Dust attenuation & $V$-band attenuation ($A_V / \text{mag}$) & (0, 8) & Uniform \\ 
Dust attenuation & Deviation from \cite{calzetti2000dust} slope ($\delta$) & $(-0.3, 0.3)$ & Gaussian \\ 
Dust attenuation & Strength of 2175~\AA \ bump ($B$) & (0, 5) & Uniform \\ \hline

\end{tabular}
\label{sedparams}
\end{table*}

The output of BAGPIPES is the star formation history and a list of physical parameters of the galaxy, such as age of the stellar population, current star formation rate, star formation history, timescale of quenching, dust, metallicity, and stellar mass.  We show the SEDs of all the selected MQGs in the Appendix \ref{sedsbagpipes}.

\begin{figure}[]
    \centering
    \includegraphics[width=0.49\textwidth]{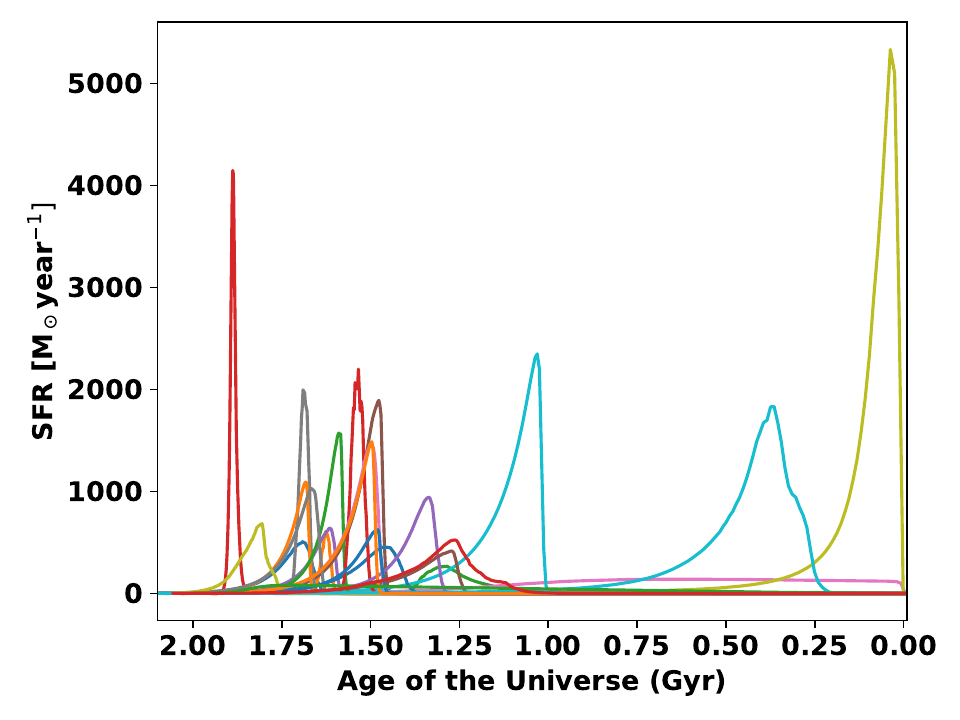}
    \caption{Star formation histories of the QGs obtained by fitting a double power law model using BAGPIPES.}
    \label{sedex}
\end{figure}

\subsection{Selection of quiescent galaxies} \label{sec3}
To separate star-forming and quiescent galaxies, we use a redshift-dependent cut in the specific star formation rate as has been widely applied in the literature \citep[e.g.,][]{pacifici2016timing,carnall2023surprising}. We define quiescent galaxies as those that have
 \begin{equation}
     ({\rm sSFR} + \sigma_{\rm sSFR}) \leq \frac{0.2{\rm yr}}{t_{age}}
    \label{eqsel}
\end{equation}

\begin{equation*}
  {\rm where  } ~~~ \frac{0.2}{t_{age}}=10^{-10.01}
    \label{eqsel1}
\end{equation*}
 where sSFR is the specific SFR, which is the SFR averaged over the last 100 Myr divided by stellar mass, $\sigma_{sSFR}$ is the uncertainty in the specific star formation rate, and $t_{age}$ is the age of the Universe at $z = 3.1$. 
 This threshold corresponds to the UVJ colour selection criteria for quiescent galaxies at $z < 0.5$ as introduced by \cite{2009ApJ...691.1879W}. 

We initially select 43 quiescent galaxy candidates by using Equation \ref{eqsel}. Then, we verify that their SEDs are reliable by visual inspection. In order to remove eventual contaminants, we exclude sources that are bright in the bands that correspond to rest-frame wavelength bluer than the Lyman break at $z\approx3.1$. 
We exclude objects that are bright in short-wavelength imaging, below the position of the Lyman break at z$\approx$3.1, at 3.7 $\mu$m. 
The SED of a MQG at $z=3.1$ is expected to be faint in the optical bands and bright in the JHK bands, showing a significant 
 Balmer break at 1.49 $\mu$m.  SED bright in the optical bands could imply that either the fitted redshift is incorrect or the star formation rate of the galaxy is not well constrained . Examples of rejected SEDs are shown in the Appendix \ref{reject}.

We also cross-match our QG candidates with the publicly available spectroscopic data. From the VIMOS Ultra Deep Survey \citep{lemaux2014vimos,le2013vimos} database, we were able to conclusively confirm the redshift of one of our QGs. The confirmed galaxy is at $z$=3.0661 and shows a clear CIV emission line \citep{lemaux2014vimos} in the spectrum that may be due to an AGN. 

To assess potential contamination from dusty star-forming galaxies, we examine the FIR/sub-mm/mm/radio flux of our QG candidates. To do so, we cross-match our QG candidates with \cite{jin2018super} and A3COSMOS \citep{liu2019automated} catalogues. \cite{jin2018super} provides photometry from Spitzer/MIPS, Herschel, SCUBA-2, AzTEC, MAMBO, and the VLA at 3 and 1.4 GHz. We find that all QG candidates in our sample have the combined 100$\mu$m-to-1.2 mm signal-to-noise ratio (S/N) $<$ 5 and non-detections (S/N $<$ 2) in all individual FIR and sub-millimetre bands. Thus our QG sample is free from significant sub-millimetre emission indicative of dust-obscured star formation. 

We have visually inspected both COSMOS-Web and Hubble image cutouts of the QGs to confirm that there are no bright contaminating sources that might affect their photometry. Among these galaxies, three have COSMOS-Web imaging available. None of the QGs are affected by contamination from nearby bright sources.

Our final sample is composed of 24 bonafide QGs. In Appendix \ref{sedsbagpipes} we show their SEDs, In Table \ref{table:galaxy_properties1} and \ref{table:galaxy_properties2}, we list their properties. We show the SFH of the QGs obtained using BAGPIPES in Figure~\ref{sedex}, where we can see that the SFHs have a starburst-like shape within the first 500~Myr followed by a rapid decline. As we can see from Table \ref{table:galaxy_properties1}, all our QG candidates have stellar masses larger than $10^{10.6} M_\odot$. Therefore, from now on we will refer to them as massive QGs (MQGs).     

According to \cite{weaver2023cosmos2020}, the outlier fraction for galaxies with magnitudes between 22 and 25 is approximately 4$\%$. Consequently, there is a 4$\%$ probability that MQGs may be misclassified at either lower or higher redshifts in the COSMOS 2020 catalogue, potentially leading to their exclusion from our sample.
Previous studies, such as \cite{leja2019measure} and \cite{Carnall_2018}, have demonstrated that standard SFH models such as exponentially declining SFH  may systematically underestimate stellar masses. 
Using LePhare sSFR in the COSMOS2020 catalogue, we successfully recover only 11 out of the 24 MQGs identified in our sample. Additionally, LePhare predicts five additional MQGs that are classified as star-forming by our Bagpipes SED fitting. A visual inspection of the SEDs of these five sources reveals that they exhibit strong optical emission, indicating that they are not true MQGs. This suggests that LePhare fails to properly constrain the SFR for these sources, leading to potential misclassification, as compared to BAGPIPES.

\subsection{Voronoi Monte Carlo map construction from the COSMOS2020 catalogue}

\label{laemap}
We adopt the Voronoi Monte Carlo (VMC) mapping technique to measure galaxy density across the entire COSMOS field within our redshift range.  This approach has been thoroughly examined and validated in previous studies in the literature \citep[eg.][]{shah2024identification,hung2020establishing,lemaux2018vimos,cucciati2018progeny,forrest2020extremely}, demonstrating its robustness and reliability. The major advantage of the VMC mapping is its ability to cover a wide dynamical range in densities, i.e. allowing the detection of both very high and very low densities, without assuming any prior on the shape of the density field. Hence, it is a non-parametric and scale-independent method.

\begin{figure*}[]
\begin{center}
\includegraphics[width=0.49\textwidth,trim=0 0 0 0, clip]{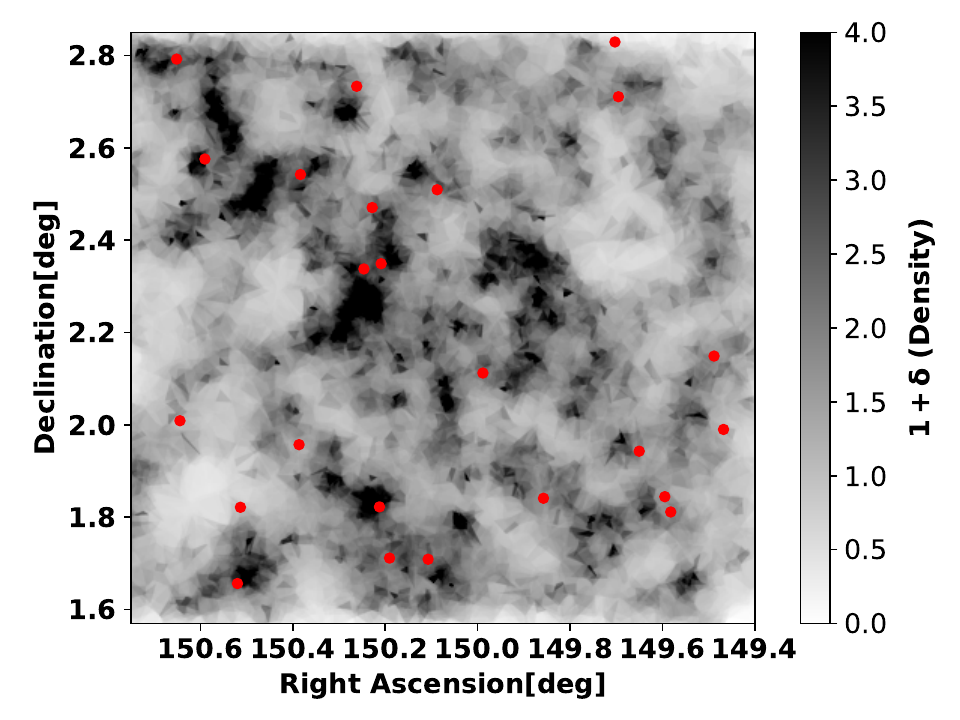}
\includegraphics[width=0.49\textwidth,trim=0 0 0 0, clip]{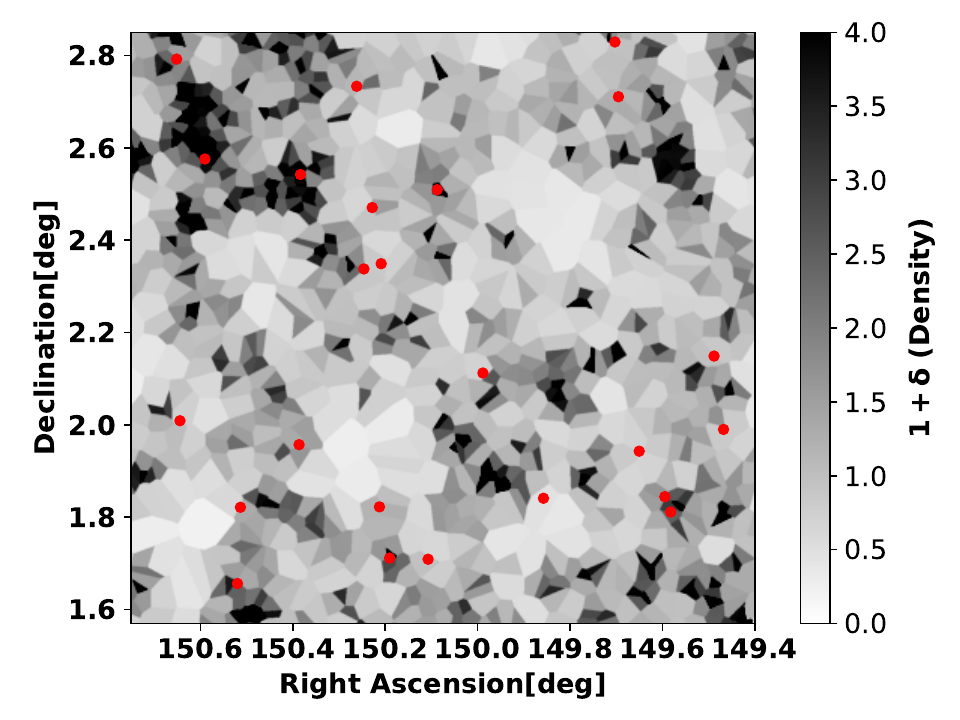}

\caption{Density map traced by the galaxies selected from the COSMOS2020 (Left) and the LAE (Right) catalogues, and obtained by using the Voronoi Monte Carlo Technique (Section \ref{sec1}). The colour bar represents the Normalised density calculated in Section~\ref{denscalc}. The red dots represent the MQGs. 
}
\label{vor1}
\end{center}
\end{figure*}

In the simple Voronoi tessellation method, a 2D plane is divided into a number of cells such that each cell is associated with an object. Each cell is defined as the collection of all the points closer to that object than to any other object. Therefore, more crowded regions will have smaller cells than less crowded regions. The density associated with each cell is calculated using the area of each cell as explained below. 
To take into account the variation of redshift within our redshift range, instead, we use the VMC technique described in \cite{hung2020establishing} representing an improvement over the standard Voronoi technique. The resultant VMC map is shown in the left panel of Figure~\ref{vor1}. The main steps are: 

\begin{itemize}
    \item We construct a series of overlapping redshift slices within our entire redshift range of 3.004 $\leq$ z $\leq$ 3.224. The width of each redshift slice (z-slice) is equal to the median uncertainty on the photometric redshift of our QG sample, i.e. 0.04.\\
    
    \item We assign a photometric redshift probability weight to each galaxy in each slice. The percentage of the redshift probability distribution function (z-pdf) lying in the z-slice is the weight of a galaxy for a given slice. For this, we assume that all galaxies have a Gaussian z-pdf. \\

    \item We generate a random number for each galaxy. The random number lies between the minimum weight and the maximum weight of all the galaxies for a slice. If the weight of a galaxy is greater than the random number, then it is accepted. Thus, we have a set of accepted galaxies for each z-slice.\\
    
    \item For each z-slice, we then generate a simple Voronoi map. The average of all 12 z-slice Voronoi maps is the final Voronoi map we use in this study. 
\end{itemize}

\subsection{Voronoi map construction from the ODIN-LAE catalogue}\label{vorlae}
In order to create the LAE Voronoi map, we assume that the redshifts of all our LAEs are fixed at $z=3.1$. This is because the narrow-band technique used to find the LAEs is very secure, with a redshift range of 0.062 \citep{firestone2024odin}.
Of the ODIN LAEs that have confident redshift measurements from DESI, $\sim 97\%$ have the correct redshift classification for being in the N501 filter. 
The LAE map and the structures discovered therein are published in \cite{Ramakrishnan_2023}. The right panel of Figure~\ref{vor1} shows the LAE Voronoi map. 

\subsection{Density estimation from the Voronoi maps}\label{denscalc}
Following the methodology described in the literature \citep[e.g.,][]{cucciati2018progeny,lemaux2018vimos,hung2020establishing}, we create a pixelated surface density map. This is done by populating the field with a uniform grid of points, where the grid spacing is set to $\approx$ 70~ckpc, which is much smaller than Voronoi cells. Each pixel within a Voronoi cell is assigned a surface density.
The surface density in a Voronoi cell (i, j), $\Sigma_{(i,j)}$,  is calculated using the area of a Voronoi cell ($A_{v}$) as follows:
\begin{equation}
    \Sigma_{(i,j)}=\dfrac{1}{A_{v}}
\end{equation}
The Normalised local density, (1+$\delta$), in a Voronoi cell (i, j) is then estimated as:
\begin{equation}
     \log(1+\delta)=\log\left(1+\dfrac{\Sigma_{(i,j)}-\tilde{\Sigma}}{\tilde{\Sigma}}\right)
     \label{denseq}
\end{equation}
where $\Sigma_{(i,j)}$ is the given Voronoi cell’s density and $\tilde{\Sigma}$ is the median density of all the Voronoi cells in the redshift slice (Figure~\ref{vor1}).  As discussed in various studies, these local density estimates correlate well with other density metrics and accurately trace known overdensity structures. 
We repeat this process for the LAE Voronoi map to construct an LAE density map. As described in \citep{Ramakrishnan_2023}, the size of a pixel is 120$\sim$ckpc on a side for the LAE density map.

\section{Results}\label{results}

\subsection{Number density and physical properties of quiescent galaxies}\label{nummass}
We identify 24 QG candidates at z$\approx$3.1. One of them has been previously identified by \cite{2023ApJ...945L...9I}. 
The remaining 23  are promising new photometric candidates. Table \ref{table:galaxy_properties1} shows their best fit physical properties.
\begin{figure*}
\begin{center}
\includegraphics[width=0.47\textwidth]{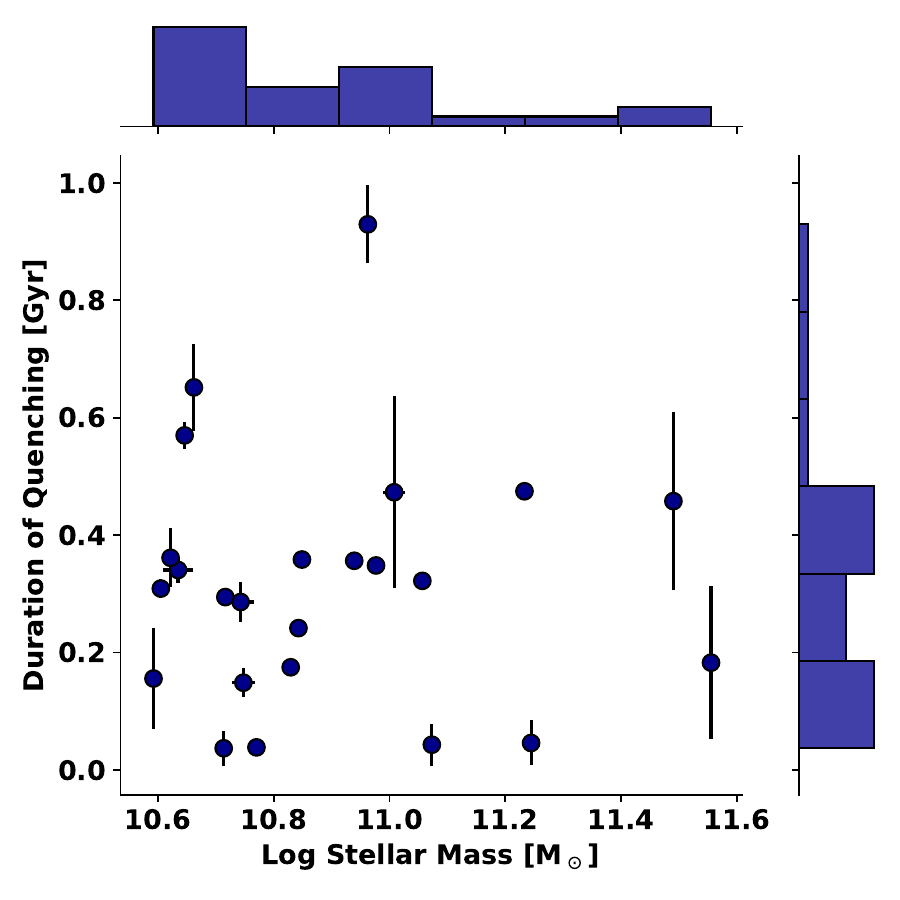}
\includegraphics[width=0.49\textwidth]{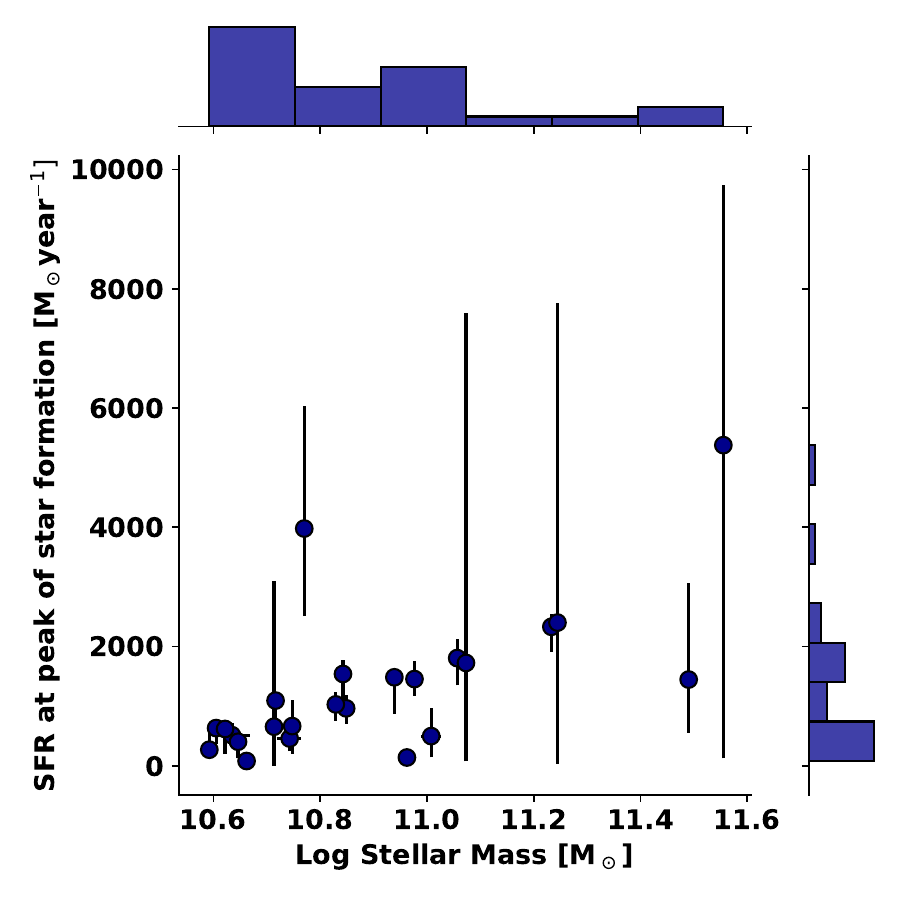}
\caption{Left: The quenching timescale of the MQGs (as defined in Equation~\ref{eqquench}) as a function of the stellar mass estimated using BAGPIPES. The median duration of quenching is $\sim$ 350~Myr and is consistent for all MQGs. The log stellar mass ($M_{\odot}$) is always $\geq$ 10.6. Right: The star formation rate at the peak of the SFH of the MQG as a function of the stellar mass. The peak SFR increases linearly with mass indicating a stronger starburst for more massive galaxies.
}
\label{figquenchmass}
\end{center}
\end{figure*}

The total volume surveyed, excluding star masks, covered by the COSMOS2020 catalogue, is 8.71$\times$10$^5$~cMpc$^3$. 
Therefore, we calculate the number density of MQGs as $ 2.87^{+0.70}_{-0.57} \times 10^{-5}
$ galaxies~cMpc$^{-3}$, consistent with the observed number densities reported in the literature \citep{straatman2014substantial,schreiber2018near,merlin2019red,carnall2023surprising}, as well as estimated from simulations, such as IllustrisTNG300 \citep{valentino2020quiescent}. We calculate the uncertainties following the method described in \cite{1986ApJ...303..336G}, which provides a procedure for estimating Poisson uncertainties in the regime of low number statistics.

The BAGPIPES code gives the SFH that best reproduce the observed photometry of a galaxy as an output.  The BAGPIPES code estimates the time of formation ($t_{form}$) and the time of quenching ($t_{quench}$) for a galaxy. $t_{form}$ is defined as 
\begin{equation*}
t_{\text{form}} = \frac{\int_{0}^{t_{\text{obs}}} t \, \text{SFR}(t) \, dt}{\int_{0}^{t_{\text{obs}}} \text{SFR}(t) \, dt},
\end{equation*}
\begin{equation*}
t_{\text{quench}} \text{ is defined as the time when } \frac{t \, \text{SFR}(t)}{M_{\text{formed}}} < 0.1.
\end{equation*}

Following the prescription of \cite{Carnall_2018}, we define the quenching timescale as follows:
\begin{equation}
    T_q=t_{quench}-t_{form}
\end{equation}\label{eqquench}
 Simulations have shown that quenching timescale is one of the critical parameters that can be used to distinguish between various quenching mechanisms\citealp{wright2019quenching,wetzel2013galaxy,walters2022quenching}. For instance, stellar feedback from supernovae can quench a galaxy within 0.1 Gyr\citep{ceverino2009role}. In contrast, merger-driven quenching is expected to have a median delay time of approximately 1.5 Gyr, though this timescale can vary significantly \citep{rodriguez2019mergers}. As shown in Figure \ref{figquenchmass}, 19 out of our 24 MQGs exhibit quenching timescales within a relatively narrow range of 300–500 Myr.  The similarity in quenching timescales among the majority of our galaxies suggests they may have been influenced by a common quenching mechanism. Simulations predict that mechanisms such as overconsumption\citep{walters2022quenching}, ram pressure stripping\citep{steinhauser2016simulations}, stellar feedback\citep{ceverino2009role}, and AGN feedback\citep[e.g,][]{hirschmann2017synthetic}operate on short timescales and are capable of rapidly quenching galaxies.

One MQG exhibits a notably longer quenching timescale ($\approx$ 0.9 Gyr) compared to the rest of the sample. The SED and SFH of this galaxy are presented in Figure \ref{SED_310229}. Unlike the rest, it could be possible that this galaxy did not undergo a strong starburst in its star-formation history and has an SFR peak of only $\sim$ 150 M$_{\odot}$yr$^{-1}$. This particular galaxy is an outlier and may have a different kind of quenching mechanism. It lies in a slight higher local density than average density, but not in a significant overdensity. It is possible that, in this case, environmental quenching processes with long timescales \cite{mao2022revealing} may have conditioned its evolution.

The quenching timescale in our analysis is derived from SED fitting of photometric data using a double power-law star formation history model, which may introduce some uncertainty in its estimation. Simulations based on the SIMBA framework have shown that this model can lead to offsets in the quenching timescale of up to 100 Myr, whereas the commonly used exponentially declining model can introduce biases as large as 400 Myr\citep{Carnall_2018}. Future studies that combine both spectroscopic and photometric observations will be crucial for placing tighter constraints on quenching timescales.

\subsection{Density maps at $z=3.1$} \label{densmaps}
The photo-z and LAE Voronoi maps are presented in Figure~\ref{vor1}. Protoclusters are overdense structures in the early universe that eventually evolve into present-day galaxy clusters. For the identification of these structures, we employ SEP \citep{Barbary2016}, a Python implementation of the SExtractor software \citep{1996A&AS..117..393B}. The number of detected structures is highly dependent on the chosen detection threshold (DETECT$\_$THRESH) and minimum area (DETECT$\_$MINAREA). According to \cite{chiang2013ancient}, a protocluster at redshift 3 is expected to have a minimum effective radius of 5 cMpc and a minimum halo mass of 10$^{12}$ M$_\odot$ to evolve into a cluster of M$_{halo} >10^{14} M_\odot$ at $z=0$.

\begin{figure*}
\begin{center}
\includegraphics[width=0.495\textwidth]{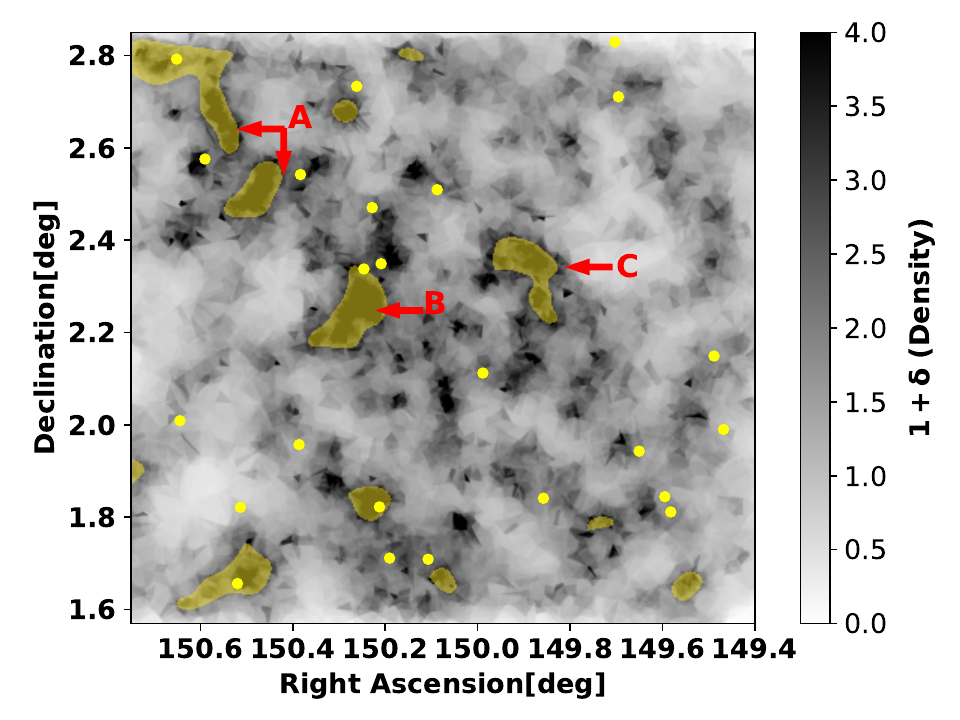}
\includegraphics[width=0.495\textwidth]{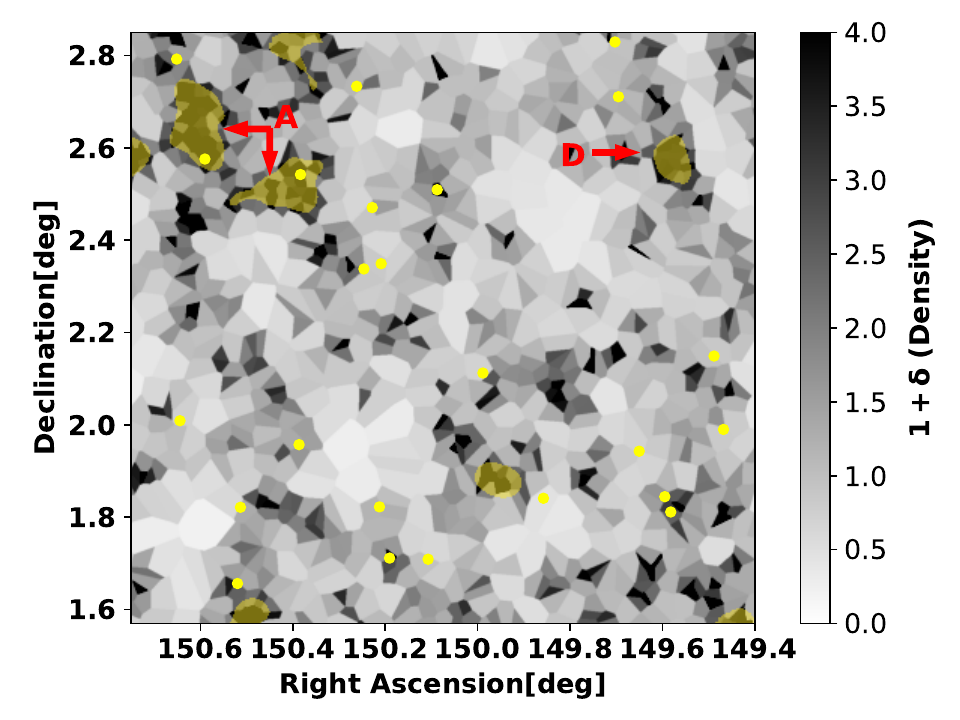}
\caption{Yellow shaded regions represent structures identified by running Source Extractor in the COSMOS2020 (left) and in the LAE (right) maps.  Yellow dots represent MQGs.  The colour bar scales the log(1+$\delta$) density for both maps as in Figure \ref{vor1}. 
The A, B, C, and D letters indicate some major overdensity groups discovered in either one or both maps. }
\label{vorcom}
\end{center}
\end{figure*}

\cite{Ramakrishnan_2023} developed specific SExtractor criteria for the ODIN survey to identify protoclusters. They find that a detection threshold (DETECT$\_$THRESH) of 4.5$\sigma$ and a minimum area (DETECT$\_$MINAREA) of 3000 pixels (corresponding to approximately 40 cMpc$^2$) results in a contamination rate of 20$\%$. We adopt these criteria from \cite{Ramakrishnan_2023} to identify protoclusters from the surface density maps, to have a consistent detection procedure for the ODIN and COSMOS2020 map. For clarity, we use the term "protocluster" exclusively for overdense structures identified using this method, as it has been optimized for the detection of protoclusters. 
  In Figure~\ref{vorcom}, we show the structures detected in the LAE and COSMOS2020 Voronoi maps, respectively.

There are some important differences between the LAE and the COSMOS2020 map. To discuss them in detail, we name some of the protoclusters in capital letters. The large protoclusters at $\delta$ $\simeq 2.3^\circ$ (B and C in Figure~\ref{vorcom}) detected in the COSMOS2020 map are not detected in the LAE map. The large protocluster at $\alpha$ $=149.6^\circ$ detected in the LAE map (D) is not detected in the COSMOS2020 map. The group of overdensities at $\alpha$ $\simeq 150.6^\circ$ (A) is detected in both maps.  The reason for the difference in the appearance of the overdensities/protoclusters may be due to the redshift ranges. The photo-z galaxies are selected to be in the range 3.004 $\leq z \leq$ 3.224, while the LAEs lie in a narrower range of $\Delta z =0.062$, centred at $z=3.124$ \citep{2024ApJ...962...36L}.
Therefore, the new structures in the photo-z map could be just outside the very narrow range of the LAE map. 
The structures detected in the density maps are promising photometric protocluster candidates. Follow-up spectroscopy of the constituent galaxies of these protoclusters could confirm their exact redshift and 3-D shape.

Only 6 of 24 MQGs are located within protocluster candidates discovered in either the LAE or photoz maps, which represents just 25$\%$ of the MQG candidates. This suggests that processes that typically occur in dense environments may not always be necessary for galaxy quenching. This includes process that occur in hot ICM such as ram-pressure stripping \citep{boselli2022ram}, thermal evaporation \citep{cowie1977thermal}, and processes that have a higher probability of occurring in dense environments such as galaxy harassment and strangulation.

\begin{figure*}
\begin{center}
\includegraphics[width=0.495\textwidth,trim=0 0 0 0, clip]{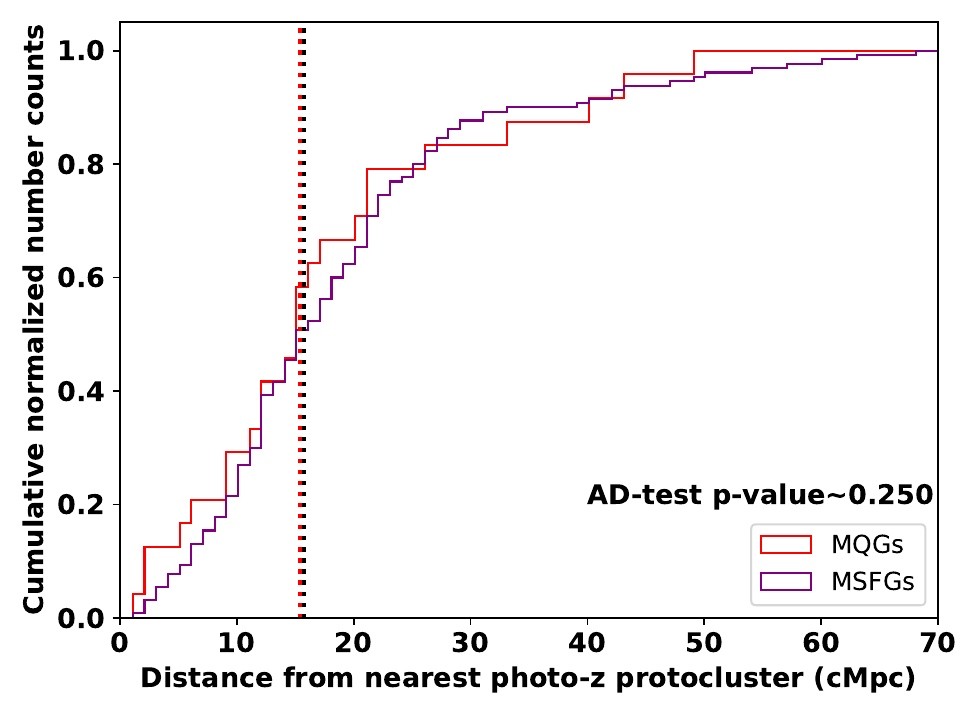}
\includegraphics[width=0.495\textwidth,trim=0 0 0 0, clip]{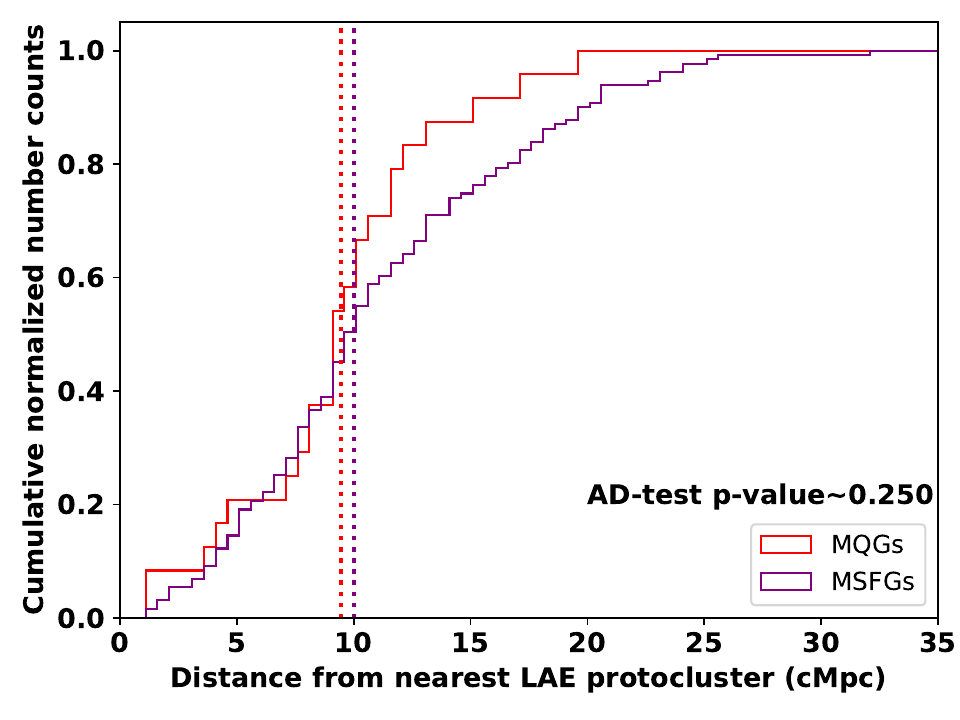}
\caption{Left: The cumulative distribution of the distances of MQGs, and massive star-forming galaxies to the photo-z protocluster centers. Right: The cumulative distribution of the distances of MQGs, and massive star-forming galaxies to the LAEs protocluster centers The dotted lines show the median distribution for each type of galaxy. The p-value of the Anderson-Darling test comparing MQGs with the all galaxy distribution is also displayed. MQGs and MSFGs are selected at the same stellar mass limit. }
\label{cumhist1}
\end{center}
\end{figure*}

\begin{figure*}[]
\begin{minipage}{0.5\textwidth}    
     \centering
         \centering
             \includegraphics[width=\textwidth]{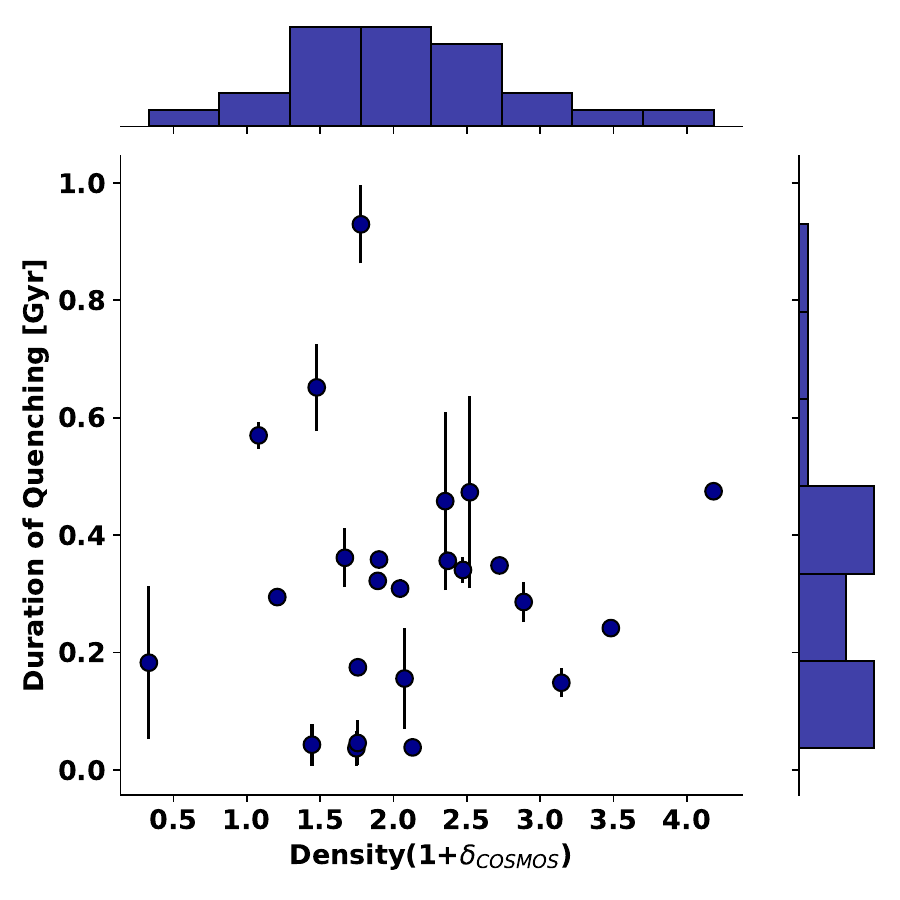}
\end{minipage}
\begin{minipage}{0.5\textwidth}    
     \centering
         \centering
             \includegraphics[width=\textwidth]{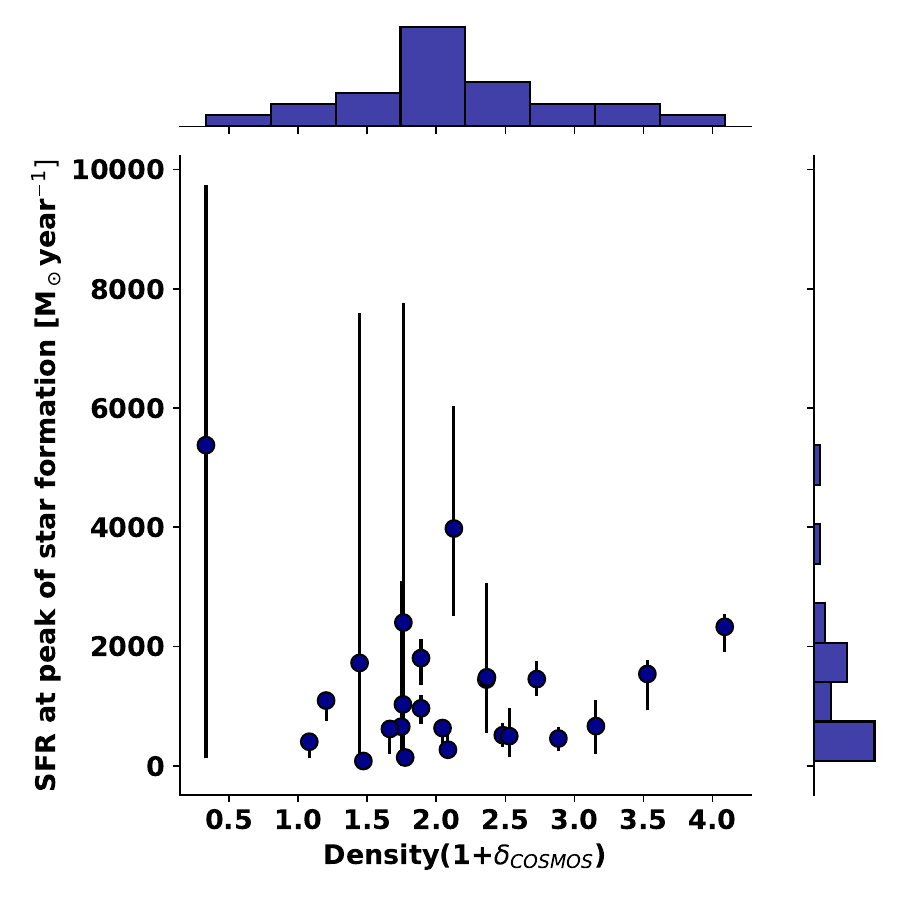}
\end{minipage}
    \caption{Left: Quenching timescale as a function of density from the COSMOS2020 VMC map. Right: SFR at peak of star formation as a function of density from the COSMOS2020 VMC map .}
    \label{coegg}
\end{figure*}

\begin{figure*}[h]
\begin{minipage}{0.5\textwidth}    
     \centering
         \centering
             \includegraphics[width=\textwidth]{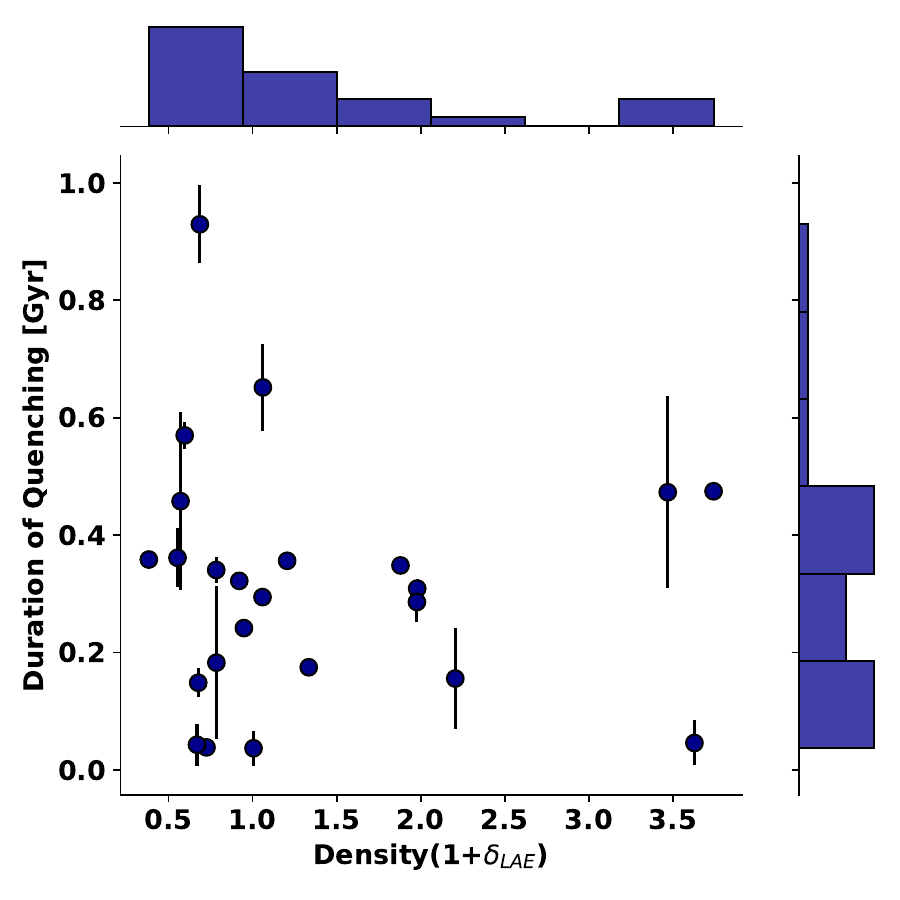}
\end{minipage}
\begin{minipage}{0.5\textwidth}    
     \centering
         \centering
             \includegraphics[width=\textwidth]{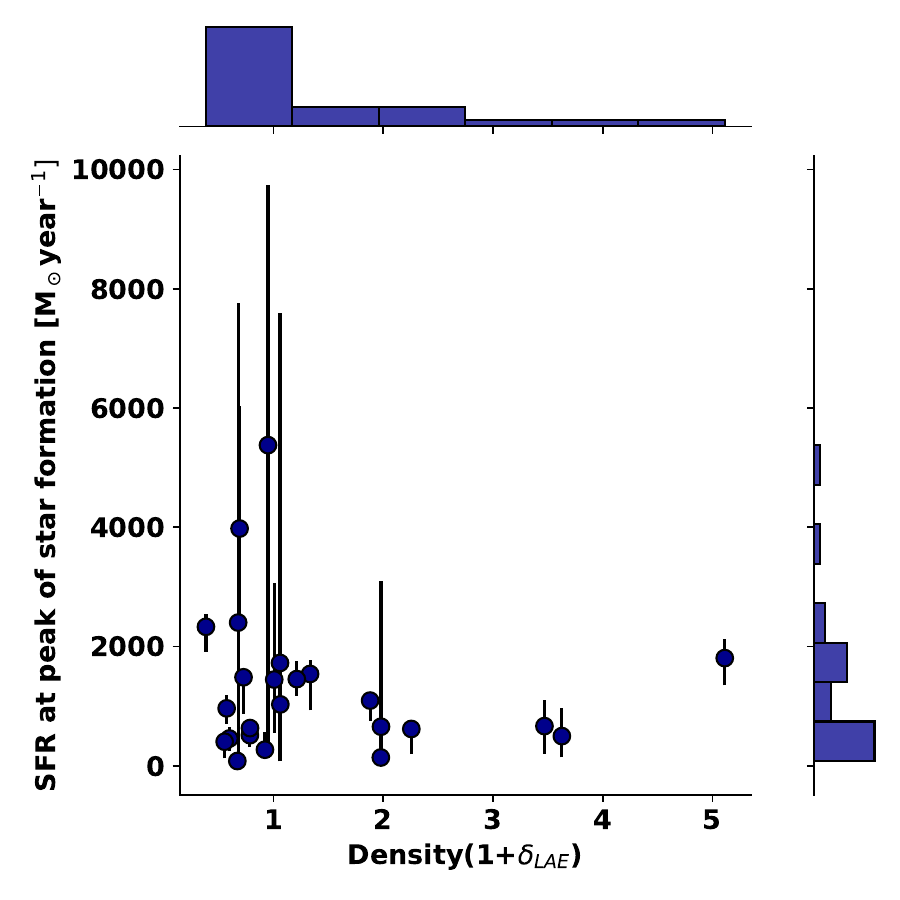}
\end{minipage}
    \caption{Left: Quenching timescale as a function of density from the LAE density map. Right: SFR at the peak of star formation as a function of density from the LAE density map.}
    \label{coegglae}
\end{figure*}

We compare the location of MQGs with that of the massive star-forming galaxies(MSFGs) in relation to the protocluster candidates. MSFGs are defined as star-forming galaxies with a stellar mass above 10$^{10.6} M_{\odot}$. We perform this comparison at a fixed limit of stellar mass, so that mass segregation doesn't bias our comparison. Figure \ref{cumhist1} displays the two distributions. To determine whether the MQGs and MSFGs follow the same distribution, we perform the Anderson-Darling test. In both cases we get a high p-value $\sim$ 0.25 which suggests that the two samples are statistically consistent with being drawn from the same distribution. This further indicates that MQGs follow the same distribution with respect to protoclusters as MSFGs.  We interpret this as evidence that environmental processes, which are more common in protoclusters, are not the dominant quenching mechanism. We also perform this test without fixing the stellar mass limits , obtaining a similar result. 
Figures \ref{coegg} and \ref{coegglae} show the quenching time scale and the SFR$_{peak}$ vs. the environmental density in the photo-z map and the LAE map, respectively. We show that neither the quenching timescale nor the SFR at the peak of the SFH depends on the local density.  This suggests that the mechanisms responsible for star formation in MQGs do not significantly depend on the environment.

\subsection{Quiescent fraction and the environment}\label{fraccor}
The quiescent fraction is described as the proportion of MQGs to the count of MSFGs present within a specified density bin. We perform this calculation at a fixed stellar mass $M > 10^{10.6} M_{\odot}$ to ensure a fair comparison between massive star-forming galaxies and massive quenched galaxies , rather than introducing a mass bias in the results.

\begin{figure*}[]
\begin{minipage}{0.5\textwidth}    
     \centering
         \centering
             \includegraphics[width=\textwidth]{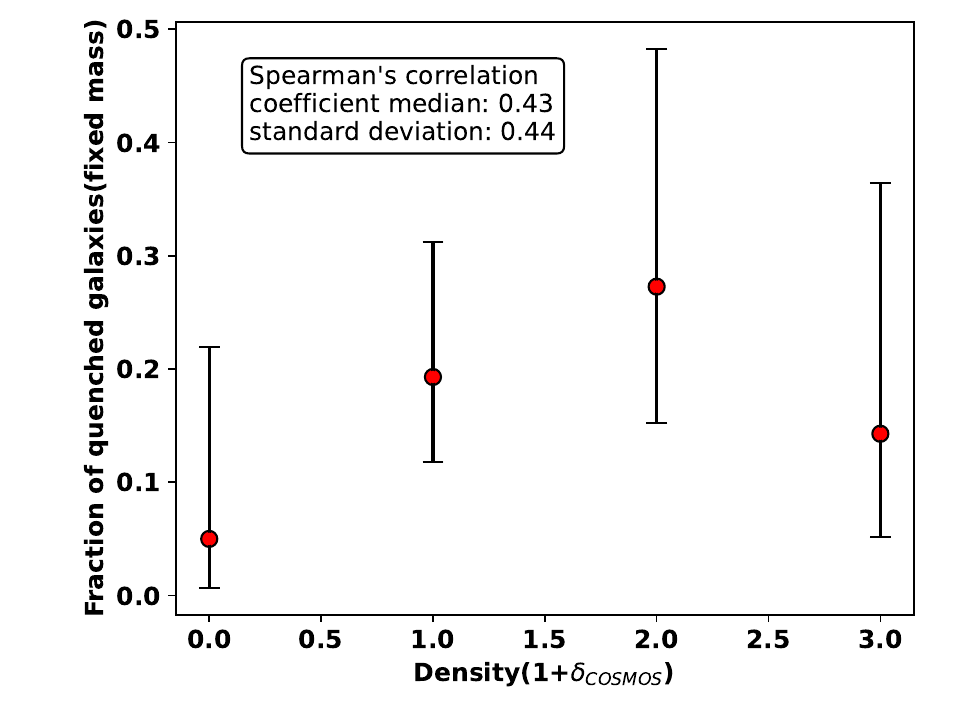}
\end{minipage}
\begin{minipage}{0.5\textwidth}    
     \centering
         \centering
             \includegraphics[width=\textwidth]{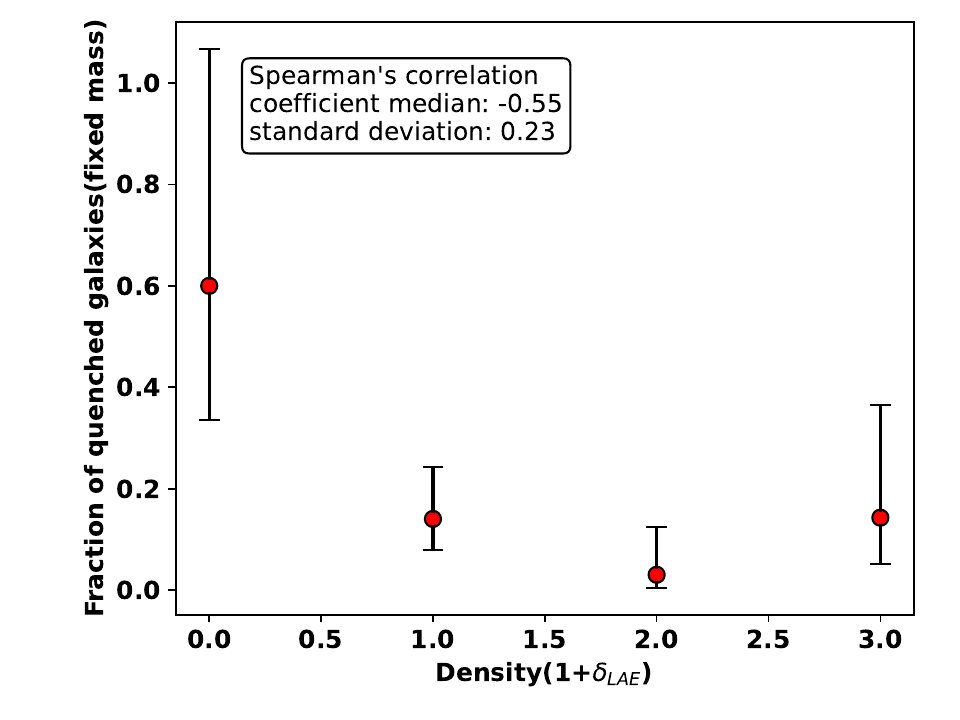}
\end{minipage}
    \caption{Left: Evolution of the quenched fraction with the local density, log(1+$\delta$) in the COSMOS2020 density map. Right: Evolution of the quenched fraction with the local density, log(1+$\delta$) in the LAE density map. }
    \label{bootstrap}
\end{figure*}

 Figure \ref{bootstrap} shows the quiescent fraction as a function of local density. By local density, we refer to the density within the pixel of the Voronoi map at which the MQG resides. The resolution of the photo-z density map and the LAE density map is 70~ckpc~pixel$^{-1}$ and 120~ckpc~pixel$^{-1}$ respectively.  The uncertainty in the quiescent fraction is obtained using the method described by \cite{1986ApJ...303..336G}, which provides convenient tables and approximate formulas to calculate confidence limits based on Poisson and binomial statistics. This method is widely used in astronomy for estimating uncertainties with small-number statistics. We use equations 10 and 12 from \cite{1986ApJ...303..336G} to calculate these uncertainties, which provide the upper ($\lambda_u$) and lower limit ($\lambda_l$) to a measured quantity according to Poisson statistics as follows:
\begin{equation}
    \lambda_u = n + S \sqrt{n + 1} + \frac{S^2 + 2}{3}
\end{equation}
\begin{equation}
    \lambda_l = n \left( 1 - \frac{1}{9n} - \frac{S}{3 \sqrt{n}} \right)^3
\end{equation}
Here n is the number of observed events (or counts). It is the observed value for which the confidence limits are calculated. S is a Poisson parameter whose value is listed in \cite{1986ApJ...303..336G} for different confidence levels.

To quantify the observed correlation between the quenched fraction and the local density in both the COSMOS2020 map and the LAE map, we employ the Spearman correlation coefficient. Taking into account the uncertainty in the quiescent fraction, we calculate a weighted Spearman coefficient. This involves bootstrapping the quenched fraction within the bounds of the uncertainty 3000 times. For each sample, we compute the Spearman correlation coefficient. Finally, we determine the mean and standard deviation of the computed coefficients, providing a robust measure of the correlation while accounting for the associated uncertainties.  We obtain a mean Spearman correlation coefficient of $\approx$ 0.55 and a standard deviation of $\approx$ 0.4. A Spearman coefficient of 0.55 signifies a weak or no correlation. To robustly confirm a correlation, the coefficient should exceed 0.7, while a value below 0.4 generally suggests the absence of correlation. In our Monte Carlo simulations, we found that 43$\%$ of the runs resulted in a correlation coefficient below 0.4, leading us to conclude that there is no dependence on the local density. To validate these statistical results, we perform simulations, generating samples with varying levels of correlation, strong and weak. Our findings indicate that a weak correlation can still be observed even with a small sample size of 24 when using weighted Spearman coefficients. Further details, including a more thorough examination of the correlation, can be found in the Appendix \ref{siulationtng}.

The independence of the quiescent fraction of the local density is in agreement with the analysis in Section~\ref{densmaps}, and further suggests that quenching mechanisms in MQGs may not be driven only by environmental processes.

 \subsection{Neighbor count} \label{nei}
 \begin{table}
    \centering
    \caption{Number of neighbours of MQG, and all galaxies}
    \label{table1}
    \begin{tabular}{lccc}
        \hline\hline
        Radius [cMpc] & QG & ALL \\
        \hline
        0.3 & $1.18 \pm 0.33$ & $1.12 \pm 0.44$  \\
        0.5 & $1.50 \pm 0.70$ & $1.42 \pm 0.71$  \\
        1.0 & $2.58 \pm 1.50$ & $2.45 \pm 1.45$  \\
        3.0 & $13.90 \pm 4.19$ & $11.90 \pm 5.90$  \\
        6.0 & $36.50 \pm 11.6$ & $29.6 \pm 13.35$  \\
        \hline
    \end{tabular}
    \tablefoot{The table shows the median$\pm$ standard deviation of the number of neighbours for circles of different radii in Mpc for MQG, all galaxies.}
\end{table}
In the local universe, environmental processes in galaxy groups and clusters play a significant role as quenching mechanism \citep[]{2006PASP..118..517B}. To investigate the environment of MQGs at different scales, we counted the number of nearest neighbours within 2-D projected distances ranging from 0.1 to 6~cMpc for all photo-z selected galaxies and all MQGs. Table~\ref{table1}
 presents the median values.
 
Figure~\ref{6nnall} in the Appendix \ref{nn} shows the distribution of the number of neighbours for MQGs and for all galaxies in the COSMOS2020 catalogue. 

From the median values, we conclude that MQGs and all galaxies have similar median values for their number of neighbours. This indicates that MQGs do not have a higher probability of residing in groups or overdense environments at various scales than star-forming galaxies. Secular quenching mechanisms, such as AGN feedback and stellar feedback, which do not necessarily depend on the environment, may play a dominant role in the quenching of galaxies at this epoch at
$z=3.1$.
\subsection{Filaments}
Filaments traced by LAEs are recognized as substantial reservoirs of cold gas that can sustain star formation in galaxies. Consequently, it is anticipated that star-forming galaxies will be located on or close to these filaments. Conversely, a negative correlation is expected between MQGs and filaments, as galaxies situated near filaments benefit from a continuous supply of cold gas conducive to star formation. 
 \cite{Ramakrishnan_2023} identified these filaments within the COSMOS field using LAEs and the Discrete Persistent Structure Extractor (DisPerSE) code \citep{sousbie2011persistent}.
\begin{figure*}
\begin{center}
\includegraphics[width=0.495\textwidth]{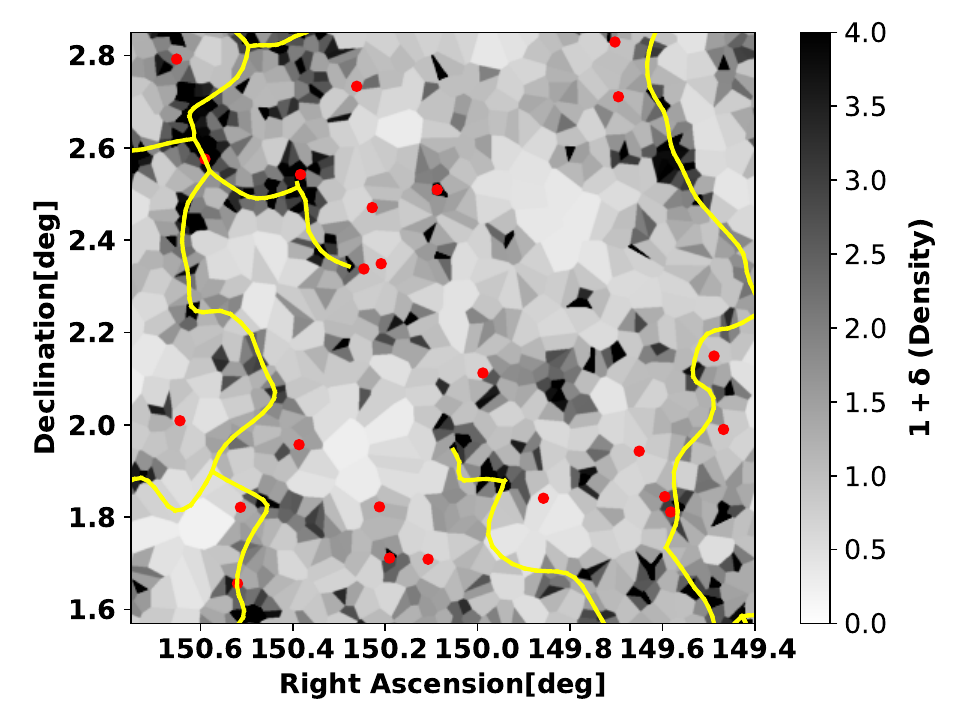}
\includegraphics[width=0.495\textwidth]{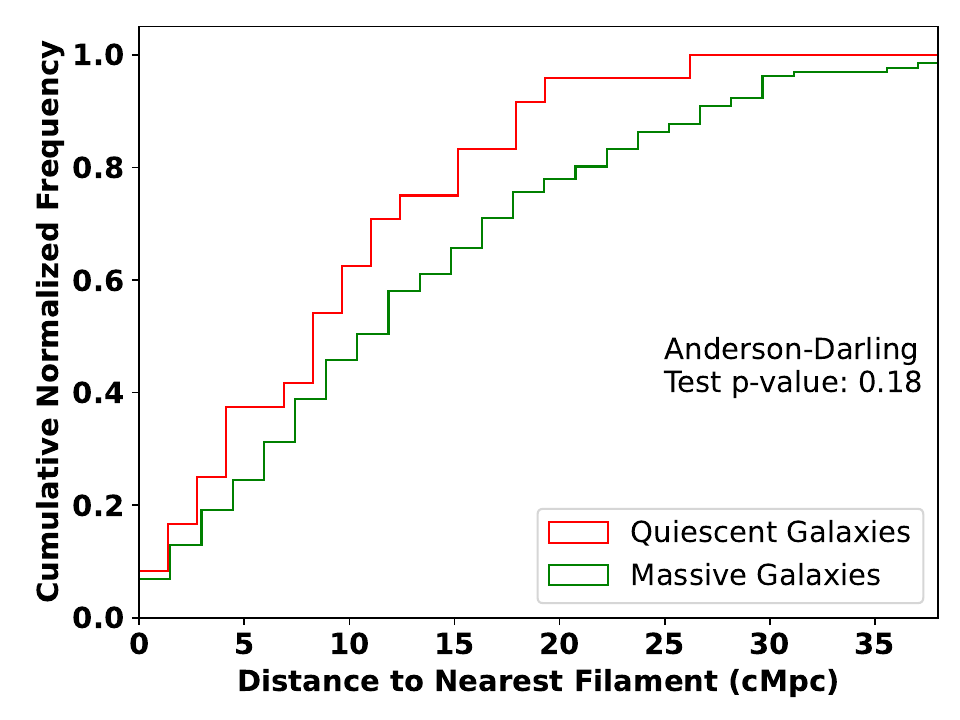}
\caption{Left: Filaments (yellow lines) discovered using DisPerSE by \cite{Ramakrishnan_2023}, overplotted on the LAE density map. The red points represent the MQGs. Right: Cumulative distribution of the distances to the nearest filaments for MQGs and MSFGs.}
\label{filaments}
\end{center}
\end{figure*}

It is crucial to note that the redshift range of the LAEs utilized to trace these filaments is very narrow, specifically $\Delta  z =0.062$. We do not have the precise spectroscopic redshift for MQGs and photo-z selected galaxies from COSMOS2020 catalogue. The redshift range for our MQGs and photo-z selected galaxies is wider and spans 3.1 $\pm$ 0.12. As a result, we can only compute the two-dimensional projected distances of our photo-z sample galaxies relative to the filaments. These distances are measured in arcseconds, where at z=3.1, 1\arcsec\ corresponds to 31.27~ckpc. We compare the distributions of MQGs and the MSFGs at 
z=3.1 using the Anderson-Darling test.
Figure \ref{filaments} displays the filaments detected in the COSMOS field with LAEs, illustrating the spatial distribution of the massive quiescent galaxies relative to these filaments.

We apply the Anderson-Darling test to assess whether the distribution of MQGs and massive star-forming galaxies with respect to the filaments differs significantly. The test yields a p-value of 0.18, which exceeds common significance thresholds (0.05, 0.01). Consequently, we find no statistically significant difference between the two samples. The median distances of MQGs and photo-z galaxies from the filaments are 9 and 11 cMpc, respectively. This similarity suggests that the probability of a galaxy to be quenched does not depend on its proximity to a filament. We also find some MQGs lying in proximity to gas-rich LAE filaments. Specifically, 9 out of 24 MQGs are located within 5 cMpc of a filament. Filaments are typically rich in cold gas and represent significant reservoirs for star formation \citep{martizzi2019baryons}. Thus, for these MQGs, their proximity to such gas-rich filaments implies that processes involving gas heating, such as virial shock heating or AGN feedback, may be actively suppressing star formation by hindering cold gas accretion.

\section{Discussion}\label{discussion}
In this work, we have searched for quiescent galaxies at $z\sim3.1$ 
, studied their physical properties and investigated their relationship with the environment.
We discovered 24 MQGs, characterized by log sSFR$<-10.01$ in a 2~deg$^2$ area of the COSMOS field. . 
The number density of the MQGs is $ 2.87^{+0.70}_{-0.57} \times 10^{-5}
$ galaxies/cMpc$^{3}$ and it is consistent with observations and simulations in the literature \citep{valentino2020quiescent}. 

We have characterised the environment in the COSMOS field, considering two independent tracers, LAEs and the general galaxy population of the COSMOS2020 catalogue, using the Voronoi Tessellation.
While there are overdense structures that are common in both these maps, there are large overdense structures which are discovered only in either of the two maps. 
The reason for this difference could be the redshift precision being different between the 2 samples. We select galaxies from the COSMOS2020 catalogue on the basis of photometric redshifts. The range of this photo-z is $\approx$ 3 to 3.2. This may lead to an overdensity detection which is not co-spatial with LAE-traced structures whose redshift range is very narrow ($\Delta$z=0.06, z$\approx$3.1).

Our study identifies two protoclusters, which we call B and C, that are present in the photo-z map but absent in the LAE map. \cite{forrest2023elentari} have discovered protoclusters at positions similar to B and C at a slightly higher redshift (z $\sim$ 3.32), named S3 and S1 in their analysis. We propose that these two protoclusters detected in our photo-z map could also be the front-end extension of the S1 and S3 protoclusters. Our LAE observations may not have fully captured certain overdensities due to the inherently narrow redshift coverage ($\Delta z \approx 0.062$) of the narrow-band selection method. Beyond observational constraints, however, there may be important astrophysical factors underlying the discrepancies observed between structures traced by LAEs and those identified using broader photometric selection methods. Specifically, the population of LAEs is known not to fully overlap with luminous Lyman-break galaxies (LBGs), MQGs, or dusty star-forming galaxies, as each population tends to occupy distinct evolutionary stages, host halo masses, and environmental contexts.

Previous studies \citep[e.g.,][]{overzier2016realm,shi2020detailed,shimakawa2017direct} have highlighted that LAEs typically trace younger, lower-mass galaxies associated with gas-rich environments or filamentary structures. In contrast, photometrically-selected galaxy populations, such as MQGs and dusty galaxies, often trace more massive halos and potentially more evolved, dynamically mature environments\citep[e.g.,][]{shi2019galaxies,calvi2023bright}. Consequently, different galaxy populations may reveal overdensities at different evolutionary stages or spatial scales within the cosmic web. Such differences emphasize the complexity of interpreting galaxy overdensities identified by varying selection methods and caution against the assumption that these methods sample identical physical environments or evolutionary phases.

Further multi-wavelength observational campaigns, particularly spectroscopic follow-up across various galaxy populations, are necessary to fully elucidate the astrophysical reasons behind these differences and to clarify the connection between different galaxy populations and their underlying large-scale structures.

LAEs are detected using the narrow-band excess created by the Lyman-$\alpha$ emission line. The absorption of the resonant Lyman-alpha line in a dusty and highly overdense region could make it harder to detect LAEs. The Lyman-alpha IGM transmission is a complex function, significantly influenced by the geometry and density of the IGM \citep{2019MNRAS.486.1882G,2020MNRAS.491.3266G}. This small-scale effect may obscure certain parts of structures, such as extremely overdense and dusty centres of some protoclusters, but it is unlikely to miss entire protoclusters, as seen in our maps. The photoz density map complements the LAE density map because the photoz-selected galaxies have a precision similar to that of the MQGs.

Our analysis shows that only 20$\%$ of our MQGs are located in protoclusters. The spatial distribution of quiescent galaxies relative to protocluster centres is similar to that of the general galaxy population. Therefore, MQGs do not show a higher tendency to be found in protoclusters than other types of galaxies. As noted in Section \ref{nei}, MQGs have a comparable number of neighbours within the entire range from  0.5 to 6 Mpc as the general galaxy population, i.e. they show a very similar environment. This suggests that environmental mechanisms prevalent in protoclusters or galaxy groups, such as ram pressure stripping, galaxy-galaxy interactions, strangulation, harassment, tidal stripping, and thermal evaporation, are not the sole factors responsible for quenching MQGs.
 
Some studies using IllustrisTNG simulation have further shown that environmental processes like mergers are not responsible for quenching galaxies at high redshift. \citet{kurinchi2024origin} found that merger events do not significantly distinguish between quiescent and star-forming galaxies in the IllustrisTNG300 simulation and are unlikely to drive galaxy quenching at z>3. Their study concluded that quenched galaxies have experienced longer and more intense AGN activity, releasing larger amounts of thermal energy than star-forming galaxies, prior to quenching.

Our findings are broadly consistent with the emerging picture that high-redshift galaxy quenching operates on rapid timescales and is driven primarily by internal processes. Recent work by \citet{Tacchella+2022} reinforced the notion that stellar feedback, alongside contributions from active galactic nuclei, can dramatically curtail star formation during the early phases of galaxy assembly. Such feedback mechanisms appear to be especially effective in massive halos, such as those of MQGs.

We find that MQGs do not exhibit a correlation with local peaks of environmental density, as illustrated in Figure~\ref{bootstrap}. The quenched fraction remains nearly constant across a wide range of environmental densities, suggesting that quenching mechanisms may operate independently of local density variations. Additionally, 7 of 24 MQGs are located in regions of exceptionally high density, characterised by $\delta>2$. 
Since high-density regions also serve as reservoirs of cold gas, mechanisms to heat the gas are necessary to prevent further accretion of cold gas that would rejuvenate the MQGs. In this context, AGN feedback is a plausible quenching mechanism.

AGN feedback is a secular quenching mechanism that can quench galaxies in a wide range of environments.
AGN outflows cause the expulsion of gas from the gravitational bounds of their host galaxies, consequently reducing the subsequent star formation activity \citep{debuhr2012galaxy,combes2017agn}. Additionally, AGNs can heat the cool gas in the interstellar medium, further preventing the formation of new stars \citep{croton2006many,man2018star,zinger2020ejective}. Recently, several observational studies using JWST and ALMA have found signs of gas outflows driven by AGN in massive quiescent galaxies at z$>3$ in morphological and spectroscopic observations \citep{2022ApJ...935...89K,2022ApJ...929...53I,park2024widespread,2024arXiv240503744}.    \cite{park2024widespread,2024IAUS..377....3G,2023arXiv230806317D,2022ApJ...929...53I,2022ApJ...935...89K} have found multiphase outflows and neutral gas outflows in massive quiescent galaxies along with evidence of AGN. \cite{2022ApJ...929...53I} find AGN activity ubiquitously in massive quiescent galaxies up to $z=5$.  
 \citet{nanayakkara2025formation} presented spectroscopic evidence linking high-redshift quiescent galaxies to powerful AGN activity, consistent with the scenario that short-lived starbursts—and subsequent quenching—may be triggered or accelerated by the presence of an AGN.
 In this context, quenching due to AGN feedback is consistent with our observation that MQGs are not correlated with protoclusters and have a similar number of neighbours as other types of galaxies.

In their study using the Magneticum simulation, \cite{kimmig2025blowing} also found that galaxies are quenched before cosmic noon ($z=3.42$) primarily because of the removal of gas by AGN feedback. However, \cite{kimmig2025blowing} predict the quenching process is also influenced by the 
environment. 
They showed that for a galaxy to be quenched, it must reside in a significant underdensity, preventing the replenishment of cold gas. In contrast to their study, we do not observe any preference for MQGs to reside in underdense regions. Instead, MQGs exhibit a random distribution with respect to protoclusters, suggesting that the effect of the environment may not be significant for quenching at a redshift of approximately 3.1.

Filaments traced by LAEs are reservoirs of neutral gas that can rejuvenate galaxies even if a starburst has consumed all of the gas. We do not observe any anti-correlation between the position of MQGs and filaments. In fact,  25$\%$ of our MQGs occur within 5 cMpc of gas-rich filaments. This could be an indication of the action of physical processes that heat the gas near the galaxies, preventing the accretion of cold gas. This could further point to AGN heating the gas around the galaxies.  \citet{Kalita+2022} highlighted the importance of internal processes such as AGN feedback and morphological stabilisation, where bulge growth can help suppress gas inflows and ultimately halt star formation. This aligns well with our inference that several of our massive quiescent galaxies reside in environments with potentially abundant gas (e.g., near filaments, in protoclusters) yet remain quenched, presumably due to heating or outflows that circumvent fresh gas accretion.

By performing SED fitting, we have obtained the star-formation history of our MQGs. It is typically characterised by a starburst phase, followed by a quenching period, with a timescale shorter than 500~Myr (Figure~\ref{figquenchmass}). The uniformity in the duration of quenching of these galaxies suggests that they were quenched by the same mechanism 
and physical processes which are unlikely strictly related to the environment.  This quenching mechanism would include a massive starburst that consumes gas rapidly, along with a mechanism to prevent the inflow of cold gas to maintain star formation.
A rapid quenching phase has been observed in other studies in the literature, such as \cite{whitaker2012large}; \cite{wild2016evolution}; \cite{schreiber2018near}; \cite{Carnall_2018}; \cite{d2020typical};
\cite{carnall2023surprising}. Quenching timescales obtained from the simulations \citep[e.g.,][]{wright2019quenching,wetzel2013galaxy,walters2022quenching} can help to distinguish between the quenching mechanisms. In general, we can distinguish quenching between fast-quenching mechanisms ($\sim$ 0.1 Gyr) and slow-quenching ($\sim$1 Gyr) mechanisms. Furthermore, any trend in the quenching timescale with environment density can indicate the dependence of the quenching mechanism on the local environment. Environmental processes that are typically much slower are hence unlikely to quench these galaxies \citep{mao2022revealing}.

From Figure~\ref{coegg} we can see that the time it takes to quench a galaxy is not strongly correlated with the density of the local environment. A similar result is found in Figure \ref{coegglae} where we have shown the quenching timescale with respect to LAE densities. No correlation between quenching timescale and the environment indicates that the environment alone may not be the significant driver of quenching. Similarly, on the right panel of Figures~\ref{coegg} and \ref{coegglae} we can see that there exists no correlation between the SFR$_{peak}$ and the environment density. We conclude that the SFH and hence the processes that quenched these MQGs do not have a strong dependence on the environment.

In order to spectroscopically confirm the redshift and obtain a more precise constraint on the SFH and stellar mass, acquiring spectra is essential. JWST \textit{Nirspec} follow-up observations can be used to spectroscopically confirm MQGs at $z\sim3.1$. Follow-up observations with the JWST \textit{Nirspec} will be instrumental in identifying AGN-driven outflows as well as neutral gas outflows, which may play a role in quenching star formation. Furthermore, morphological analysis using NIRCam would provide valuable insight into the impact of mergers within dense environments.
\section{Conclusions}
We present a detailed analysis of 24 quiescent galaxies identified at $z\approx3.1$ within the $2-deg^2$ COSMOS field, using deep photometric data from the COSMOS2020 catalogue. These galaxies are characterized by high stellar masses exceeding $10^{10.6}M_\odot$.  Our spectral energy distribution fitting reveals that these MQGs exhibit remarkably consistent star-formation histories, characterized by intense early starburst episodes rapidly followed by quenching within short timescales of $\leq$ 400 Myr. The similarity in quenching timescales across the entire sample strongly suggests the presence of a universal and highly efficient quenching mechanism operating at cosmic noon.

By examining galaxy environments using Voronoi-based density maps constructed independently from photometrically-selected COSMOS2020 galaxies and ODIN-selected LAEs, we find no significant correlation between galaxy quenching parameters (quenching duration, quenched fraction, or timing) and local environmental density. Furthermore, MQGs do not exhibit a preferential distribution with respect to protoclusters or cosmic filaments compared to similarly massive star-forming galaxies, indicating that purely environmental processes alone—such as ram pressure stripping, galaxy mergers, or harassment—may not dominate galaxy quenching at this epoch. The presence of MQGs within environments expected to be rich in cold gas, such as protoclusters and filaments, further implies that simple gas exhaustion might not be the sole contributor to quenching, and gas heating mechanisms induced by active galactic nuclei  or stellar feedback could also play significant roles.

Our findings support the hypothesis that internal mechanisms-such as AGN feedback, stellar feedback-driven gas heating, virial shock heating, or morphological quenching—could play a more dominant role in galaxy quenching at these redshifts. Future spectroscopic confirmation of the precise redshifts of these galaxies will be crucial to securely associate them with identified protoclusters and filaments. Additionally, detailed studies of gas dynamics and gas ionization states within and around MQGs using integral-field spectroscopy (e.g., MUSE) and interferometric observations (e.g., ALMA) across diverse environments will be essential. Such observations can confirm the availability, temperature, and ionization state of the gas reservoirs, clarifying the contributions of gas heating and exhaustion. Spectroscopic follow-up to confirm the presence and strength of AGN activity in MQGs will further test our proposed hypothesis regarding dominant internal quenching mechanisms operating during this critical epoch of galaxy evolution.

\begin{acknowledgements}
      AS and LG acknowledge the FONDECYT regular project number 1230591 for financial support. EG and KSL acknowledge support from NSF grants AST-2206222, AST-2206705, and AST-2408359.
     Based on observations at Cerro Tololo Inter-American
 Observatory, NSF's NOIRLab (NOIRLab Prop. ID 2020B0201; PI: K.-S. Lee), which is managed by the Association of Universities for Research in Astronomy (AURA) under a cooperative agreement with the National Science Foundation.
      LG also gratefully acknowledges financial support from ANID - MILENIO - NCN2024$\_$112 and the ANID BASAL project FB210003.
      This work made use of the computer server RAGNAR at Universidad Andres Bello.
\end{acknowledgements}

\bibliographystyle{aa}
\bibliography{aa52406-24}

\begin{thebibliography}{105}
\expandafter\ifx\csname natexlab\endcsname\relax\def\natexlab#1{#1}\fi

\bibitem[{Aihara {et~al.}(2019)Aihara, AlSayyad, Ando, Armstrong, Bosch, Egami,
  Furusawa, Furusawa, Goulding, Harikane, {et~al.}}]{aihara2019second}
Aihara, H., AlSayyad, Y., Ando, M., {et~al.} 2019, \pasj, 71, 114

\bibitem[{Arnouts {et~al.}(1999)Arnouts, Cristiani, Moscardini, Matarrese,
  Lucchin, Fontana, \& Giallongo}]{arnouts1999}
Arnouts, S., Cristiani, S., Moscardini, L., {et~al.} 1999, \mnras, 310, 540

\bibitem[{Barbary(2016)}]{Barbary2016}
Barbary, K. 2016, Journal of Open Source Software, 1, 58

\bibitem[{{Bertin} \& {Arnouts}(1996)}]{1996A&AS..117..393B}
{Bertin}, E. \& {Arnouts}, S. 1996, \aaps, 117, 393

\bibitem[{Boselli {et~al.}(2022)Boselli, Fossati, \& Sun}]{boselli2022ram}
Boselli, A., Fossati, M., \& Sun, M. 2022, Astronomy and Astrophysics Reviews,
  30, 3

\bibitem[{{Boselli} \& {Gavazzi}(2006)}]{2006PASP..118..517B}
{Boselli}, A. \& {Gavazzi}, G. 2006, \pasp, 118, 517

\bibitem[{Bressan {et~al.}(2012)Bressan, Marigo, Girardi, Salasnich, Dal~Cero,
  Rubele, \& Nanni}]{bressan2012parsec}
Bressan, A., Marigo, P., Girardi, L., {et~al.} 2012, \mnras, 427, 127

\bibitem[{Bruzual \& Charlot(2003)}]{bruzual2003stellar}
Bruzual, G. \& Charlot, S. 2003, \mnras, 344, 1000

\bibitem[{Calvi {et~al.}(2023)Calvi, Castignani, \&
  Dannerbauer}]{calvi2023bright}
Calvi, R., Castignani, G., \& Dannerbauer, H. 2023, \aap, 678, A15

\bibitem[{Calzetti {et~al.}(2000)Calzetti, Armus, Bohlin, Kinney, Koornneef, \&
  Storchi-Bergmann}]{calzetti2000dust}
Calzetti, D., Armus, L., Bohlin, R.~C., {et~al.} 2000, \apj, 533, 682

\bibitem[{Carnall {et~al.}(2023)Carnall, McLeod, McLure, Dunlop, Begley,
  Cullen, Donnan, Hamadouche, Jewell, Jones, {et~al.}}]{carnall2023surprising}
Carnall, A., McLeod, D., McLure, R., {et~al.} 2023, \mnras, 520, 3974

\bibitem[{Carnall {et~al.}(2018)Carnall, McLure, Dunlop, \&
  Davé}]{Carnall_2018}
Carnall, A.~C., McLure, R.~J., Dunlop, J.~S., \& Davé, R. 2018, \mnras, 480,
  4379–4401

\bibitem[{Casey(2016)}]{casey2016ubiquity}
Casey, C.~M. 2016, \apj, 824, 36

\bibitem[{Ceverino \& Klypin(2009)}]{ceverino2009role}
Ceverino, D. \& Klypin, A. 2009, \apj, 695, 292

\bibitem[{Chevallard \& Charlot(2016)}]{chevallard2016modelling}
Chevallard, J. \& Charlot, S. 2016, \mnras, 462, 1415

\bibitem[{Chiang {et~al.}(2013)Chiang, Overzier, \&
  Gebhardt}]{chiang2013ancient}
Chiang, Y.-K., Overzier, R., \& Gebhardt, K. 2013, \apj, 779, 127

\bibitem[{Cicone {et~al.}(2014)Cicone, Maiolino, Sturm, Graci{\'a}-Carpio,
  Feruglio, Neri, Aalto, Davies, Fiore, Fischer, {et~al.}}]{cicone2014massive}
Cicone, C., Maiolino, R., Sturm, E., {et~al.} 2014, \aap, 562, A21

\bibitem[{Combes(2017)}]{combes2017agn}
Combes, F. 2017, Frontiers in Astronomy and Space Sciences, 4, 10

\bibitem[{Cowie \& Songaila(1977)}]{cowie1977thermal}
Cowie, L.~L. \& Songaila, A. 1977, \nat, 266, 501

\bibitem[{Croton {et~al.}(2006)Croton, Springel, White, De~Lucia, Frenk, Gao,
  Jenkins, Kauffmann, Navarro, \& Yoshida}]{croton2006many}
Croton, D.~J., Springel, V., White, S.~D., {et~al.} 2006, \mnras, 365, 11

\bibitem[{Cucciati {et~al.}(2018)Cucciati, Lemaux, Zamorani, Le~F{\`e}vre,
  Tasca, Hathi, Lee, Bardelli, Cassata, Garilli,
  {et~al.}}]{cucciati2018progeny}
Cucciati, O., Lemaux, B., Zamorani, G., {et~al.} 2018, \aap, 619, A49

\bibitem[{Dav{\'e} {et~al.}(2016)Dav{\'e}, Thompson, \&
  Hopkins}]{dave2016mufasa}
Dav{\'e}, R., Thompson, R., \& Hopkins, P.~F. 2016, \mnras, 462, 3265

\bibitem[{DeBuhr {et~al.}(2012)DeBuhr, Quataert, \& Ma}]{debuhr2012galaxy}
DeBuhr, J., Quataert, E., \& Ma, C.-P. 2012, \mnras, 420, 2221

\bibitem[{{D'Eugenio} {et~al.}(2023){D'Eugenio}, {Perez-Gonzalez}, {Maiolino},
  {Scholtz}, {Perna}, {Circosta}, {Uebler}, {Arribas}, {Boeker}, {Bunker},
  {Carniani}, {Charlot}, {Chevallard}, {Cresci}, {Curtis-Lake}, {Jones},
  {Kumari}, {Lamperti}, {Looser}, {Parlanti}, {Rix}, {Robertson}, {Rodriguez
  Del Pino}, {Tacchella}, {Venturi}, \& {Willott}}]{2023arXiv230806317D}
{D'Eugenio}, F., {Perez-Gonzalez}, P., {Maiolino}, R., {et~al.} 2023, arXiv
  e-prints, arXiv:2308.06317

\bibitem[{{Dressler}(1980)}]{1980ApJ...236..351D}
{Dressler}, A. 1980, \apj, 236, 351

\bibitem[{Dunlop {et~al.}(2023)Dunlop, Bowler, Franx, Fynbo, McCracken,
  Milvang-Jensen, \& Moneti}]{dunlop2023ultravista}
Dunlop, J.~S., Bowler, R.~A., Franx, M., {et~al.} 2023, A Decade of ESO
  Wide-field Imaging Surveys (surveys2023, 10

\bibitem[{D’Eugenio {et~al.}(2020)D’Eugenio, Daddi, Gobat, Strazzullo,
  Lustig, Delvecchio, Jin, Puglisi, Calabr{\'o}, Mancini,
  {et~al.}}]{d2020typical}
D’Eugenio, C., Daddi, E., Gobat, R., {et~al.} 2020, \apjl, 892, L2

\bibitem[{Fabian(2012)}]{fabian2012observational}
Fabian, A.~C. 2012, \araa, 50, 455

\bibitem[{Falc{\'o}n-Barroso {et~al.}(2011)Falc{\'o}n-Barroso,
  S{\'a}nchez-Bl{\'a}zquez, Vazdekis, Ricciardelli, Cardiel, Cenarro, Gorgas,
  \& Peletier}]{falcon2011updated}
Falc{\'o}n-Barroso, J., S{\'a}nchez-Bl{\'a}zquez, P., Vazdekis, A., {et~al.}
  2011, \aap, 532, A95

\bibitem[{Firestone {et~al.}(2024)Firestone, Gawiser, Ramakrishnan, Lee,
  Valdes, Park, Yang, Ciardullo, Artale, Benda, {et~al.}}]{firestone2024odin}
Firestone, N.~M., Gawiser, E., Ramakrishnan, V., {et~al.} 2024, \apj, 974, 217

\bibitem[{Forrest {et~al.}(2020)Forrest, Annunziatella, Wilson, Marchesini,
  Muzzin, Cooper, Marsan, McConachie, Chan, Gomez,
  {et~al.}}]{forrest2020extremely}
Forrest, B., Annunziatella, M., Wilson, G., {et~al.} 2020, \apjl, 890, L1

\bibitem[{Forrest {et~al.}(2023)Forrest, Lemaux, Shah, Staab, McConachie,
  Cucciati, Gal, Hung, Lubin, Cassar{\`a}, {et~al.}}]{forrest2023elentari}
Forrest, B., Lemaux, B.~C., Shah, E., {et~al.} 2023, \mnras, 526, L56

\bibitem[{Gawiser {et~al.}(2007)Gawiser, Francke, Lai, Schawinski, Gronwall,
  Ciardullo, Quadri, Orsi, Barrientos, Blanc, {et~al.}}]{gawiser2007lyalpha}
Gawiser, E., Francke, H., Lai, K., {et~al.} 2007, \apj, 671, 278

\bibitem[{{Gehrels}(1986)}]{1986ApJ...303..336G}
{Gehrels}, N. 1986, \apj, 303, 336

\bibitem[{{Girelli} {et~al.}(2019){Girelli}, {Bolzonella}, \&
  {Cimatti}}]{2019A&A...632A..80G}
{Girelli}, G., {Bolzonella}, M., \& {Cimatti}, A. 2019, \aap, 632, A80

\bibitem[{Glazebrook {et~al.}(2022)Glazebrook, Nanayakkara, Marchesini,
  Kacprzak, \& Jacobs}]{glazebrook2022revelation}
Glazebrook, K., Nanayakkara, T., Marchesini, D., Kacprzak, G., \& Jacobs, C.
  2022, Proceedings of the International Astronomical Union, 18, 3

\bibitem[{{Glazebrook} {et~al.}(2024){Glazebrook}, {Nanayakkara}, {Marchesini},
  {Kacprzak}, \& {Jacobs}}]{2024IAUS..377....3G}
{Glazebrook}, K., {Nanayakkara}, T., {Marchesini}, D., {Kacprzak}, G., \&
  {Jacobs}, C. 2024, in Early Disk-Galaxy Formation from JWST to the Milky Way,
  ed. F.~{Tabatabaei}, B.~{Barbuy}, \& Y.-S. {Ting}, Vol. 377, 3--8

\bibitem[{Glazebrook {et~al.}(2017)Glazebrook, Schreiber, Labb{\'e},
  Nanayakkara, Kacprzak, Oesch, Papovich, Spitler, Straatman, Tran,
  {et~al.}}]{glazebrook2017massive}
Glazebrook, K., Schreiber, C., Labb{\'e}, I., {et~al.} 2017, \nat, 544, 71

\bibitem[{Gunn \& Gott~III(1972)}]{gunn1972infall}
Gunn, J.~E. \& Gott~III, J.~R. 1972, \apj, 176, 1

\bibitem[{{Gurung-L{\'o}pez} {et~al.}(2019){Gurung-L{\'o}pez}, {Orsi},
  {Bonoli}, {Baugh}, \& {Lacey}}]{2019MNRAS.486.1882G}
{Gurung-L{\'o}pez}, S., {Orsi}, {\'A}.~A., {Bonoli}, S., {Baugh}, C.~M., \&
  {Lacey}, C.~G. 2019, \mnras, 486, 1882

\bibitem[{{Gurung-L{\'o}pez} {et~al.}(2020){Gurung-L{\'o}pez}, {Orsi},
  {Bonoli}, {Padilla}, {Lacey}, \& {Baugh}}]{2020MNRAS.491.3266G}
{Gurung-L{\'o}pez}, S., {Orsi}, {\'A}.~A., {Bonoli}, S., {et~al.} 2020, \mnras,
  491, 3266

\bibitem[{Hartley {et~al.}(2023)Hartley, Nelson, Suess, Garcia, Park,
  Hernquist, Bezanson, Nevin, Pillepich, Schechter,
  {et~al.}}]{hartley2023first}
Hartley, A.~I., Nelson, E.~J., Suess, K.~A., {et~al.} 2023, \mnras, 522, 3138

\bibitem[{Hirschmann {et~al.}(2017)Hirschmann, Charlot, Feltre, Naab, Choi,
  Ostriker, \& Somerville}]{hirschmann2017synthetic}
Hirschmann, M., Charlot, S., Feltre, A., {et~al.} 2017, \mnras, 472, 2468

\bibitem[{Hu \& McMahon(1996)}]{hu1996detection}
Hu, E.~M. \& McMahon, R.~G. 1996, \nat, 382, 231

\bibitem[{Hung {et~al.}(2020)Hung, Lemaux, Gal, Tomczak, Lubin, Cucciati,
  Pelliccia, Shen, Le~F{\`e}vre, Wu, {et~al.}}]{hung2020establishing}
Hung, D., Lemaux, B., Gal, R., {et~al.} 2020, \mnras, 491, 5524

\bibitem[{{Ito} {et~al.}(2022){Ito}, {Tanaka}, {Miyaji}, {Ilbert}, {Kauffmann},
  {Koekemoer}, {Marchesi}, {Shuntov}, {Toft}, {Valentino}, \&
  {Weaver}}]{2022ApJ...929...53I}
{Ito}, K., {Tanaka}, M., {Miyaji}, T., {et~al.} 2022, \apj, 929, 53

\bibitem[{Ito {et~al.}(2023)Ito, Tanaka, Valentino, Toft, Brammer, Gould,
  Ilbert, Kashikawa, Kubo, Liang, {et~al.}}]{ito2023cosmos2020}
Ito, K., Tanaka, M., Valentino, F., {et~al.} 2023, \apjl, 945, L9

\bibitem[{{Ito} {et~al.}(2023){Ito}, {Tanaka}, {Valentino}, {Toft}, {Brammer},
  {Gould}, {Ilbert}, {Kashikawa}, {Kubo}, {Liang}, {McCracken}, \&
  {Weaver}}]{2023ApJ...945L...9I}
{Ito}, K., {Tanaka}, M., {Valentino}, F., {et~al.} 2023, \apjl, 945, L9

\bibitem[{Jin {et~al.}(2018)Jin, Daddi, Liu, Smol{\v{c}}i{\'c}, Schinnerer,
  Calabr{\`o}, Gu, Delhaize, Delvecchio, Gao, {et~al.}}]{jin2018super}
Jin, S., Daddi, E., Liu, D., {et~al.} 2018, \apj, 864, 56

\bibitem[{Jin {et~al.}(2024)Jin, Sillassen, Magdis, Brinch, Shuntov, Brammer,
  Gobat, Valentino, Carnall, Lee, {et~al.}}]{jin2024cosmic}
Jin, S., Sillassen, N.~B., Magdis, G.~E., {et~al.} 2024, \aap, 683, L4

\bibitem[{{Kalita} {et~al.}(2021)}]{Kalita+2022}
{Kalita}, B. {et~al.} 2021, ApJ, 917, L17

\bibitem[{Kimmig {et~al.}(2025)Kimmig, Remus, Seidel, Valenzuela, Dolag, \&
  Burkert}]{kimmig2025blowing}
Kimmig, L.~C., Remus, R.-S., Seidel, B., {et~al.} 2025, \apj, 979, 15

\bibitem[{Kova{\v{c}} {et~al.}(2007)Kova{\v{c}}, Somerville, Rhoads, Malhotra,
  \& Wang}]{kovavc2007clustering}
Kova{\v{c}}, K., Somerville, R.~S., Rhoads, J.~E., Malhotra, S., \& Wang, J.
  2007, \apj, 668, 15

\bibitem[{{Kubo} {et~al.}(2022){Kubo}, {Umehata}, {Matsuda}, {Kajisawa},
  {Steidel}, {Yamada}, {Tanaka}, {Hatsukade}, {Tamura}, {Nakanishi}, {Kohno},
  {Lee}, {Matsuda}, {Ao}, {Nagao}, \& {Yun}}]{2022ApJ...935...89K}
{Kubo}, M., {Umehata}, H., {Matsuda}, Y., {et~al.} 2022, \apj, 935, 89

\bibitem[{Kurinchi-Vendhan {et~al.}(2024)Kurinchi-Vendhan, Farcy, Hirschmann,
  \& Valentino}]{kurinchi2024origin}
Kurinchi-Vendhan, S., Farcy, M., Hirschmann, M., \& Valentino, F. 2024, \mnras,
  534, 3974

\bibitem[{Larson(1990)}]{larson1990galaxy}
Larson, R.~B. 1990, \pasp, 102, 709

\bibitem[{Le~F{\`e}vre {et~al.}(2013)Le~F{\`e}vre, Cassata, Cucciati, Garilli,
  Ilbert, Le~Brun, Maccagni, Moreau, Scodeggio, Tresse, {et~al.}}]{le2013vimos}
Le~F{\`e}vre, O., Cassata, P., Cucciati, O., {et~al.} 2013, \aap, 559, A14

\bibitem[{{Lee} {et~al.}(2024){Lee}, {Gawiser}, {Park}, {Yang}, {Valdes},
  {Lang}, {Ramakrishnan}, {Moon}, {Firestone}, {Appleby}, {Artale}, {Andrews},
  {Bauer}, {Benda}, {Broussard}, {Chiang}, {Ciardullo}, {Dey}, {Farooq},
  {Gronwall}, {Guaita}, {Huang}, {Hwang}, {Im}, {Jeong}, {Karthikeyan}, {Kim},
  {Kim}, {Kumar}, {Nagaraj}, {Nantais}, {Padilla}, {Park}, {Pope}, {Popescu},
  {Schlegel}, {Seo}, {Singh}, {Song}, {Troncoso}, {Vivas}, {Zabludoff}, \&
  {Zenteno}}]{2024ApJ...962...36L}
{Lee}, K.-S., {Gawiser}, E., {Park}, C., {et~al.} 2024, \apj, 962, 36

\bibitem[{Leja {et~al.}(2019)Leja, Carnall, Johnson, Conroy, \&
  Speagle}]{leja2019measure}
Leja, J., Carnall, A.~C., Johnson, B.~D., Conroy, C., \& Speagle, J.~S. 2019,
  \apj, 876, 3

\bibitem[{Lemaux {et~al.}(2014)Lemaux, Cucciati, Tasca, Le~F{\`e}vre, Zamorani,
  Cassata, Garilli, Le~Brun, Maccagni, Pentericci, {et~al.}}]{lemaux2014vimos}
Lemaux, B., Cucciati, O., Tasca, L., {et~al.} 2014, \aap, 572, A41

\bibitem[{Lemaux {et~al.}(2018)Lemaux, Le~F{\`e}vre, Cucciati, Ribeiro, Tasca,
  Zamorani, Ilbert, Thomas, Bardelli, Cassata, {et~al.}}]{lemaux2018vimos}
Lemaux, B.~C., Le~F{\`e}vre, O., Cucciati, O., {et~al.} 2018, \aap, 615, A77

\bibitem[{Liu {et~al.}(2019)Liu, Lang, Magnelli, Schinnerer, Leslie, Fudamoto,
  Bondi, Groves, Jim{\'e}nez-Andrade, Harrington, {et~al.}}]{liu2019automated}
Liu, D., Lang, P., Magnelli, B., {et~al.} 2019, \apjs, 244, 40

\bibitem[{Man \& Belli(2018)}]{man2018star}
Man, A. \& Belli, S. 2018, Nature Astronomy, 2, 695

\bibitem[{Mao {et~al.}(2022)Mao, Kodama, P{\'e}rez-Mart{\'\i}nez, Suzuki,
  Yamamoto, \& Adachi}]{mao2022revealing}
Mao, Z., Kodama, T., P{\'e}rez-Mart{\'\i}nez, J.~M., {et~al.} 2022, \aap, 666,
  A141

\bibitem[{Marigo {et~al.}(2017)Marigo, Girardi, Bressan, Rosenfield, Aringer,
  Chen, Dussin, Nanni, Pastorelli, Rodrigues, {et~al.}}]{marigo2017new}
Marigo, P., Girardi, L., Bressan, A., {et~al.} 2017, \apj, 835, 77

\bibitem[{Martizzi {et~al.}(2019)Martizzi, Vogelsberger, Artale, Haider,
  Torrey, Marinacci, Nelson, Pillepich, Weinberger, Hernquist,
  {et~al.}}]{martizzi2019baryons}
Martizzi, D., Vogelsberger, M., Artale, M.~C., {et~al.} 2019, \mnras, 486, 3766

\bibitem[{McCracken {et~al.}(2012)McCracken, Milvang-Jensen, Dunlop, Franx,
  Fynbo, Le~F{\`e}vre, Holt, Caputi, Goranova, Buitrago,
  {et~al.}}]{mccracken2012ultravista}
McCracken, H., Milvang-Jensen, B., Dunlop, J., {et~al.} 2012, \aap, 544, A156

\bibitem[{Merlin {et~al.}(2019)Merlin, Fortuni, Torelli, Santini, Castellano,
  Fontana, Grazian, Pentericci, Pilo, \& Schmidt}]{merlin2019red}
Merlin, E., Fortuni, F., Torelli, M., {et~al.} 2019, \mnras, 490, 3309

\bibitem[{Moore {et~al.}(1996)Moore, Katz, Lake, Dressler, \&
  Oemler}]{moore1996galaxy}
Moore, B., Katz, N., Lake, G., Dressler, A., \& Oemler, A. 1996, \nat, 379, 613

\bibitem[{Nanayakkara {et~al.}(2024)Nanayakkara, Glazebrook, Jacobs,
  Kawinwanichakij, Schreiber, Brammer, Esdaile, Kacprzak, Labbe, Lagos,
  {et~al.}}]{nanayakkara2024population}
Nanayakkara, T., Glazebrook, K., Jacobs, C., {et~al.} 2024, Scientific Reports,
  14, 3724

\bibitem[{Nanayakkara {et~al.}(2025)Nanayakkara, Glazebrook, Schreiber,
  Chittenden, Brammer, Esdaile, Jacobs, Kacprzak, Kawinwanichakij, Kimmig,
  {et~al.}}]{nanayakkara2025formation}
Nanayakkara, T., Glazebrook, K., Schreiber, C., {et~al.} 2025, \apj, 981, 78

\bibitem[{Ouchi {et~al.}(2020)Ouchi, Ono, \& Shibuya}]{ouchi2020observations}
Ouchi, M., Ono, Y., \& Shibuya, T. 2020, \araa, 58, 617

\bibitem[{Ouchi {et~al.}(2003)Ouchi, Shimasaku, Furusawa, Miyazaki, Doi,
  Hamabe, Hayashino, Kimura, Kodaira, Komiyama, {et~al.}}]{ouchi2003subaru}
Ouchi, M., Shimasaku, K., Furusawa, H., {et~al.} 2003, \apj, 582, 60

\bibitem[{Overzier(2016)}]{overzier2016realm}
Overzier, R.~A. 2016, Astronomy and Astrophysics Reviews, 24, 14

\bibitem[{Pacifici {et~al.}(2016)Pacifici, Oh, Oh, Lee, \&
  Sukyoung}]{pacifici2016timing}
Pacifici, C., Oh, S., Oh, K., Lee, J., \& Sukyoung, K.~Y. 2016, \apj, 824, 45

\bibitem[{{Park} \& {Hwang}(2009)}]{2009ApJ...699.1595P}
{Park}, C. \& {Hwang}, H.~S. 2009, \apj, 699, 1595

\bibitem[{Park {et~al.}(2024)Park, Belli, Conroy, Johnson, Davies, Leja,
  Tacchella, Mendel, Benton, Bugiani, {et~al.}}]{park2024widespread}
Park, M., Belli, S., Conroy, C., {et~al.} 2024, \apj, 976, 72

\bibitem[{Peng {et~al.}(2010)}]{peng2010environmental}
Peng, Y. {et~al.} 2010, \apj, 721, 193

\bibitem[{Ramakrishnan {et~al.}(2023)Ramakrishnan, Moon, Im, Farooq, Lee,
  Gawiser, Yang, Park, Hwang, Valdes, Artale, Ciardullo, Dey, Gronwall, Guaita,
  Jeong, Padilla, Singh, \& Zabludoff}]{Ramakrishnan_2023}
Ramakrishnan, V., Moon, B., Im, S.~H., {et~al.} 2023, \apj, 951, 119

\bibitem[{Rodr{\'\i}guez~Montero {et~al.}(2019)Rodr{\'\i}guez~Montero,
  Dav{\'e}, Wild, Angl{\'e}s-Alc{\'a}zar, \& Narayanan}]{rodriguez2019mergers}
Rodr{\'\i}guez~Montero, F., Dav{\'e}, R., Wild, V., Angl{\'e}s-Alc{\'a}zar, D.,
  \& Narayanan, D. 2019, \mnras, 490, 2139

\bibitem[{Salim \& Narayanan(2020)}]{salim2020dust}
Salim, S. \& Narayanan, D. 2020, \araa, 58, 529

\bibitem[{Sanchez-Blazquez {et~al.}(2006)Sanchez-Blazquez, Peletier,
  Jimenez-Vicente, Cardiel, Cenarro, Falcon-Barroso, Gorgas, Selam, \&
  Vazdekis}]{sanchez2006medium}
Sanchez-Blazquez, P., Peletier, R., Jimenez-Vicente, J., {et~al.} 2006, \mnras,
  371, 703

\bibitem[{{Sawicki} {et~al.}(2019){Sawicki}, {Arnouts}, {Huang}, {Coupon},
  {Golob}, {Gwyn}, {Foucaud}, {Moutard}, {Iwata}, {Liu}, {Chen}, {Desprez},
  {Harikane}, {Ono}, {Strauss}, {Tanaka}, {Thibert}, {Balogh}, {Bundy},
  {Chapman}, {Gunn}, {Hsieh}, {Ilbert}, {Jing}, {LeF{\`e}vre}, {Li}, {Matsuda},
  {Miyazaki}, {Nagao}, {Nishizawa}, {Ouchi}, {Shimasaku}, {Silverman}, {de la
  Torre}, {Tresse}, {Wang}, {Willott}, {Yamada}, {Yang}, \&
  {Yee}}]{2019MNRAS.489.5202S}
{Sawicki}, M., {Arnouts}, S., {Huang}, J., {et~al.} 2019, \mnras, 489, 5202

\bibitem[{Schreiber {et~al.}(2018)Schreiber, Glazebrook, Nanayakkara, Kacprzak,
  Labb{\'e}, Oesch, Yuan, Tran, Papovich, Spitler,
  {et~al.}}]{schreiber2018near}
Schreiber, C., Glazebrook, K., Nanayakkara, T., {et~al.} 2018, \aap, 618, A85

\bibitem[{Scoville {et~al.}(2007)Scoville, Aussel, Brusa, Capak, Carollo,
  Elvis, Giavalisco, Guzzo, Hasinger, Impey, {et~al.}}]{scoville2007cosmic}
Scoville, N., Aussel, H., Brusa, M., {et~al.} 2007, \apjs, 172, 1

\bibitem[{Shah {et~al.}(2024)Shah, Lemaux, Forrest, Cucciati, Hung, Staab,
  Hathi, Lubin, Gal, Shen, {et~al.}}]{shah2024identification}
Shah, E.~A., Lemaux, B., Forrest, B., {et~al.} 2024, \mnras, stae519

\bibitem[{Shi {et~al.}(2019)Shi, Huang, Lee, Toshikawa, Bowen, Malavasi,
  Lemaux, Cucciati, Le~Fevre, \& Dey}]{shi2019galaxies}
Shi, K., Huang, Y., Lee, K.-S., {et~al.} 2019, \apj, 879, 9

\bibitem[{Shi {et~al.}(2020)Shi, Toshikawa, Cai, Lee, \&
  Fang}]{shi2020detailed}
Shi, K., Toshikawa, J., Cai, Z., Lee, K.-S., \& Fang, T. 2020, \apj, 899, 79

\bibitem[{Shimakawa {et~al.}(2017)Shimakawa, Kodama, Hayashi, Tanaka, Matsuda,
  Kashikawa, Shibuya, Tadaki, Koyama, Suzuki, {et~al.}}]{shimakawa2017direct}
Shimakawa, R., Kodama, T., Hayashi, M., {et~al.} 2017, \mnras: Letters, 468,
  L21

\bibitem[{Sousbie(2011)}]{sousbie2011persistent}
Sousbie, T. 2011, \mnras, 414, 350

\bibitem[{Spitler {et~al.}(2014)Spitler, Straatman, Labb{\'e}, Glazebrook,
  Tran, Kacprzak, Quadri, Papovich, Persson, Van~Dokkum,
  {et~al.}}]{spitler2014exploring}
Spitler, L.~R., Straatman, C.~M., Labb{\'e}, I., {et~al.} 2014, \apjl, 787, L36

\bibitem[{Steinhauser {et~al.}(2016)Steinhauser, Schindler, \&
  Springel}]{steinhauser2016simulations}
Steinhauser, D., Schindler, S., \& Springel, V. 2016, \aap, 591, A51

\bibitem[{Straatman {et~al.}(2014)Straatman, Labb{\'e}, Spitler, Allen,
  Altieri, Brammer, Dickinson, Van~Dokkum, Inami, Glazebrook,
  {et~al.}}]{straatman2014substantial}
Straatman, C.~M., Labb{\'e}, I., Spitler, L.~R., {et~al.} 2014, \apjl, 783, L14

\bibitem[{{Tacchella} {et~al.}(2022)}]{Tacchella+2022}
{Tacchella}, S. {et~al.} 2022, ApJ, 927, 170

\bibitem[{Tanaka {et~al.}(2024)Tanaka, Onodera, Shimakawa, Ito, Kakimoto, Kubo,
  Morishita, Toft, Valentino, \& Wu}]{tanaka2024protocluster}
Tanaka, M., Onodera, M., Shimakawa, R., {et~al.} 2024, \apj, 970, 59

\bibitem[{Valentino {et~al.}(2020)Valentino, Tanaka, Davidzon, Toft,
  G{\'o}mez-Guijarro, Stockmann, Onodera, Brammer, Ceverino, Faisst,
  {et~al.}}]{valentino2020quiescent}
Valentino, F., Tanaka, M., Davidzon, I., {et~al.} 2020, \apj, 889, 93

\bibitem[{Walters {et~al.}(2022)Walters, Woo, \&
  Ellison}]{walters2022quenching}
Walters, D., Woo, J., \& Ellison, S.~L. 2022, \mnras, 511, 6126

\bibitem[{Weaver {et~al.}(2023)Weaver, Davidzon, Toft, Ilbert, McCracken,
  Gould, Jespersen, Steinhardt, Lagos, Capak, {et~al.}}]{weaver2023cosmos2020}
Weaver, J., Davidzon, I., Toft, S., {et~al.} 2023, \aap, 677, A184

\bibitem[{{Weaver} {et~al.}(2021){Weaver}, {Kauffmann}, {Shuntov}, {Davidzon},
  {Ilbert}, {Brammer}, {Hsieh}, {Capak}, {Moneti}, {McCracken}, {Toft}, \& {The
  Cosmos Team}}]{2021AAS...23721506W}
{Weaver}, J.~R., {Kauffmann}, O., {Shuntov}, M., {et~al.} 2021, in American
  Astronomical Society Meeting Abstracts, Vol.~53, American Astronomical
  Society Meeting Abstracts, 215.06

\bibitem[{Wetzel {et~al.}(2013)Wetzel, Tinker, Conroy, \& Van
  Den~Bosch}]{wetzel2013galaxy}
Wetzel, A.~R., Tinker, J.~L., Conroy, C., \& Van Den~Bosch, F.~C. 2013, \mnras,
  432, 336

\bibitem[{Whitaker {et~al.}(2012)Whitaker, Kriek, van Dokkum, Bezanson,
  Brammer, Franx, \& Labb{\'e}}]{whitaker2012large}
Whitaker, K.~E., Kriek, M., van Dokkum, P.~G., {et~al.} 2012, \apj, 745, 179

\bibitem[{Wild {et~al.}(2016)Wild, Almaini, Dunlop, Simpson, Rowlands, Bowler,
  Maltby, \& McLure}]{wild2016evolution}
Wild, V., Almaini, O., Dunlop, J., {et~al.} 2016, \mnras, 463, 832

\bibitem[{{Williams} {et~al.}(2009){Williams}, {Quadri}, {Franx}, {van Dokkum},
  \& {Labb{\'e}}}]{2009ApJ...691.1879W}
{Williams}, R.~J., {Quadri}, R.~F., {Franx}, M., {van Dokkum}, P., \&
  {Labb{\'e}}, I. 2009, \apj, 691, 1879

\bibitem[{Wright {et~al.}(2019)Wright, Lagos, Davies, Power, Trayford, \&
  Wong}]{wright2019quenching}
Wright, R.~J., Lagos, C. d.~P., Davies, L.~J., {et~al.} 2019, \mnras, 487, 3740

\bibitem[{Zinger {et~al.}(2020)Zinger, Pillepich, Nelson, Weinberger, Pakmor,
  Springel, Hernquist, Marinacci, \& Vogelsberger}]{zinger2020ejective}
Zinger, E., Pillepich, A., Nelson, D., {et~al.} 2020, \mnras, 499, 768

\end{thebibliography}
\begin{appendix}
\section{Galaxy properties}
\begin{table*}[h]
    \caption{Galaxy Properties}
    \label{table:galaxy_properties1}
    \centering
    \begin{tabular}{|>{\centering\arraybackslash}p{2.5cm}|c|c|c|c|c|c|}
        \hline\hline
COSMOS-ID ('FARMER')   &   $z_{phot}$   &    log($Mass / M_\odot$)     &  log($sSFR yr^{-1}$)  &    Age(Gyr)     &    Dust($A_V$)     &    Metallicity    \\ \hline
13648   &     3.060   $\pm$  0.080   &  10.635   $\pm$    0.026   &    -10.144    $\pm$    0.098  &   0.390   $\pm$   0.020   &    0.774   $\pm$   0.075   &    0.809  $\pm$   0.097       \\ \hline
20998   &     3.176   $\pm$  0.059   &  10.714   $\pm$    0.001   &    -41.305    $\pm$    19.354 &    0.400  $\pm$    0.017   &    0.086  $\pm$    0.007   &    1.739 $\pm$    0.041    \\ \hline
50248   &     3.149   $\pm$  0.065   &  10.592   $\pm$    0.008   &    -31.406    $\pm$    33.488 &    0.684  $\pm$    0.032   &    0.009  $\pm$    0.011   &    2.040 $\pm$    0.085    \\ \hline
110859  &     3.149   $\pm$   0.035  &  10.770   $\pm$    0.002   &    -12.015    $\pm$    2.075  &   0.144   $\pm$   0.003   &    0.613   $\pm$   0.008   &    0.401  $\pm$   0.003   \\ \hline
278297  &     3.174   $\pm$   0.053  &  10.605   $\pm$    0.005   &    -10.194    $\pm$    0.061  &   0.377   $\pm$   0.013   &    0.307   $\pm$   0.018   &    0.445  $\pm$   0.040   \\ \hline
282964  &     3.180   $\pm$   0.029  &  11.057   $\pm$    0.002   &    -10.573    $\pm$    0.022  &   0.468   $\pm$   0.012   &    0.171   $\pm$   0.007   &    0.426  $\pm$   0.021   \\ \hline
310229  &     3.096   $\pm$   0.040  &  10.963   $\pm$    0.008   &    -10.815    $\pm$    0.040  &   1.419   $\pm$   0.039   &    0.084   $\pm$   0.052   &    2.227  $\pm$   0.072   \\ \hline
338392  &     3.102   $\pm$   0.039  &  11.073   $\pm$    0.001   &    -31.811    $\pm$    19.397 &    0.394  $\pm$    0.017   &    0.163  $\pm$    0.002   &    0.597 $\pm$    0.011    \\ \hline
341682  &     3.081   $\pm$   0.072  &  11.555   $\pm$    0.004   &    -11.258    $\pm$    0.062  &   2.005   $\pm$   0.038   &    0.347   $\pm$   0.034   &    2.107  $\pm$   0.050   \\ \hline
343373   &    3.092   $\pm$   0.043  &  11.233   $\pm$    0.001   &    -11.126    $\pm$    0.012  &   0.956   $\pm$   0.008   &    0.000   $\pm$   0.000   &    0.169  $\pm$   0.002   \\ \hline
383298   &    3.090   $\pm$    0.101 &  10.743   $\pm$    0.023   &    -9.999     $\pm$    0.100  &   0.471    $\pm$  0.033   &    0.792    $\pm$  0.097   &    0.657   $\pm$  0.137   \\ \hline
405872   &    3.079   $\pm$    0.051 &  10.717   $\pm$    0.002   &    -10.332    $\pm$    0.030  &   0.382   $\pm$   0.005   &    0.114   $\pm$   0.008   &    1.318  $\pm$   0.047   \\ \hline
412439   &    3.192   $\pm$    0.044 &  10.843   $\pm$    0.002   &    -10.765    $\pm$    0.057  &   0.382   $\pm$   0.008   &    0.163   $\pm$   0.005   &    0.685  $\pm$   0.018   \\ \hline
489177   &    3.147   $\pm$    0.061 &  11.245   $\pm$    0.003   &    -37.269    $\pm$    21.645 &    0.408  $\pm$    0.022   &    0.670  $\pm$    0.016   &    1.159 $\pm$    0.045    \\ \hline
642338   &    3.093   $\pm$    0.066 &  10.849   $\pm$    0.002   &    -10.808    $\pm$    0.022  &   0.579   $\pm$   0.013   &    0.010   $\pm$   0.010   &    2.182  $\pm$   0.038   \\ \hline
658452   &    3.054   $\pm$    0.063 &  10.646   $\pm$    0.005   &    -10.371    $\pm$    0.016  &   0.738   $\pm$   0.019   &    0.010   $\pm$   0.012   &    2.158  $\pm$   0.050   \\ \hline
681407   &    3.062   $\pm$    0.022 &  10.977   $\pm$    0.002   &    -10.736    $\pm$    0.032  &   0.537   $\pm$   0.011   &    0.182   $\pm$   0.008   &    0.101  $\pm$   0.000   \\ \hline
706985   &    3.085   $\pm$    0.054 &  10.829   $\pm$    0.005   &    -10.121    $\pm$    0.041  &   0.387   $\pm$   0.011   &    0.245   $\pm$   0.021   &    1.688  $\pm$   0.059    \\ \hline
779869   &    3.093   $\pm$    0.066 &  10.748   $\pm$    0.020   &    -9.651     $\pm$   0.451   &  0.239    $\pm$  0.028   &    0.696    $\pm$  0.071   &    1.672   $\pm$  0.309    \\ \hline
803389   &    3.097   $\pm$    0.088 &  11.490   $\pm$    0.011   &    -11.317    $\pm$   0.088   &  1.695   $\pm$   0.092   &    1.182   $\pm$   0.063   &    0.121  $\pm$   0.012    \\ \hline
881980   &    3.082   $\pm$    0.056 &  10.622   $\pm$    0.013   &    -10.355    $\pm$   0.024   &  0.470   $\pm$   0.010   &    0.006   $\pm$   0.006   &    1.277  $\pm$   0.091    \\ \hline
911001   &    3.175   $\pm$    0.055 &  10.939   $\pm$    0.003   &    -10.436    $\pm$   0.027   &  0.482   $\pm$   0.006   &    0.141   $\pm$   0.008   &    0.809  $\pm$   0.043    \\ \hline
961549   &    3.175   $\pm$    0.069 &  10.662   $\pm$    0.013   &    -10.048    $\pm$   0.118   &  0.709   $\pm$   0.071   &    0.089   $\pm$   0.049   &    1.746  $\pm$   0.100    \\ \hline
962569   &    3.114   $\pm$    0.078 &  11.008   $\pm$    0.019   &    -11.144    $\pm$   0.131   &  0.797   $\pm$   0.150   &    0.172   $\pm$   0.082   &    2.292  $\pm$   0.102    \\ \hline
\hline
    \end{tabular}-
\end{table*}

\begin{table*}[H]
    \caption{Galaxy Properties-2}
    \label{table:galaxy_properties2}
    \begin{tabular}{|>{\centering\arraybackslash}p{2.5cm}|c|c|c|}
        \hline\hline
 COSMOS-ID ('FARMER')   &    Quenching time(Gyr) & Density($1+\delta_{COSMOS}$) & Density($1+\delta_{LAE}$)  \\ \hline
 13648  &     0.341   $\pm$    0.021  &   2.473   &   0.783     \\ \hline
20998   &      0.037  $\pm$     0.030  &   1.747   &   1.005     \\ \hline
50248   &      0.156  $\pm$     0.085  &   2.075   &   2.205     \\ \hline
110859  &     0.038   $\pm$    0.015  &   2.131   &   0.726   \\ \hline
278297  &     0.309   $\pm$    0.017  &   2.045   &   1.979   \\ \hline
282964  &     0.322   $\pm$    0.013  &   1.893   &   0.921   \\ \hline
310229  &     0.929   $\pm$    0.067  &   1.778   &   0.686   \\ \hline
338392  &      0.043  $\pm$     0.036  &   1.444   &   0.669     \\ \hline
341682  &     0.183   $\pm$    0.130  &   0.332   &   0.784   \\ \hline
343373   &    0.475   $\pm$    0.010   &   4.182   &   3.742   \\ \hline
383298   &    0.286    $\pm$  0.034   &   2.888   &   1.977   \\ \hline
405872   &    0.294   $\pm$    0.006   &   1.208   &   1.059   \\ \hline
412439   &    0.242   $\pm$    0.009   &   3.482   &   0.948   \\ \hline
489177   &     0.046  $\pm$     0.038   &   1.756   &   3.628     \\ \hline
642338   &    0.358   $\pm$    0.013   &   1.901   &   0.383   \\ \hline
658452   &    0.570   $\pm$    0.023   &   1.081   &   0.596   \\ \hline
681407   &    0.348   $\pm$    0.012   &   2.723   &   1.879   \\ \hline
706985   &    0.175   $\pm$    0.013   &   1.757   &   1.334    \\ \hline
779869   &   0.148    $\pm$   0.025   &   3.144   &   0.676  \\ \hline
803389   &   0.458   $\pm$     0.151   &   2.353   &   0.571  \\ \hline
881980   &   0.361   $\pm$     0.050   &   1.669   &   0.553  \\ \hline
911001   &   0.356   $\pm$     0.009   &   2.370   &   1.205  \\ \hline
961549   &   0.652   $\pm$     0.074   &   1.477   &   1.061   \\ \hline
962569   &  0.473   $\pm$     0.163    &   2.520   &   3.469 \\ \hline
\hline
    \end{tabular}-
\end{table*}
\end{appendix}
\begin{appendix}
    
\section{Simulations to check the validity of correlation}\label{siulationtng}
In Figure~\ref{bootstrap}, we plot the quenched fraction versus the local density on the COSMOS2020 and LAE maps. 
We perform a test to investigate whether our current data are sufficient to detect the presence (or absence) of a correlation between the quenched fraction and the local density.

To quantify the discriminating power of our dataset, we select galaxies and designate them as 'MQGs'. We select 6 sets of galaxies, each containing 25, 50, 100, 200, 250, and 300 galaxies, respectively.  These galaxies are not selected entirely randomly. We first select sets with a "strong correlation" with the local density. They are selected such that:

\begin{itemize}
    \item 70 $\%$ of the set lies in highly overdense regions ($\delta$>3)
    \item 20 $\%$ of the set lies in overdense region ($2<\delta$<3)
    \item 5 $\%$ of the set lies in average density region ($1<\delta<2$)
    \item 5 $\%$ of the set lies in underdense region ($\delta$<1)
\end{itemize}
We then calculate the 'quenched fraction' using the selected galaxies. We calculate the Spearman correlation coefficient using the Monte Carlo technique defined in Section \ref{fraccor}. We display the results of the exercise in Table \ref{tab:sim}.
\begin{table*}[h]
\centering
\begin{tabular}{|c|c|c|c|c|}

\hline
\textbf{Number of "MQGs"} & \textbf{mean SC} & \textbf{standard deviation SC} & \textbf{f$_{SC<=0.4}$} & \textbf{f$_{SC>=0.7}$} \\ \hline
25  & 0.87  & 0.15  & 5.6  & 94.4 \\ \hline
50  & 0.86  & 0.10  & 0.67 & 99.33 \\ \hline
75  & 0.84  & 0.08  & 0.0  & 100.0 \\ \hline
100 & 0.81  & 0.04  & 0.0  & 100.0 \\ \hline
150 & 0.80  & 0.11e-16  & 0.0  & 100.0 \\ \hline
200 & 0.80  & 0.11e-16  & 0.0  & 100.0 \\ \hline
250 & 0.80  & 0.11e-16  & 0.0  & 100.0 \\ \hline
300 & 0.80  & 0.11e-16  & 0.0  & 100.0 \\ \hline
\end{tabular}
\caption{The statistics obtained for the strongly correlated simulated sample. Columns 2 and 3 display the mean and standard deviation of the 3000 Spearman coefficient (SC) Monte Carlo runs. Columns 4 and 5 show the fraction as percentage when the SC is < 0.4 and the SC is > 0.7 respectively.}
\label{tab:sim}
\end{table*}
From this simulation, we conclude that if a strong correlation existed, we would be able to find it using our analysis, even with a relatively small sample of 24 galaxies. However, for a smaller sample of galaxies (up to 100), the Spearman correlation coefficient is overestimated and finally converges for 150 or more sources. The mean correlation coefficient calculated for our sample in Section~\ref{fraccor} is 0.55 and therefore we can confidently rule out the presence of a strong correlation between the quenched fraction and the environmental density.

We repeat this analysis for a weakly correlated sample  generated such that:
\begin{itemize}
    \item 15 $\%$ of the set lies in highly overdense regions ($\delta$>3)
    \item 20 $\%$ of the set lies in overdense region ($2<\delta$<3)
    \item 45 $\%$ of the set lies in average density region ($1<\delta<2$)
    \item 20 $\%$ of the set lies in underdense region ($\delta$<1)
   
\end{itemize}
\begin{table*}[ht]
\centering
\begin{tabular}{|c|c|c|c|c|}
\hline
\textbf{Number of ``MQGs''} & \textbf{mean SC} & \textbf{standard deviation SC} & \textbf{f$_{SC<=0.4}$} & \textbf{f$_{SC>=0.7}$} \\ \hline
25  & 0.586  & 0.357  & 21.9  & 49.3  \\ \hline
50  & 0.553  & 0.283  & 17.9  & 29.5  \\ \hline
75  & 0.564  & 0.259  & 14.9  & 27.4  \\ \hline
100 & 0.558  & 0.232  & 13.3  & 18.9  \\ \hline
150 & 0.601  & 0.165  & 5.6   & 17.5  \\ \hline
200 & 0.614  & 0.127  & 2.8   & 15.3  \\ \hline
250 & 0.614  & 0.094  & 1.3   & 11.1  \\ \hline
300 & 0.610  & 0.082  & 1.0   & 8.1   \\ \hline
\end{tabular}
\caption{The statistics obtained for weakly correlated samples. Columns 2 and 3 display the mean and standard deviation of the 3000 Spearman coefficient (SC) Monte Carlo runs. Columns 4 and 5 show the percentage of runs in which the SC is $<$ 0.4 and $>$ 0.7, respectively.}
\end{table*}

In this weakly correlated sample, we can also see a correlation at a median $\approx$ 0.6. The Spearman correlation coefficient increases with an increase in the number of selected sources.
From these simulations, we conclude that if a correlation existed, we would be able to find it using our analysis, even with a relatively small sample of 24 galaxies, as in the main part of this paper. Further notable is that even for a small number of sources, no more than 22$\%$ of the Monte Carlo runs show no correlation (SC<0.4). In this case too, for a small number of sources, the correlation coefficient can be overestimated, with nearly 50$\%$ of the Monte Carlo predictions incorrectly showing a strong correlation. This is different from our calculations in Section~\ref{fraccor} where nearly 45$\%$ of Monte Carlo, where SC<0.4. We conclude from this that there exists  no correlation between quenched fraction and the environment density. 

\section{SEDs and SFH of MQGs}\label{sedsbagpipes}
\begin{figure*}[!h]
    \centering
    \includegraphics[width=0.49\textwidth]{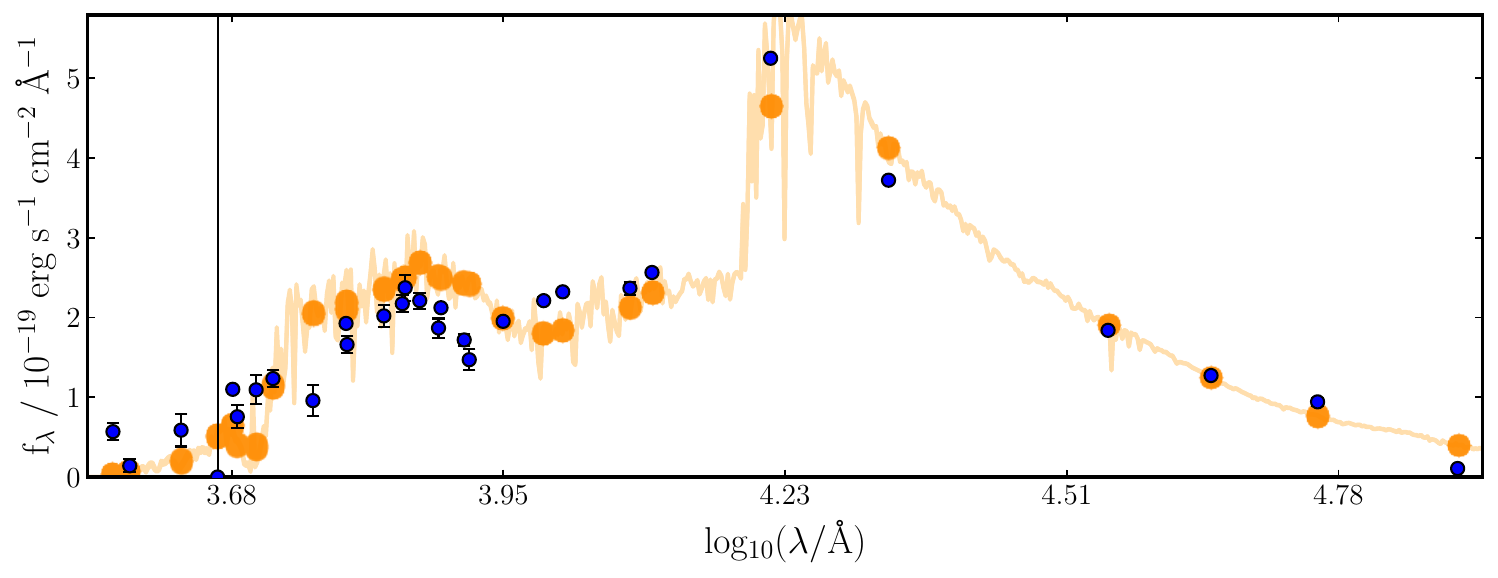}
    \includegraphics[width=0.49\textwidth]{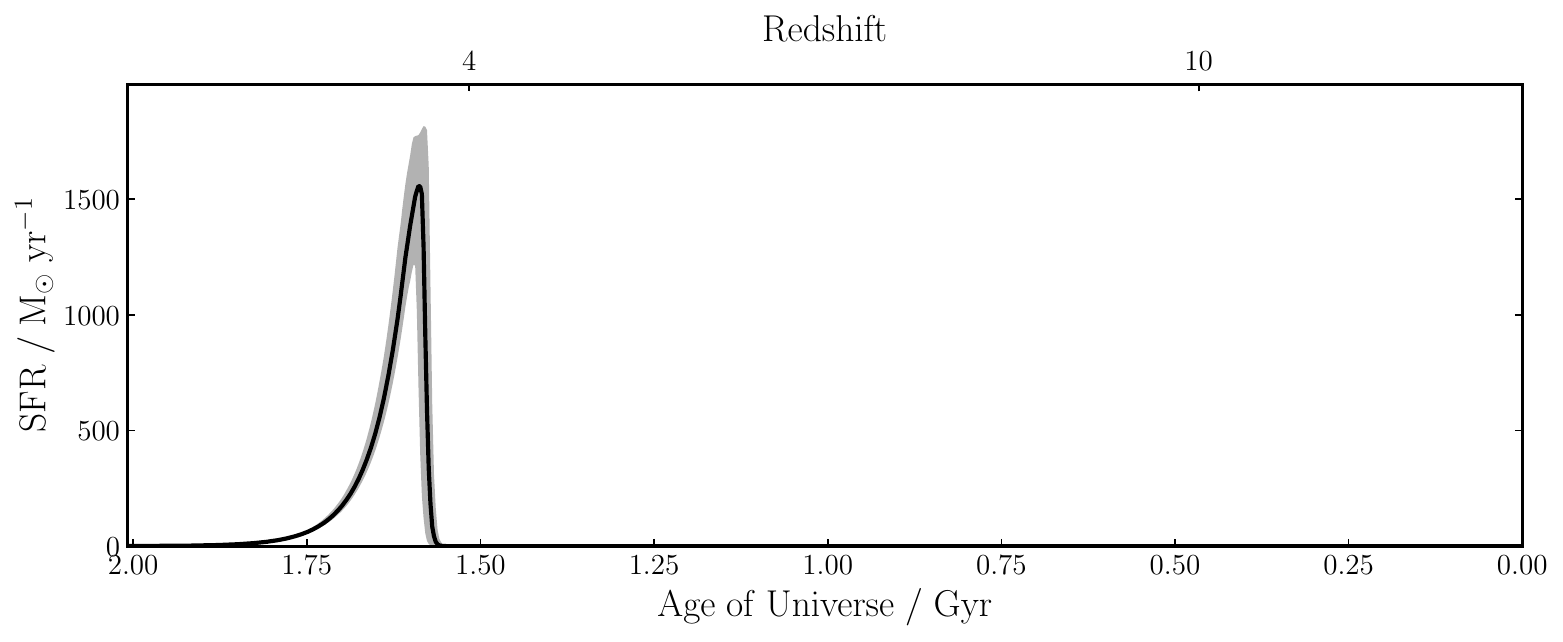}
    \caption{SED and SFH for COSMOS ID 412439}
    \label{SED_ID2}
\end{figure*}
\begin{figure*}[!h]
    \centering
    \includegraphics[width=0.49\textwidth]{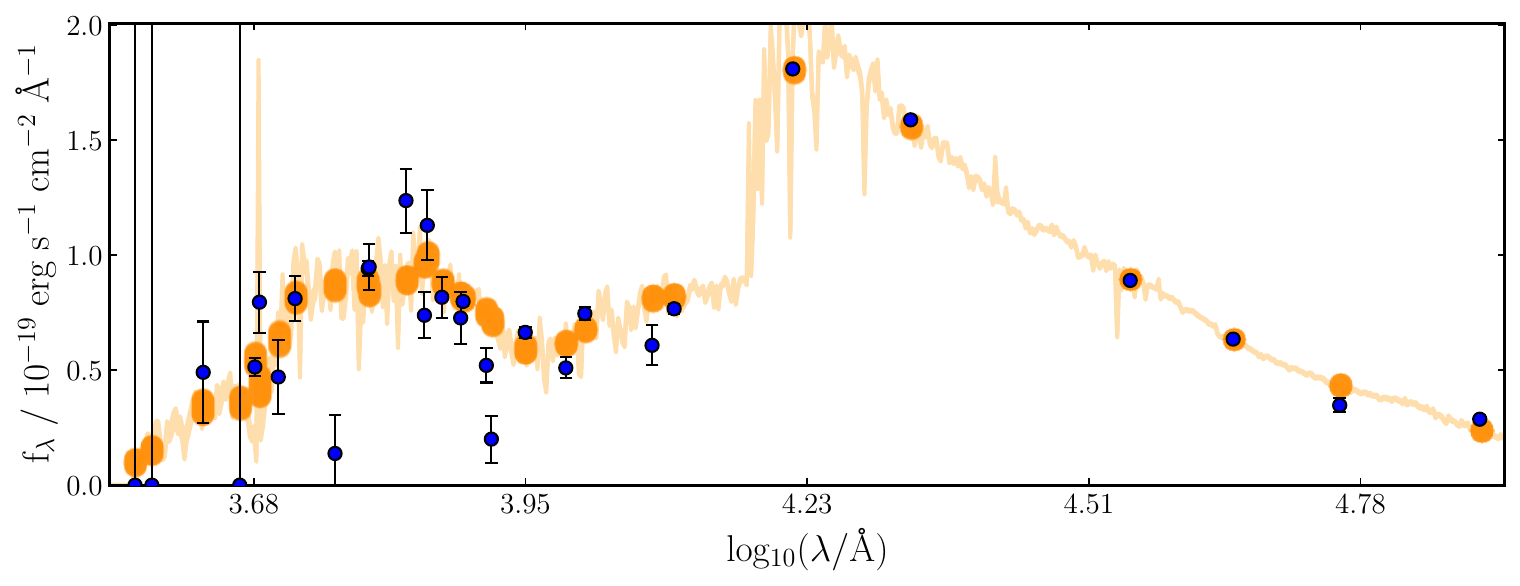}
    \includegraphics[width=0.49\textwidth]{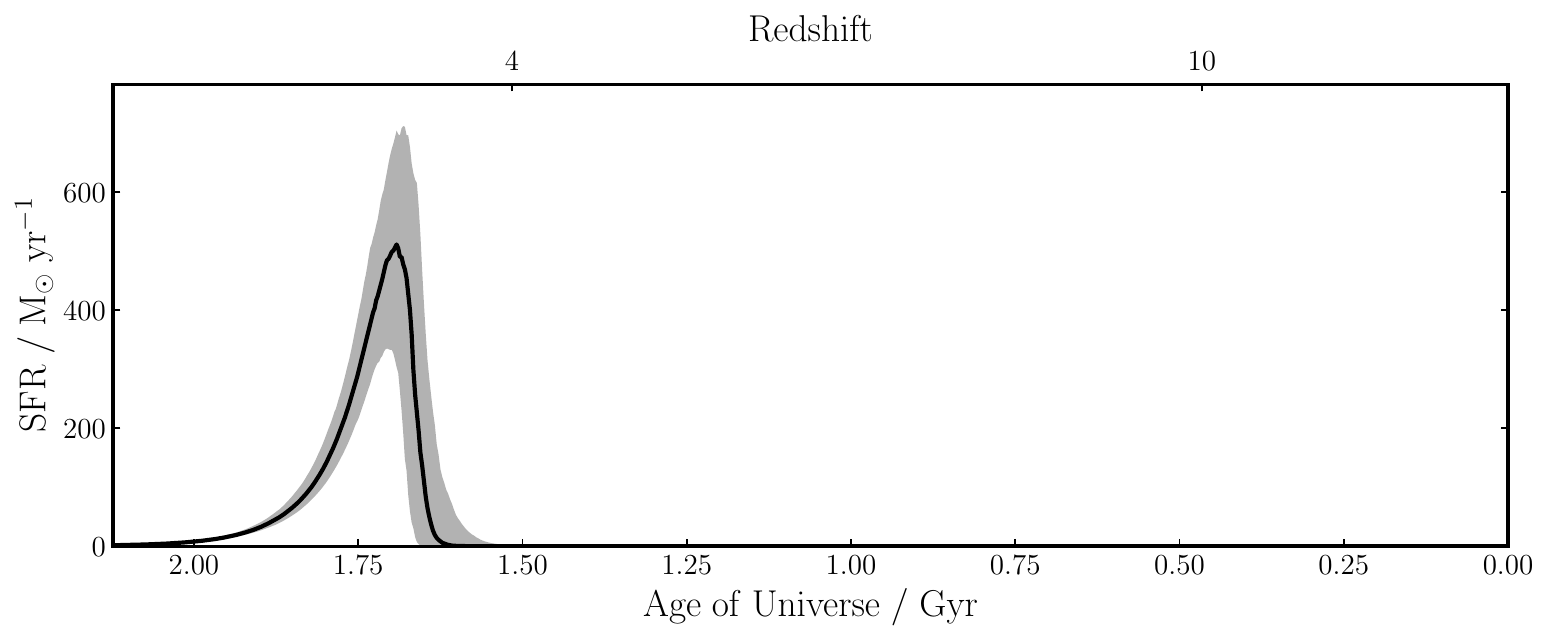}
    \caption{SED and SFH for COSMOS ID 13648}
    \label{SED_13648}
\end{figure*}

\begin{figure*}[!h]
    \centering
    \includegraphics[width=0.49\textwidth]{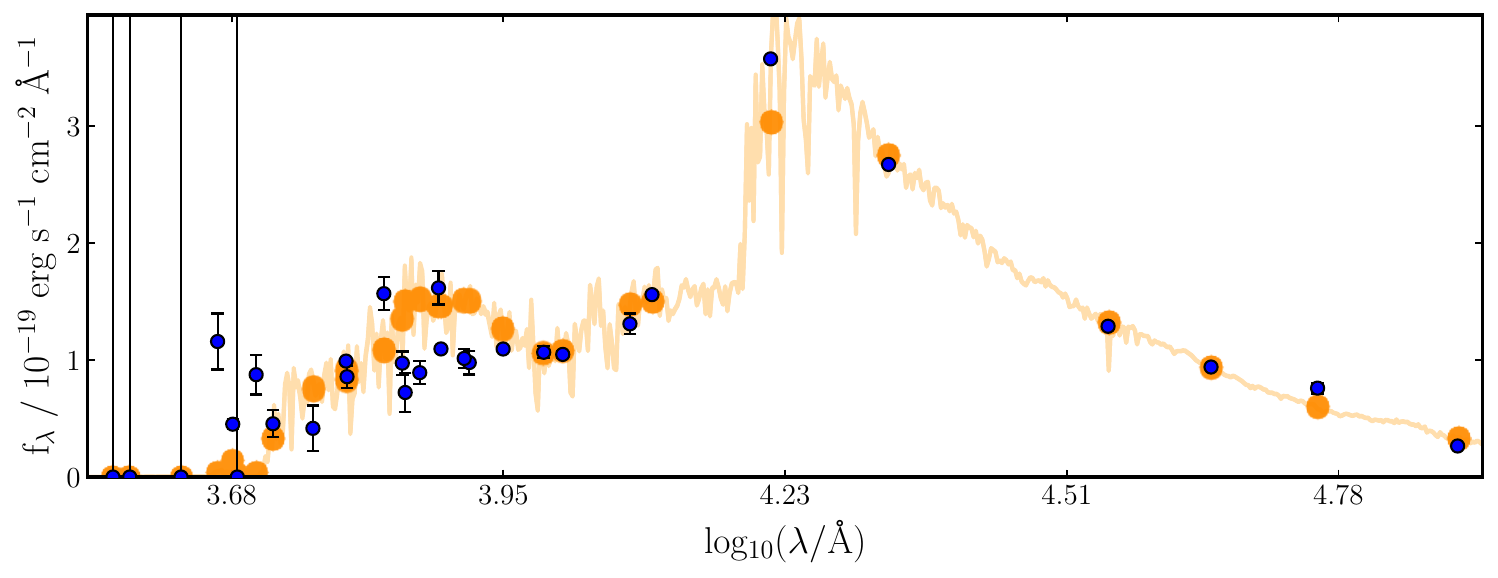}
    \includegraphics[width=0.49\textwidth]{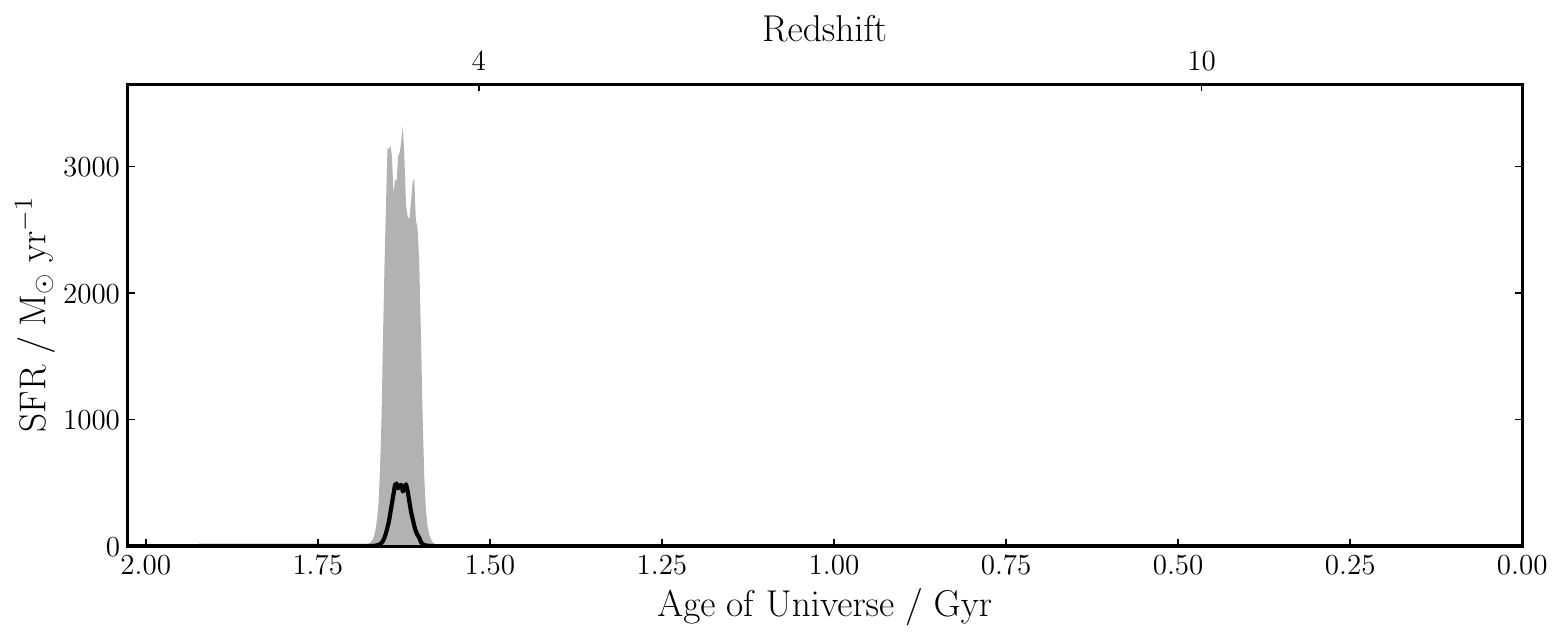}
    \caption{SED and SFH for COSMOS ID 20998}
    \label{SED_20998}
\end{figure*}

\begin{figure*}[!h]
    \centering
    \includegraphics[width=0.49\textwidth]{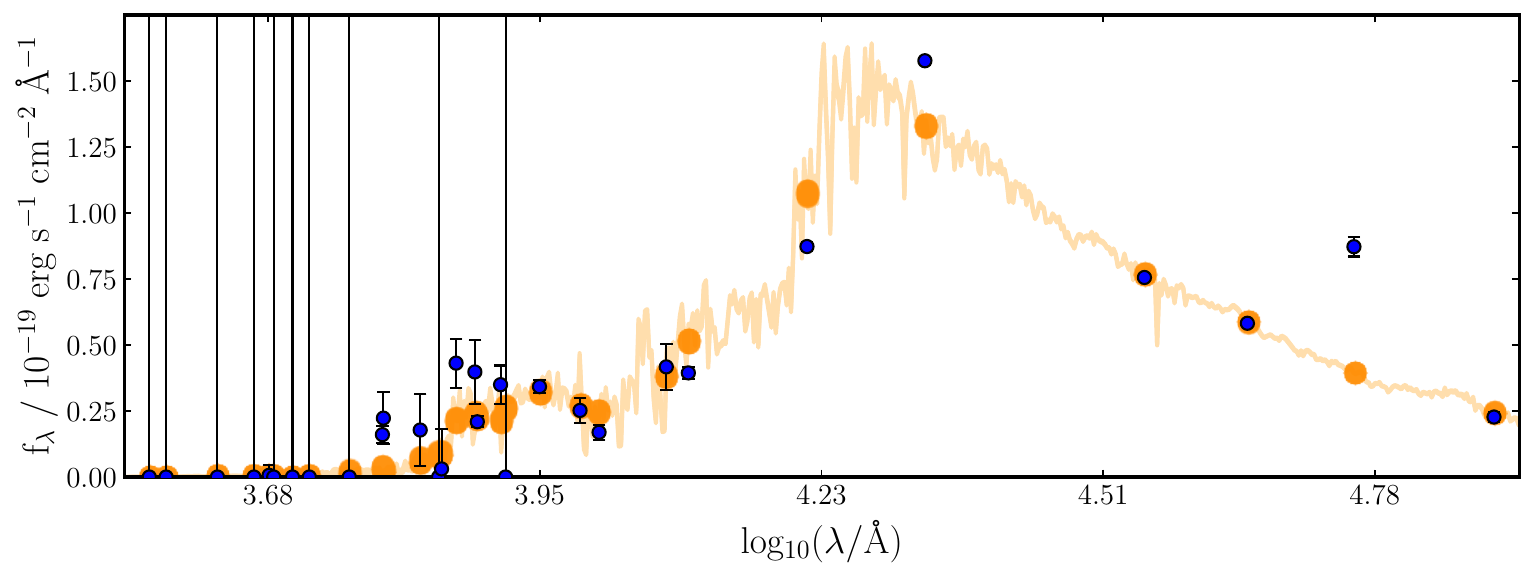}
    \includegraphics[width=0.49\textwidth]{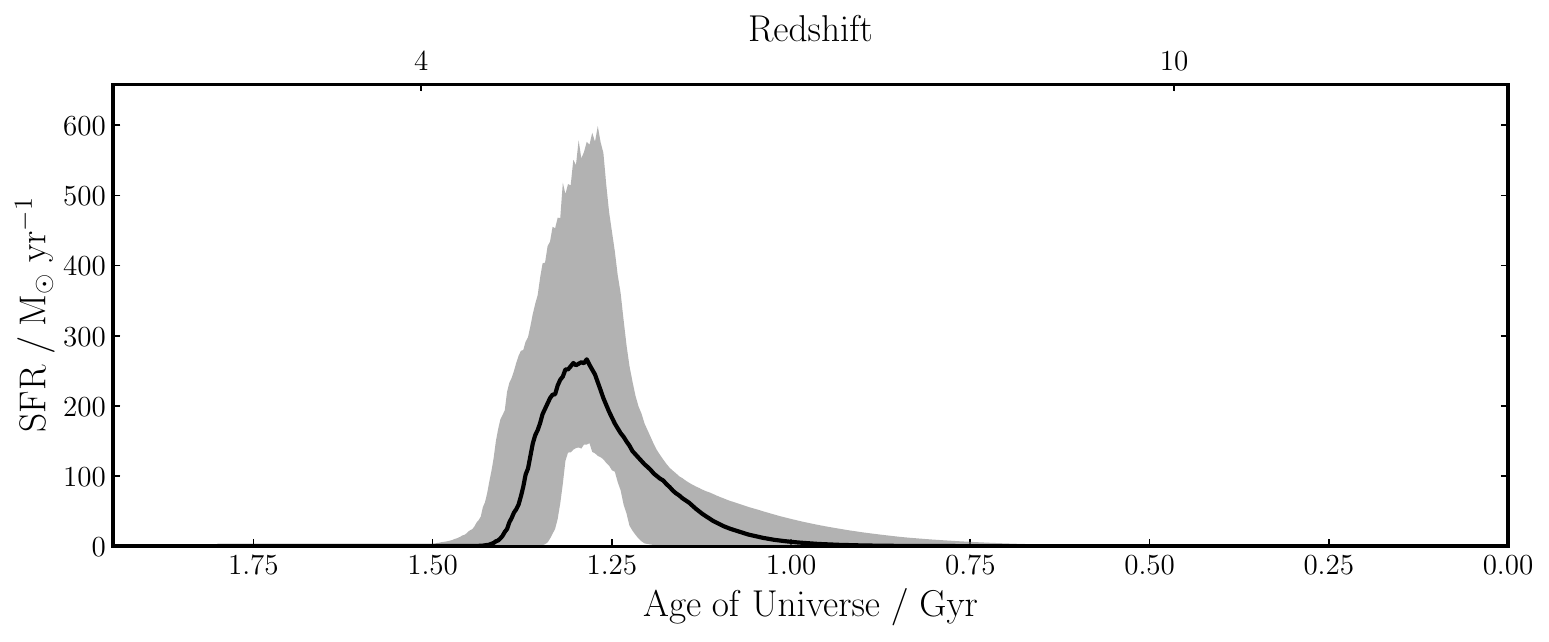}
    \caption{SED and SFH for COSMOS ID 50248}
    \label{SED_50248}
\end{figure*}

\begin{figure*}[!h]
    \centering
    \includegraphics[width=0.49\textwidth]{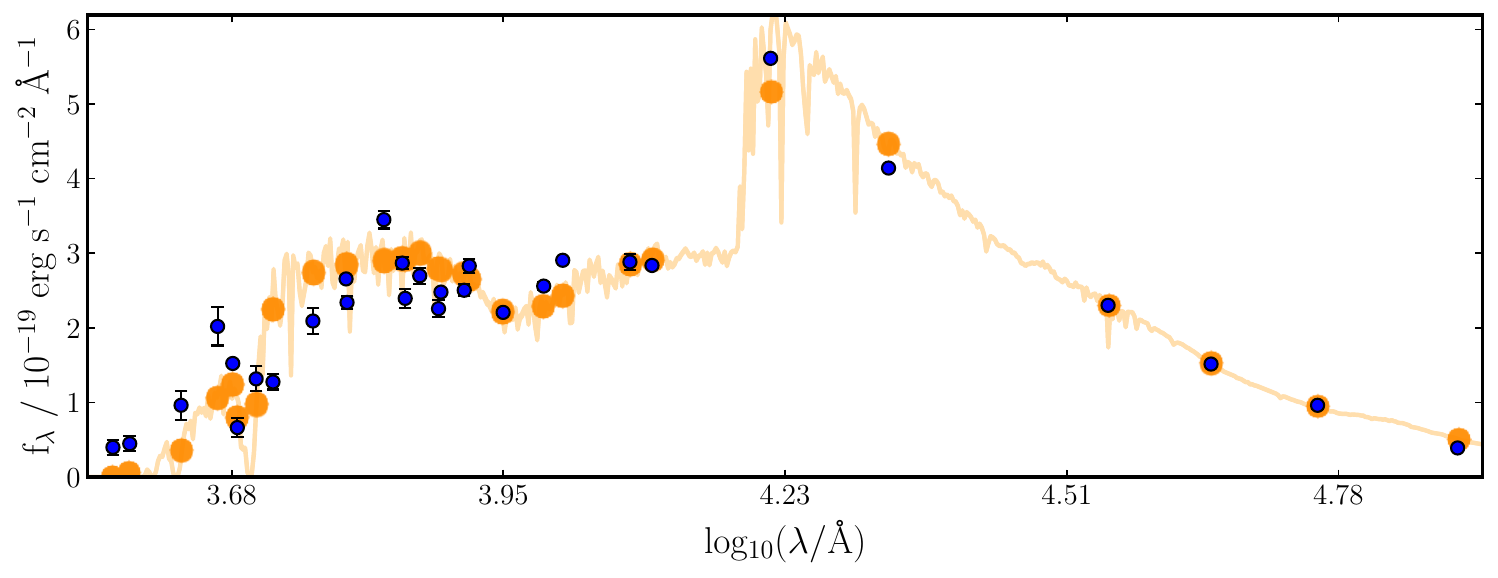}
    \includegraphics[width=0.49\textwidth]{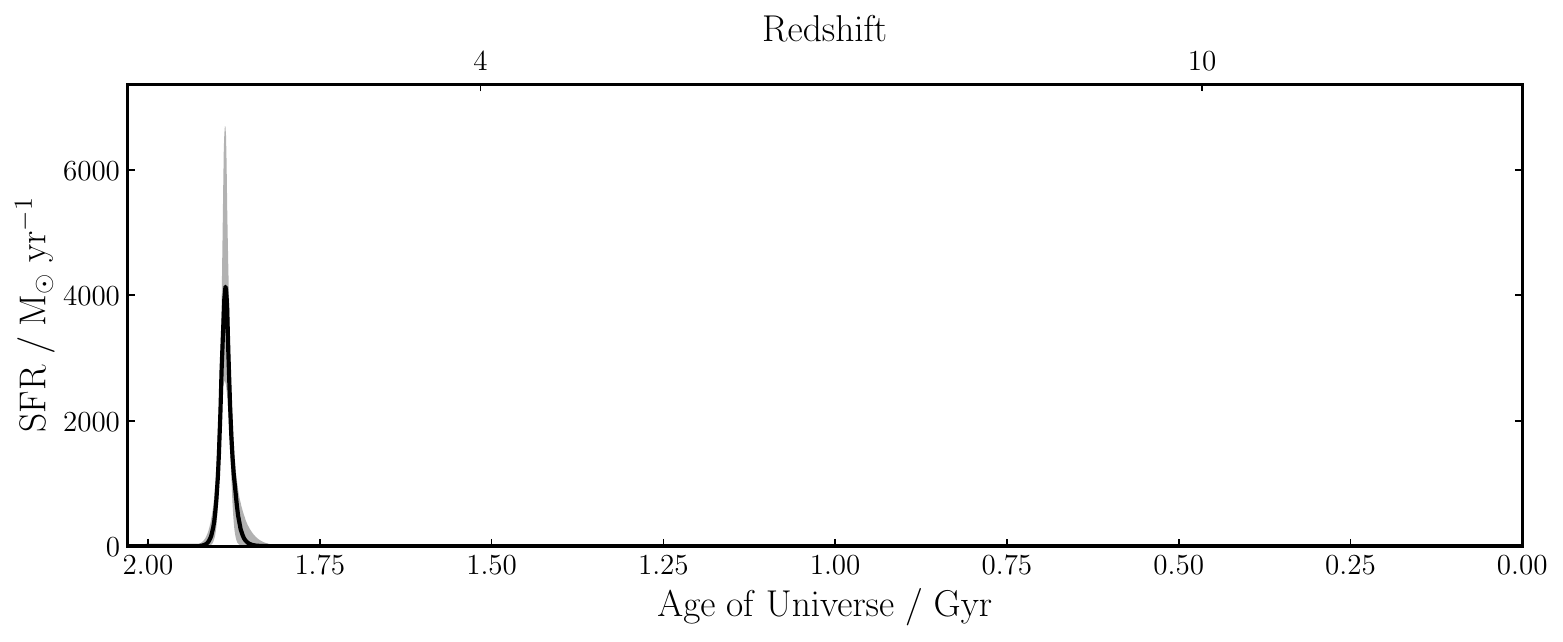}
    \caption{SED and SFH for COSMOS ID 110859}
    \label{SED_110859}
\end{figure*}

\begin{figure*}[!h]
    \centering
    \includegraphics[width=0.49\textwidth]{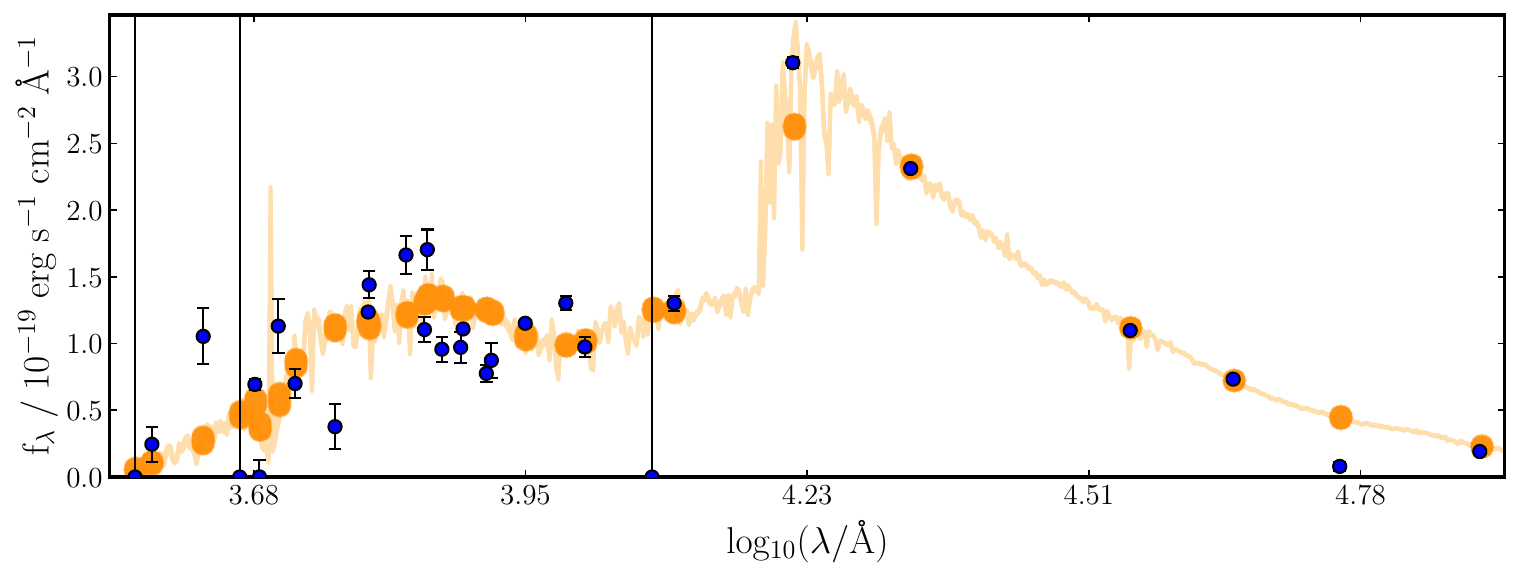}
    \includegraphics[width=0.49\textwidth]{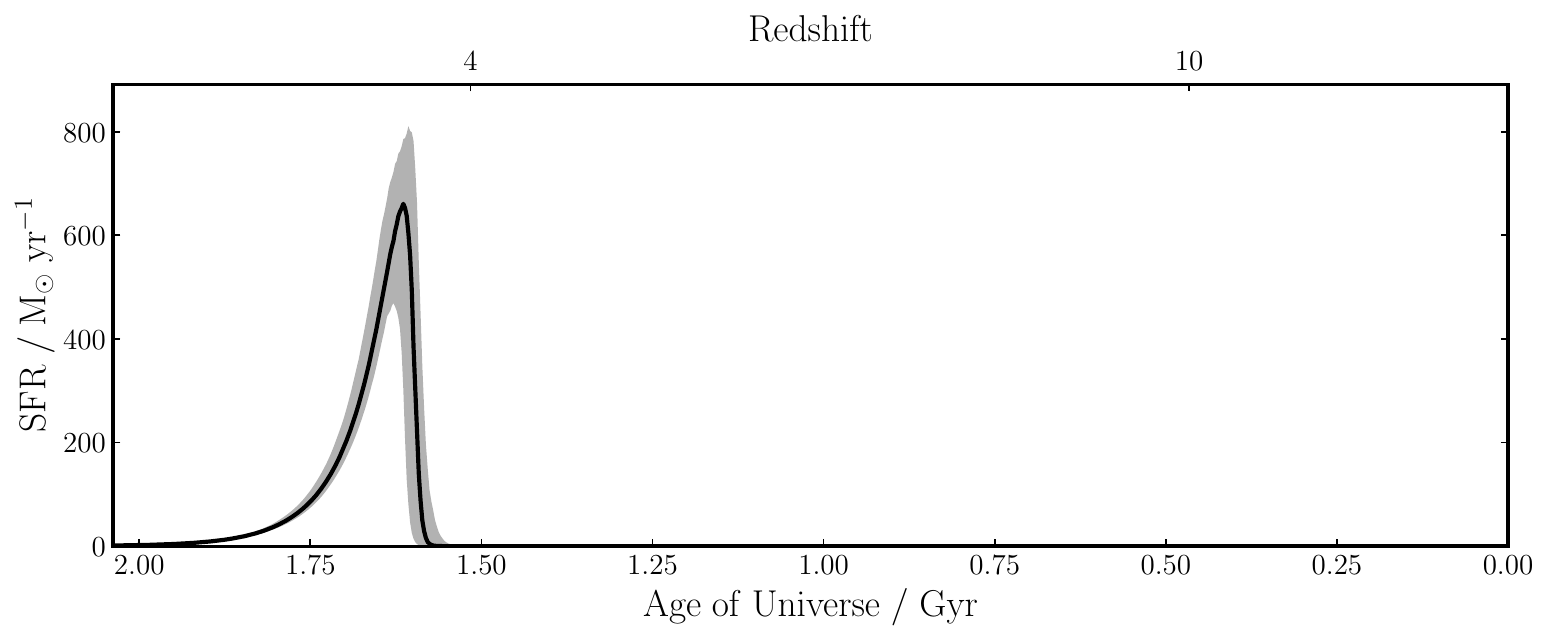}
    \caption{SED and SFH for COSMOS ID 278297}
    \label{SED_278297}
\end{figure*}

\begin{figure*}[!h]
    \centering
    \includegraphics[width=0.49\textwidth]{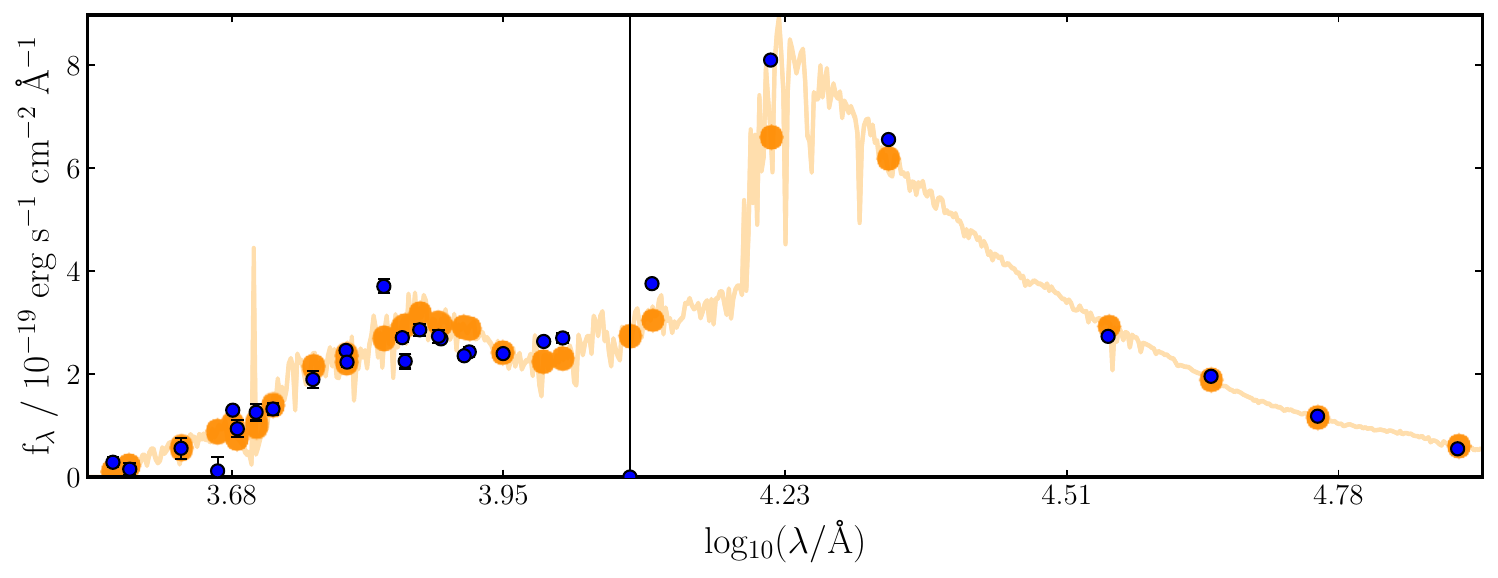}
    \includegraphics[width=0.49\textwidth]{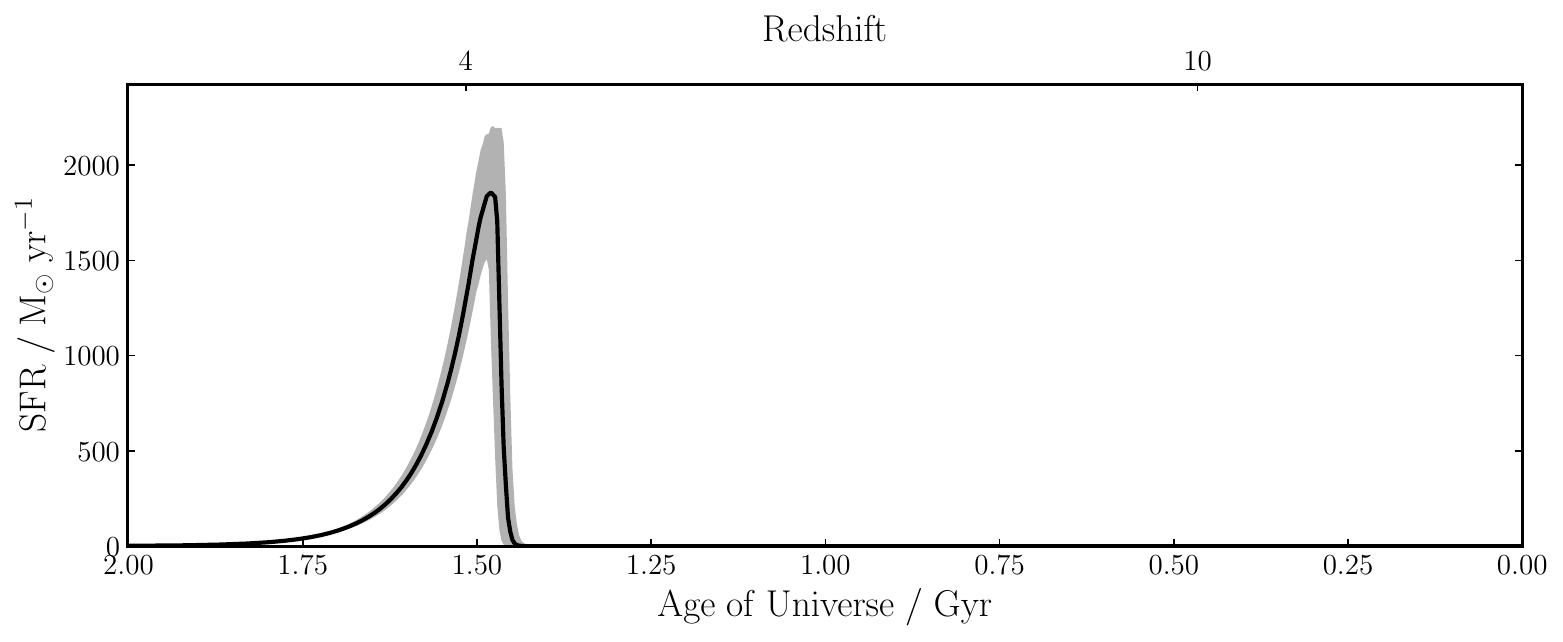}
    \caption{SED and SFH for COSMOS ID 282964}
    \label{SED_282964}
\end{figure*}

\begin{figure*}[!h]
    \centering
    \includegraphics[width=0.49\textwidth]{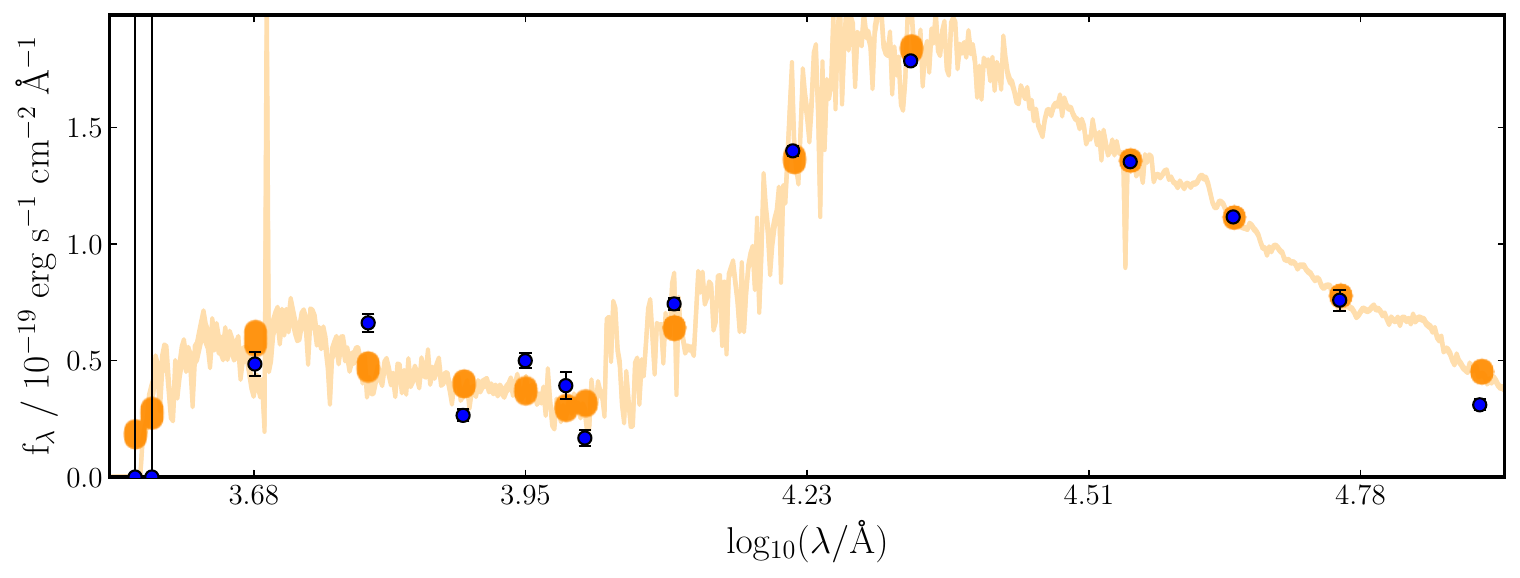}
    \includegraphics[width=0.49\textwidth]{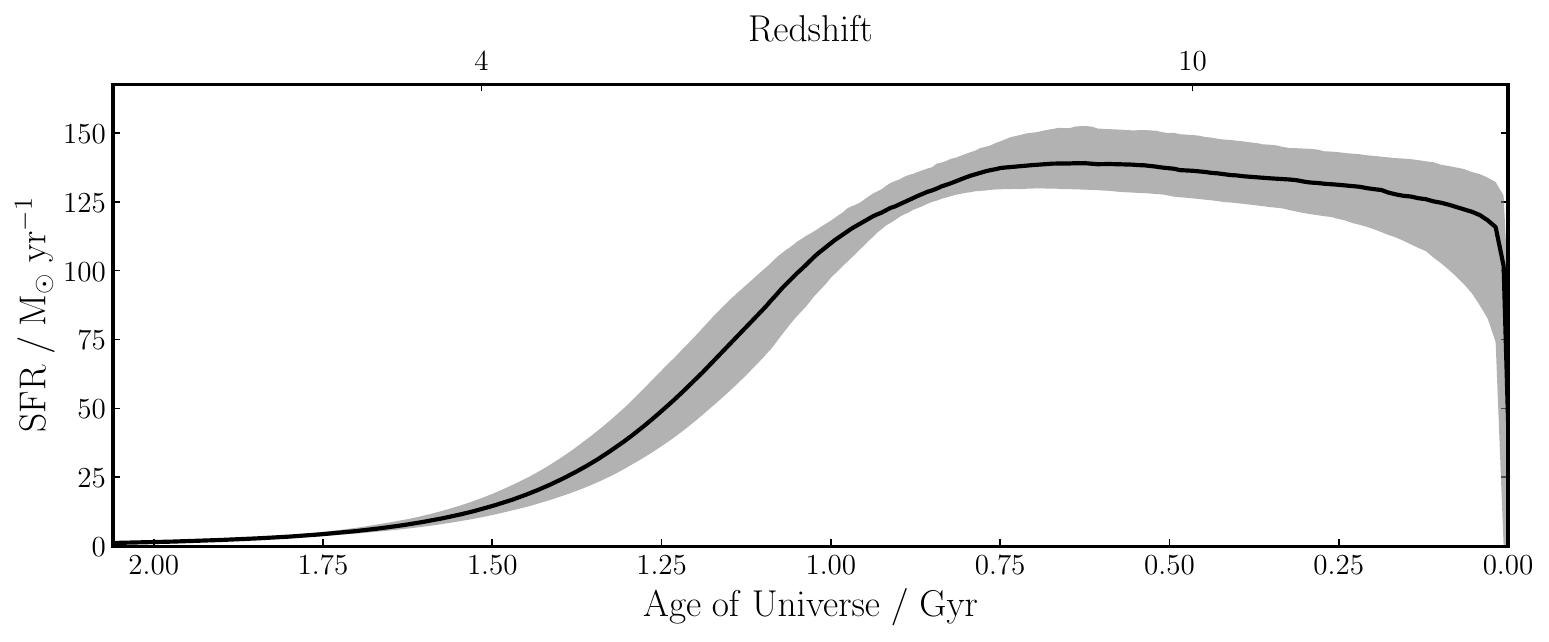}
    \caption{SED and SFH for COSMOS ID 310229}
    \label{SED_310229}
\end{figure*}

\begin{figure*}[!h]
    \centering
    \includegraphics[width=0.49\textwidth]{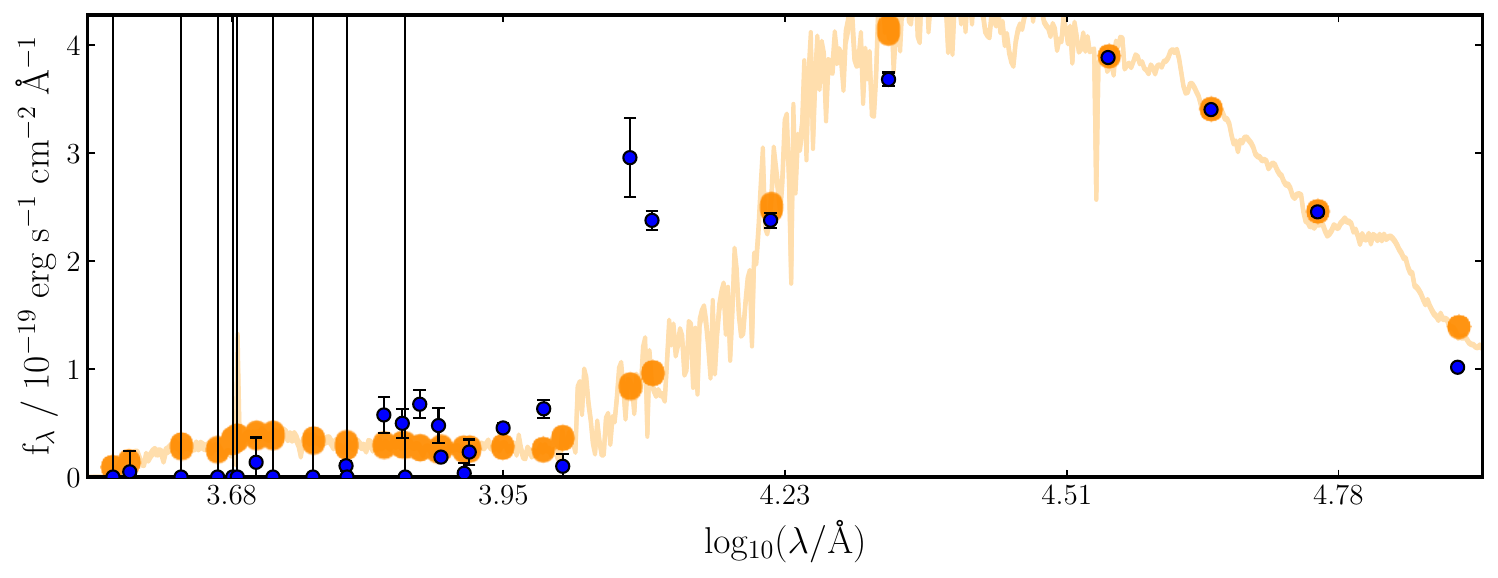}
    \includegraphics[width=0.49\textwidth]{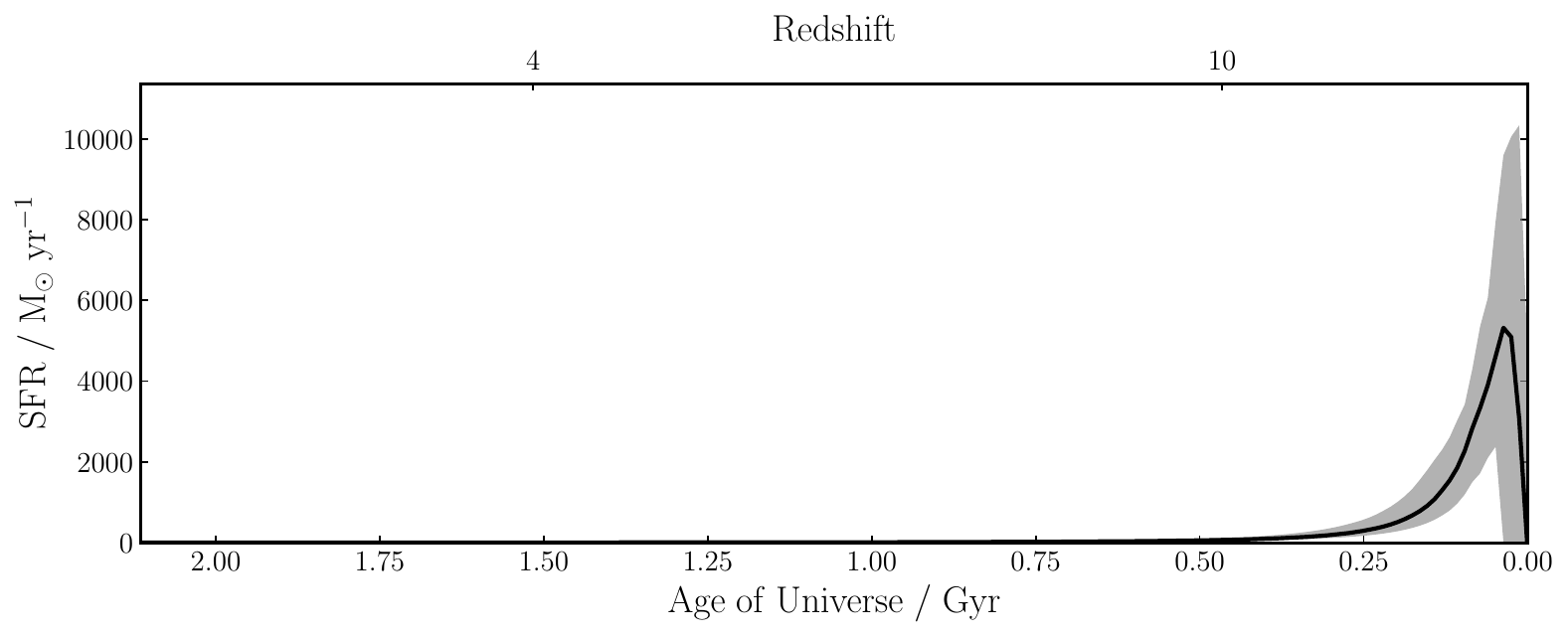}
    \caption{SED and SFH for COSMOS ID 341682}
    \label{SED_341682}
\end{figure*}

\begin{figure*}[!h]
    \centering
    \includegraphics[width=0.49\textwidth]{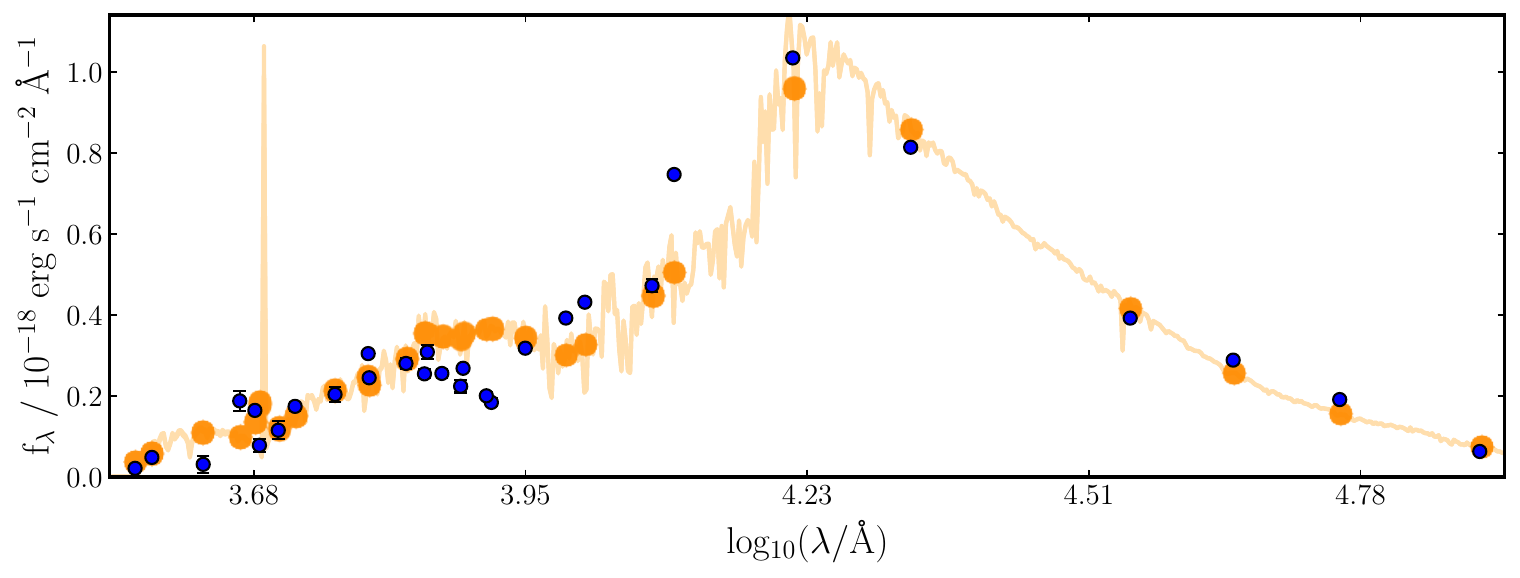}
    \includegraphics[width=0.49\textwidth]{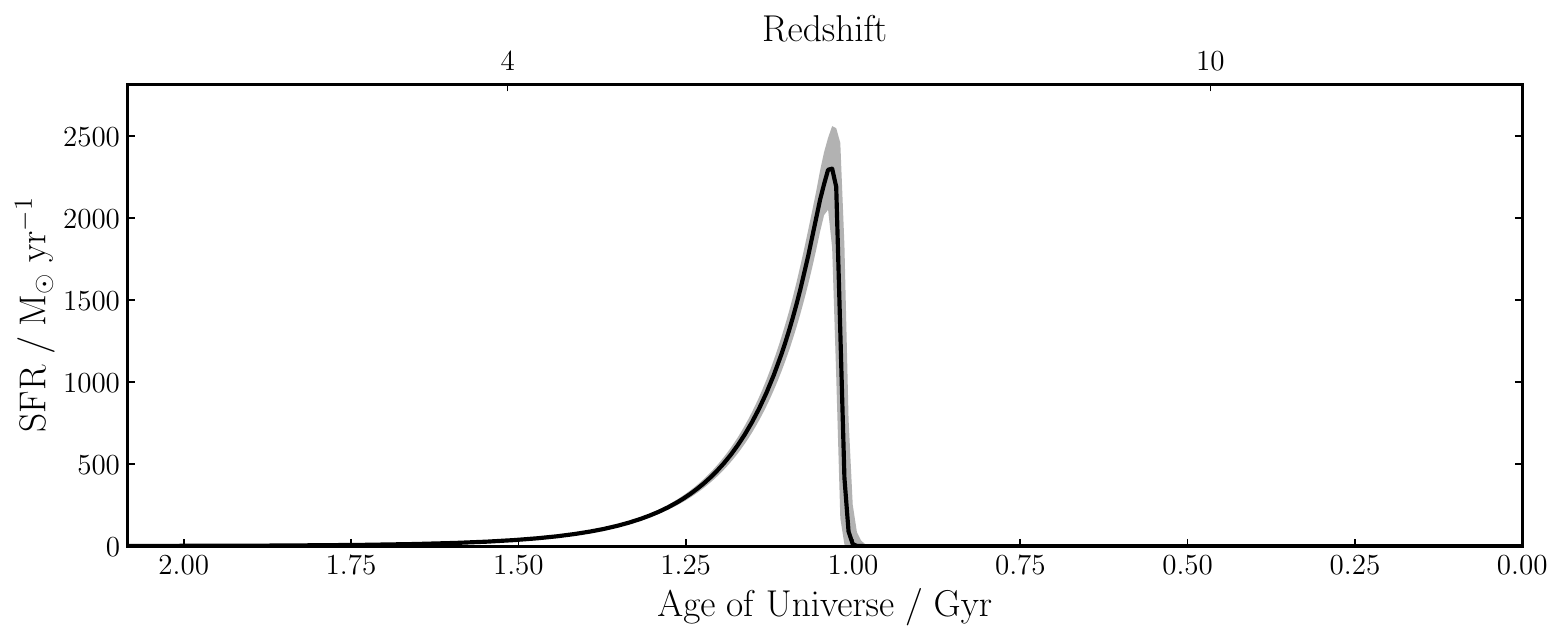}
    \caption{SED and SFH for COSMOS ID 343373}
    \label{SED_343373}
\end{figure*}

\begin{figure*}[!h]
    \centering
    \includegraphics[width=0.49\textwidth]{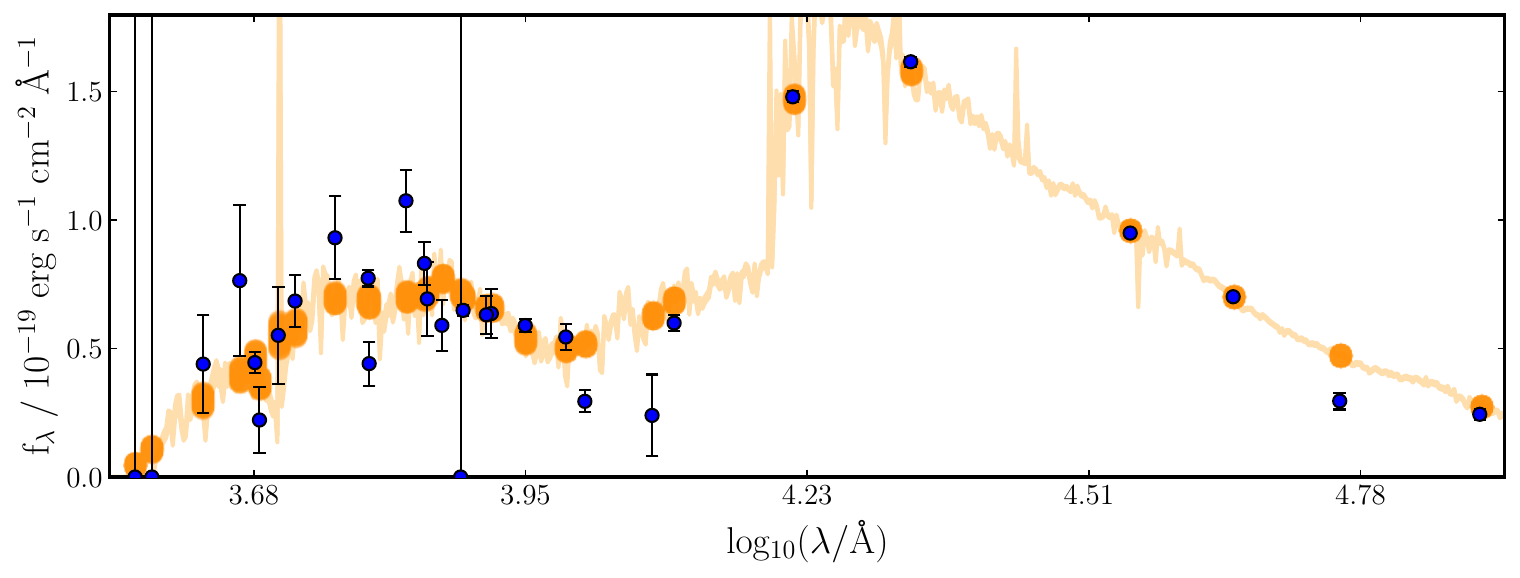}
    \includegraphics[width=0.49\textwidth]{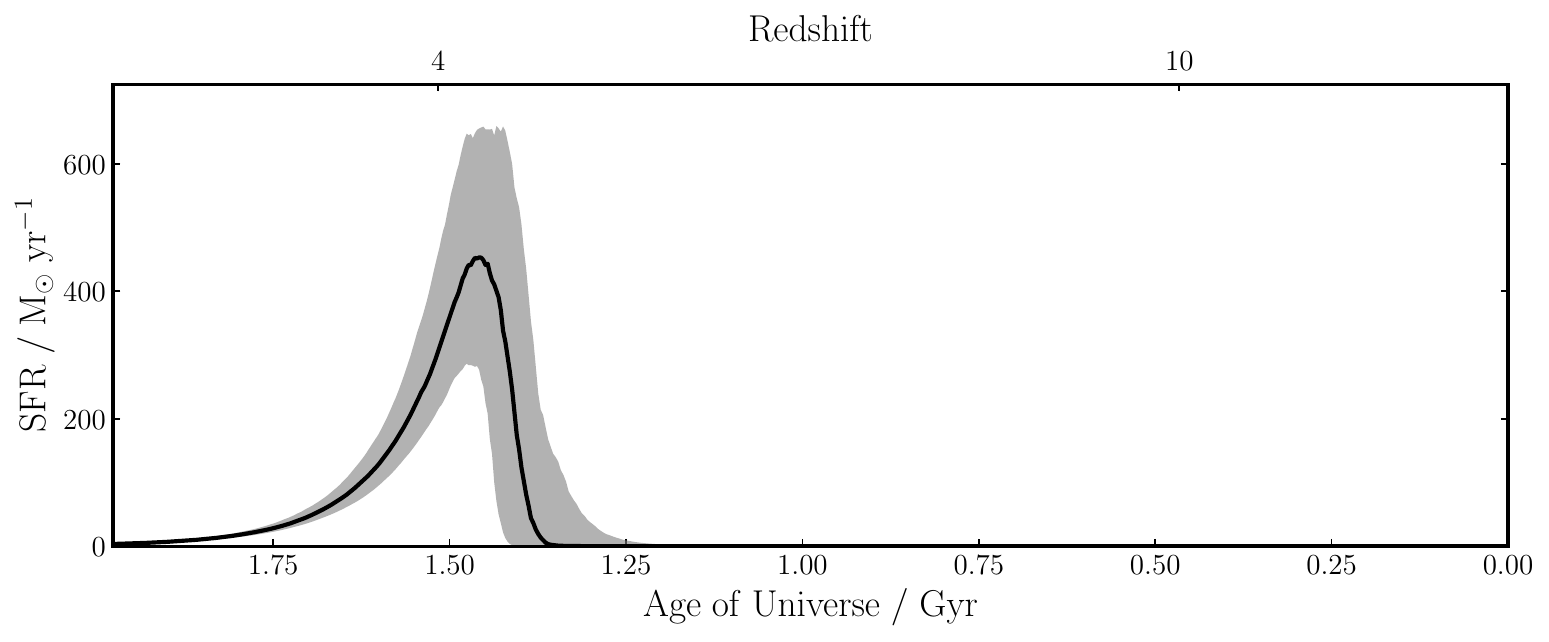}
    \caption{SED and SFH for COSMOS ID 383298}
    \label{SED_383298}
\end{figure*}

\begin{figure*}[!h]
    \centering
    \includegraphics[width=0.49\textwidth]{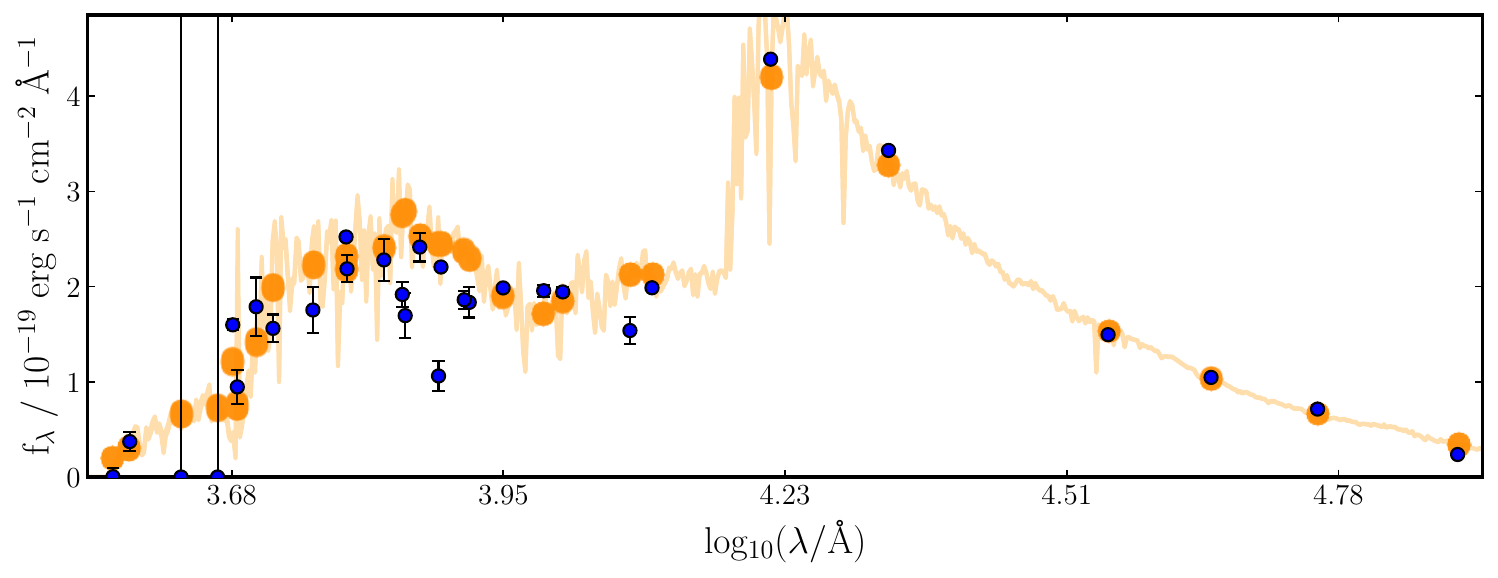}
    \includegraphics[width=0.49\textwidth]{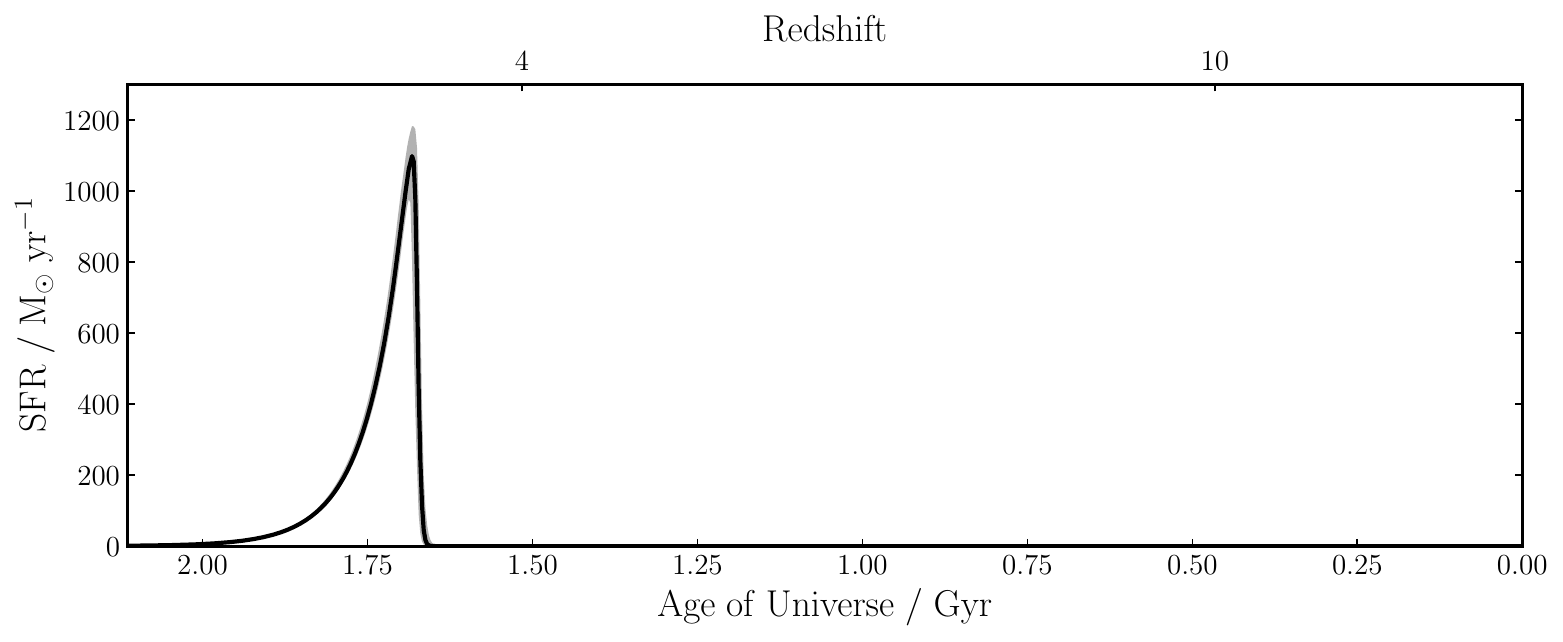}
    \caption{SED and SFH for COSMOS ID 405872}
    \label{SED_405872}
\end{figure*}

\begin{figure*}[!h]
    \centering
    \includegraphics[width=0.49\textwidth]{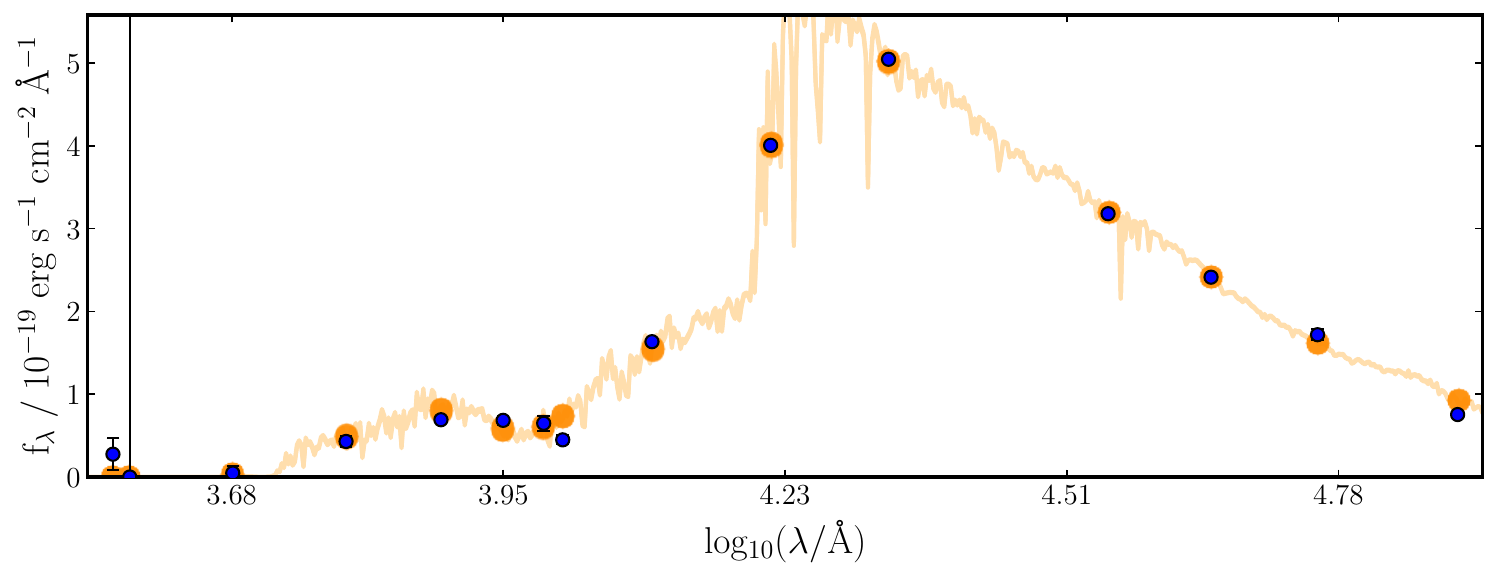}
    \includegraphics[width=0.49\textwidth]{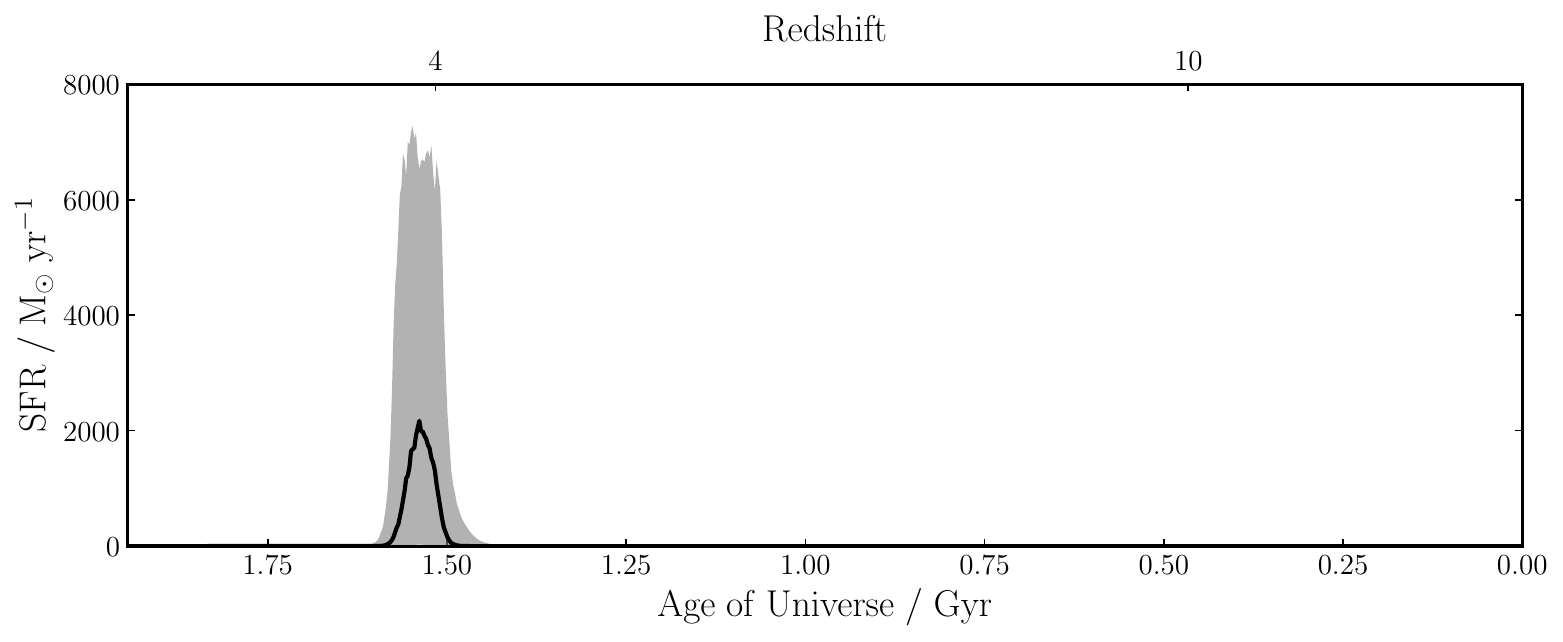}
    \caption{SED and SFH for COSMOS ID 489177}
    \label{SED_489177}
\end{figure*}

\begin{figure*}[!h]
    \centering
    \includegraphics[width=0.49\textwidth]{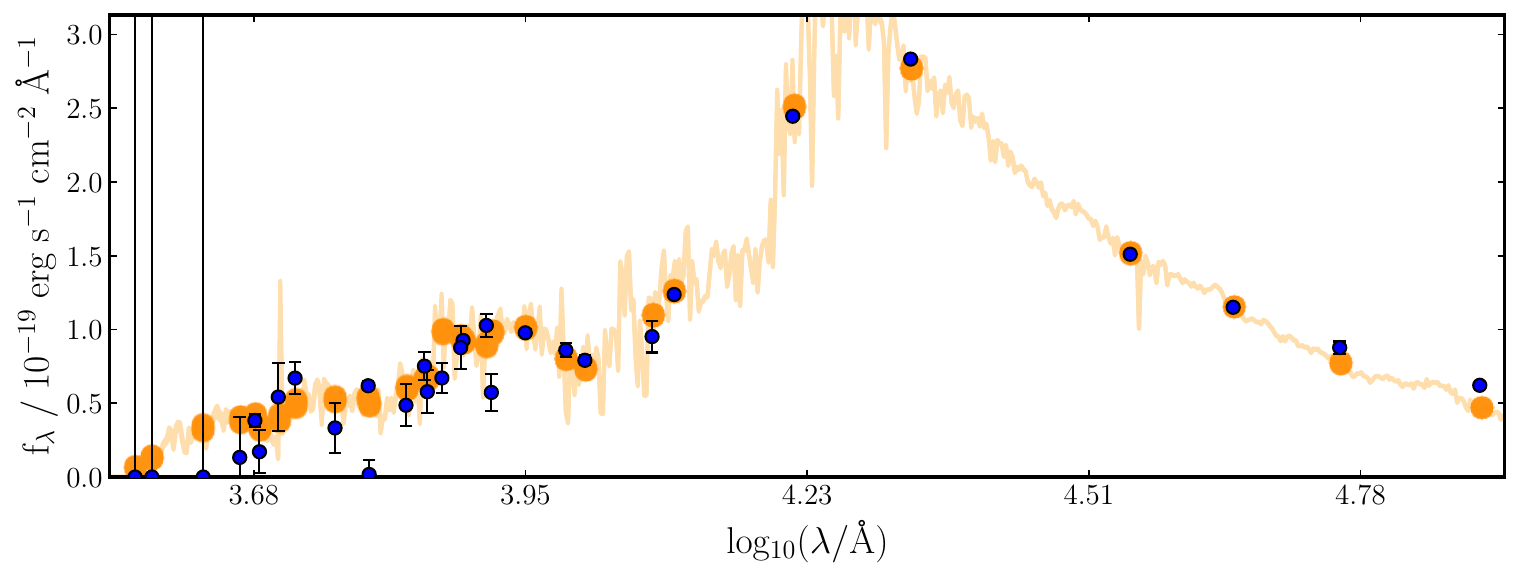}
    \includegraphics[width=0.49\textwidth]{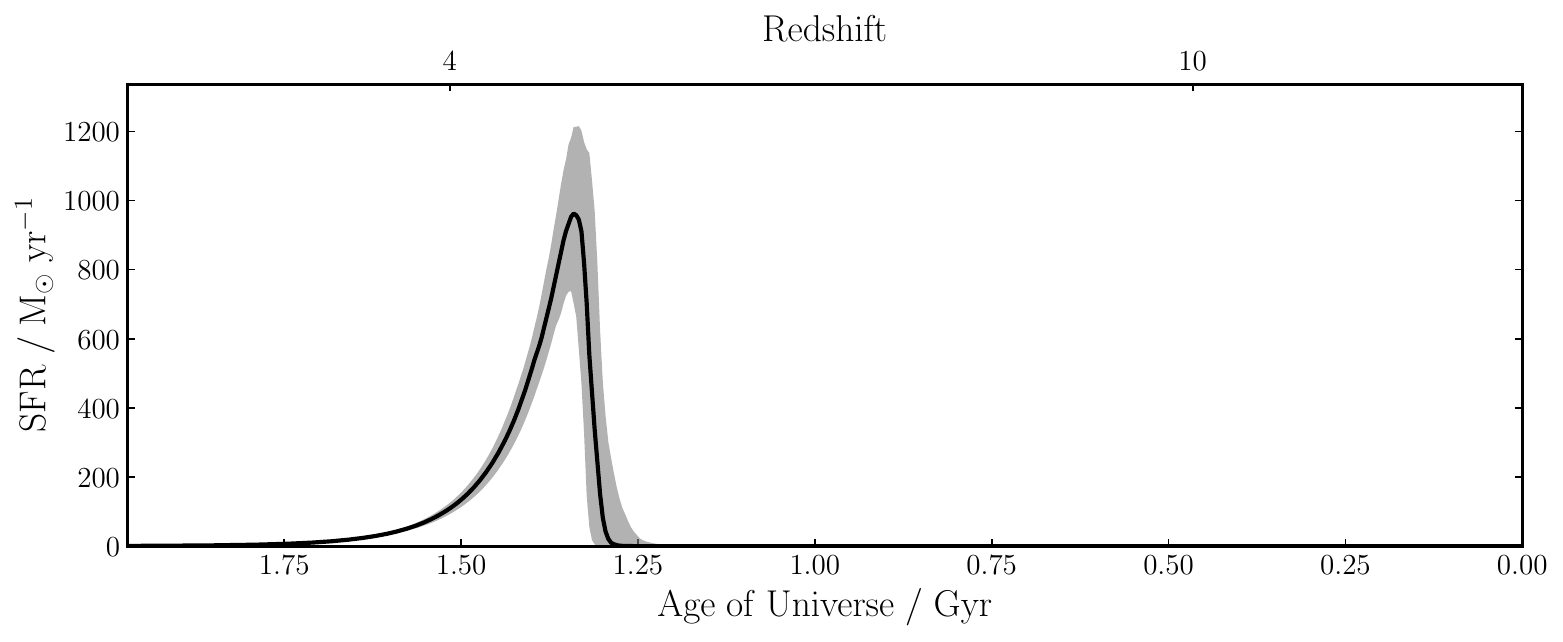}
    \caption{SED and SFH for COSMOS ID 642338}
    \label{SED_642338}
\end{figure*}

\begin{figure*}[!h]
    \centering
    \includegraphics[width=0.49\textwidth]{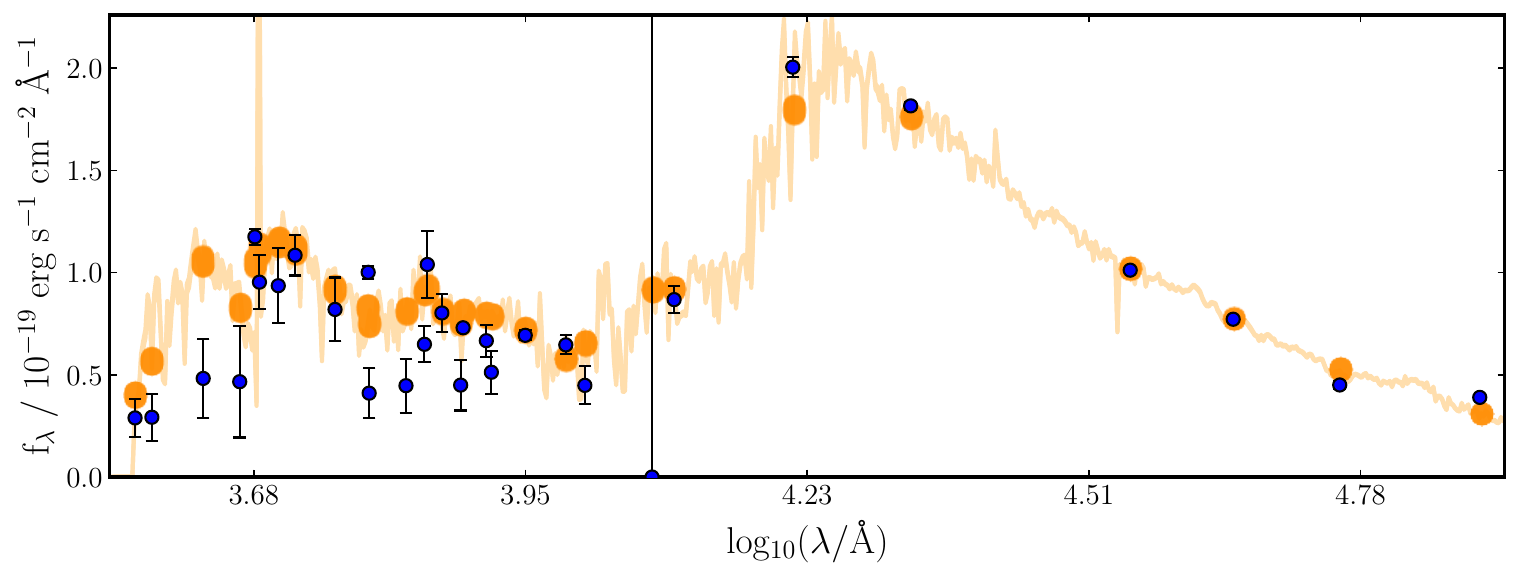}
    \includegraphics[width=0.49\textwidth]{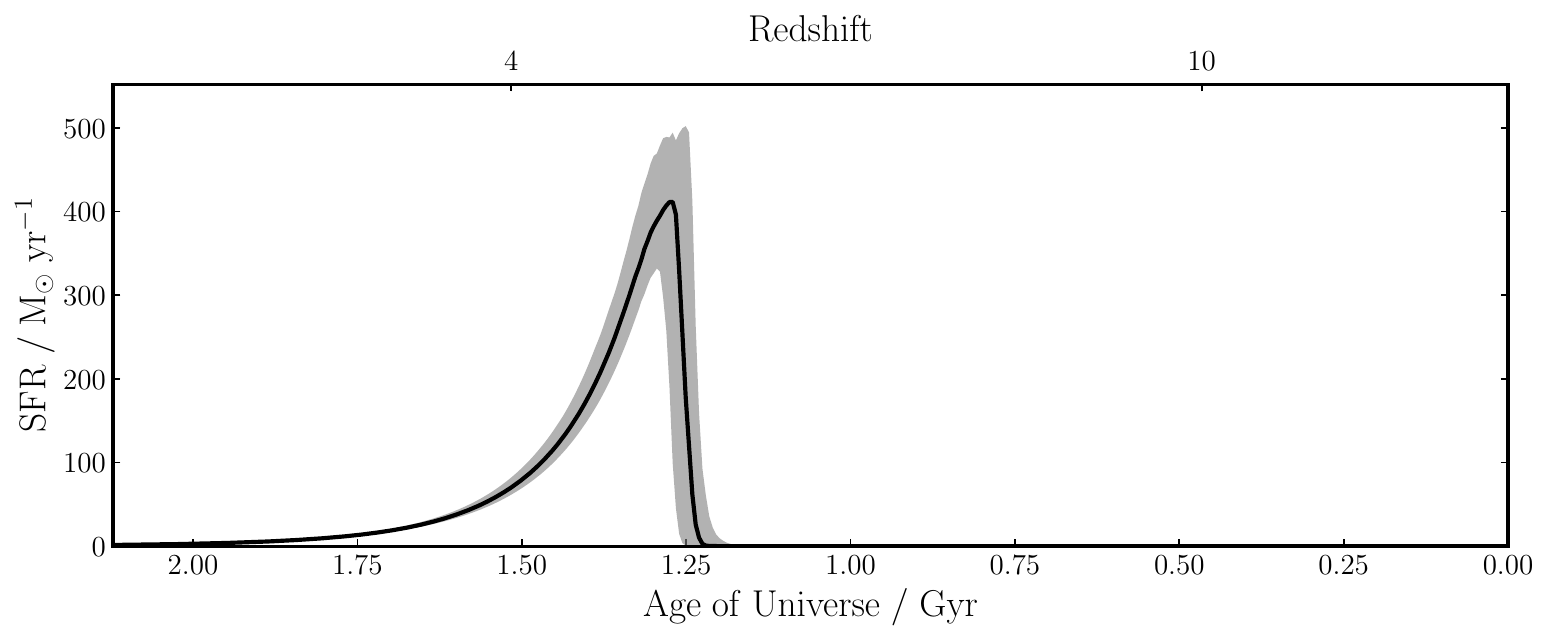}
    \caption{SED and SFH for COSMOS ID 658452}
    \label{SED_658452}
\end{figure*}

\begin{figure*}[!h]
    \centering
    \includegraphics[width=0.49\textwidth]{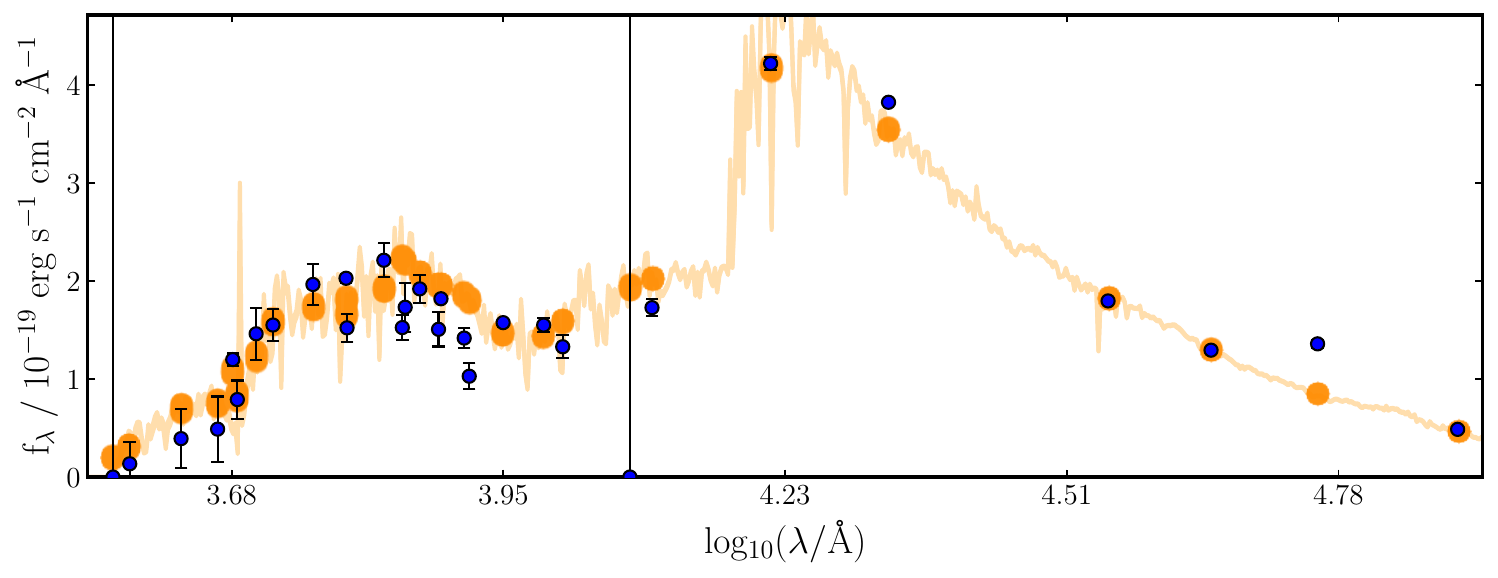}
    \includegraphics[width=0.49\textwidth]{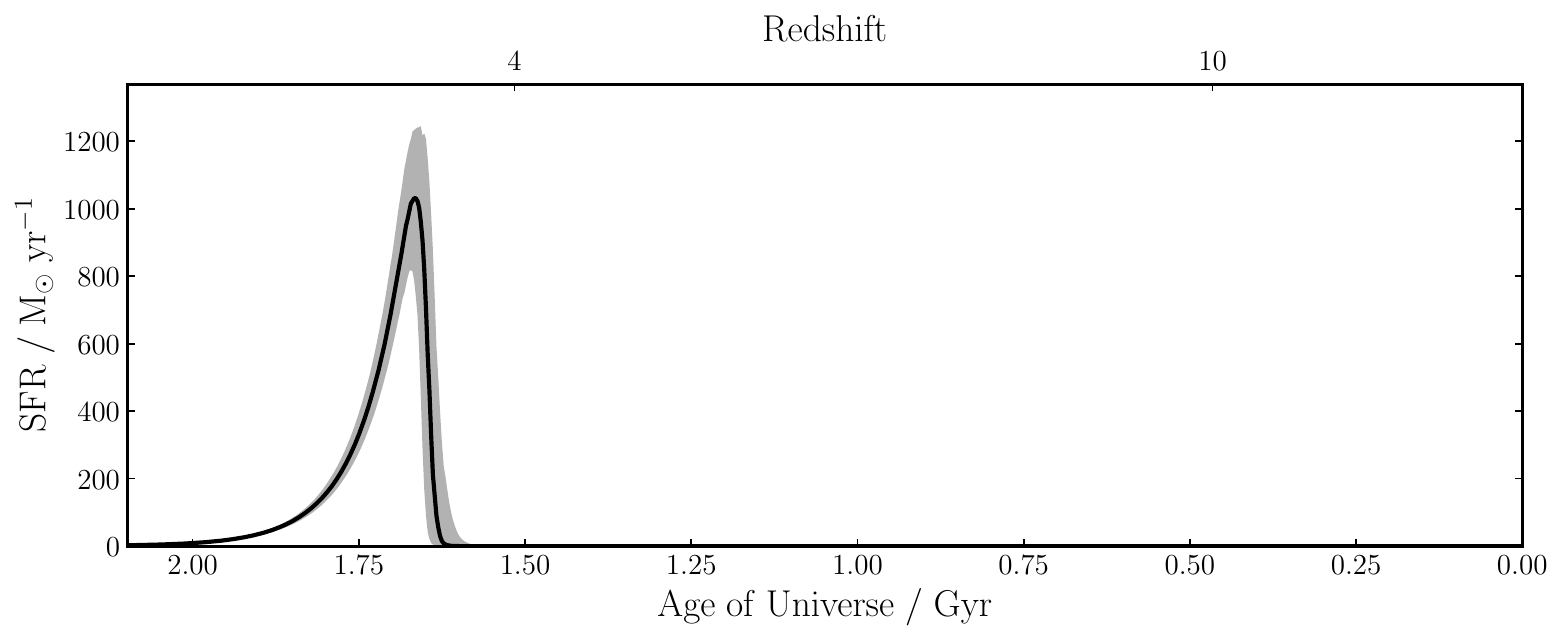}
    \caption{SED and SFH for COSMOS ID 706985}
    \label{SED_706985}
\end{figure*}

\begin{figure*}[!h]
    \centering
    \includegraphics[width=0.49\textwidth]{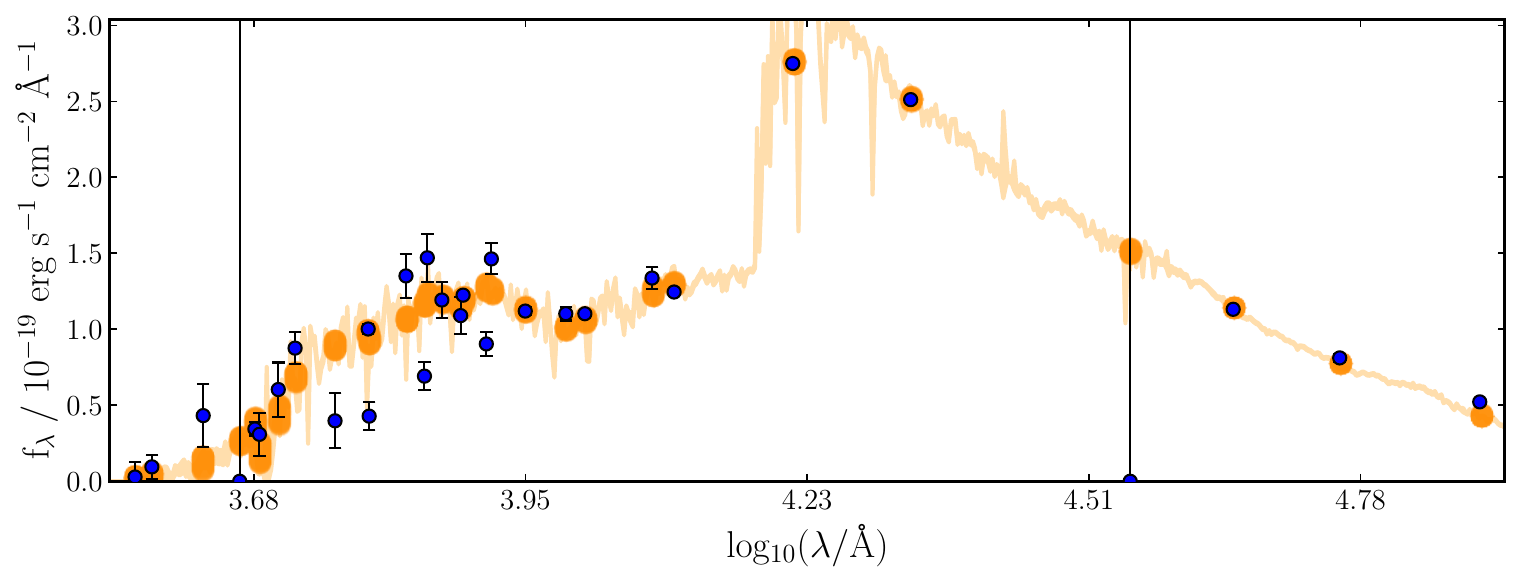}
    \includegraphics[width=0.49\textwidth]{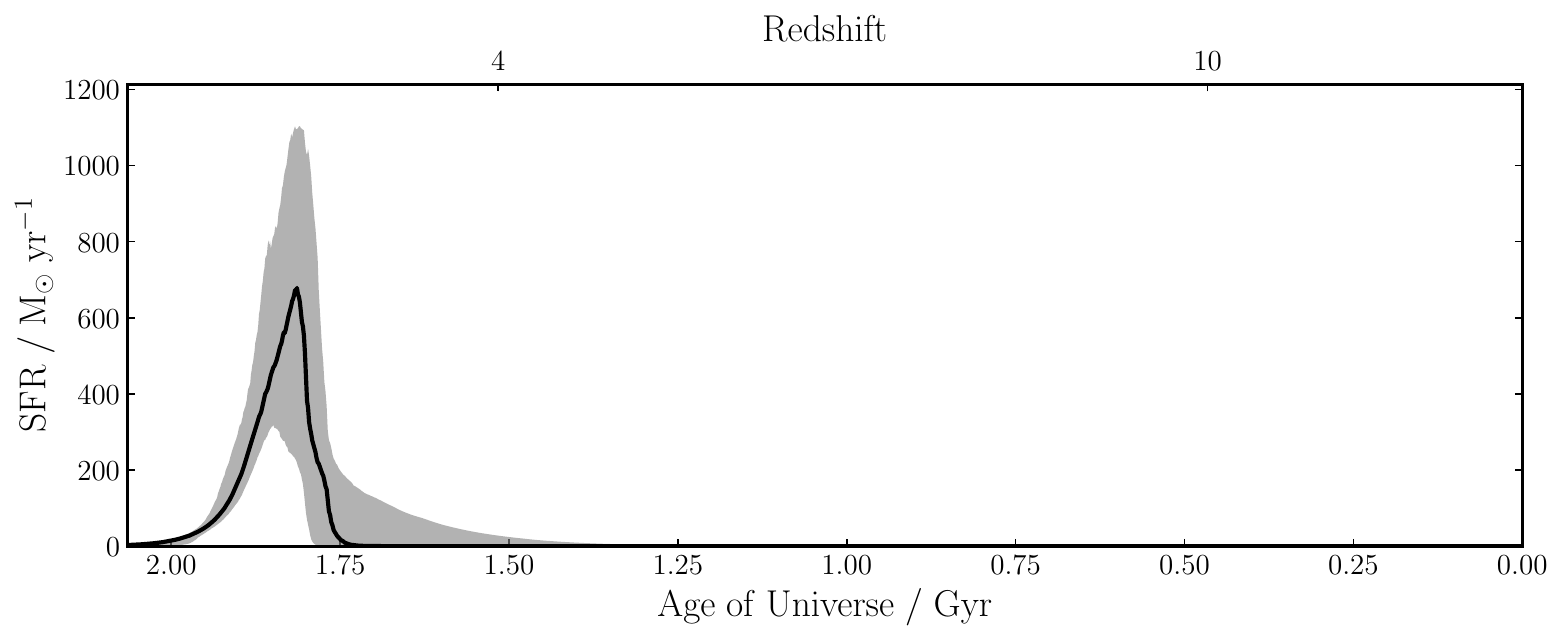}
    \caption{SED and SFH for COSMOS ID 779869}
    \label{SED_779869}
\end{figure*}

\begin{figure*}[!h]
    \centering
    \includegraphics[width=0.49\textwidth]{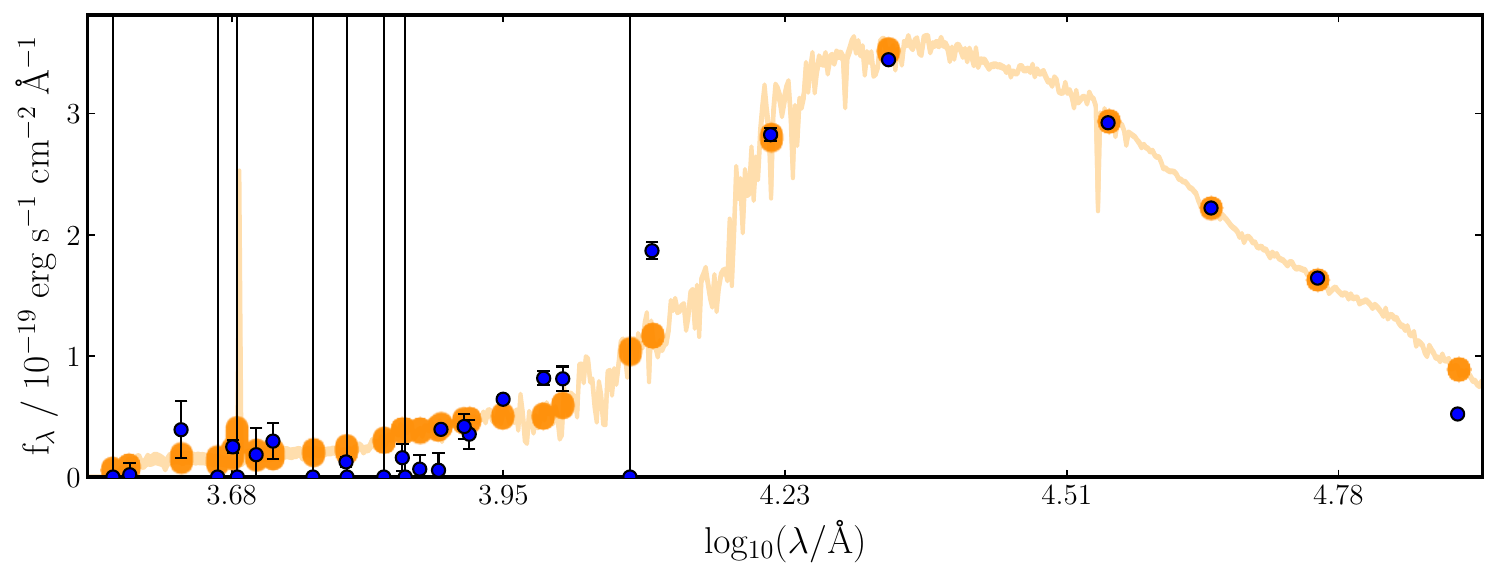}
    \includegraphics[width=0.49\textwidth]{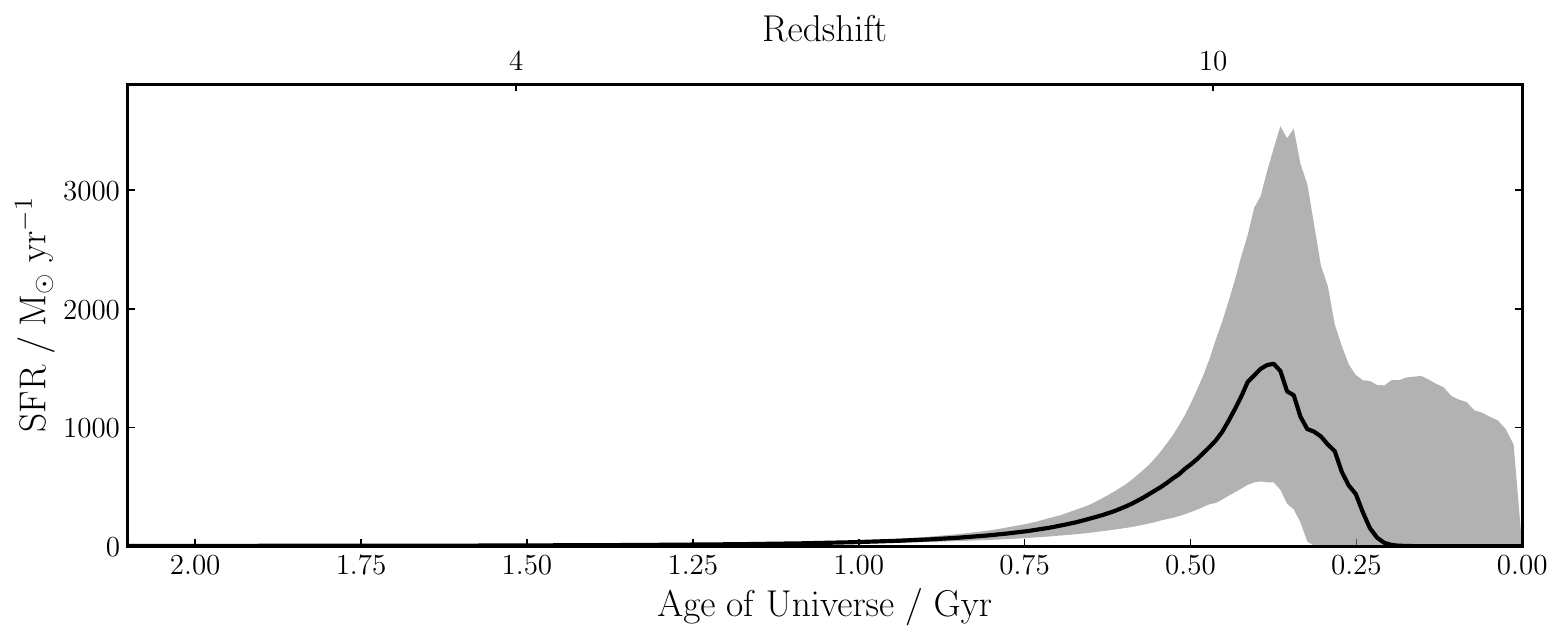}
    \caption{SED and SFH for COSMOS ID 803389}
    \label{SED_803389}
\end{figure*}

\begin{figure*}[!h]
    \centering
    \includegraphics[width=0.49\textwidth]{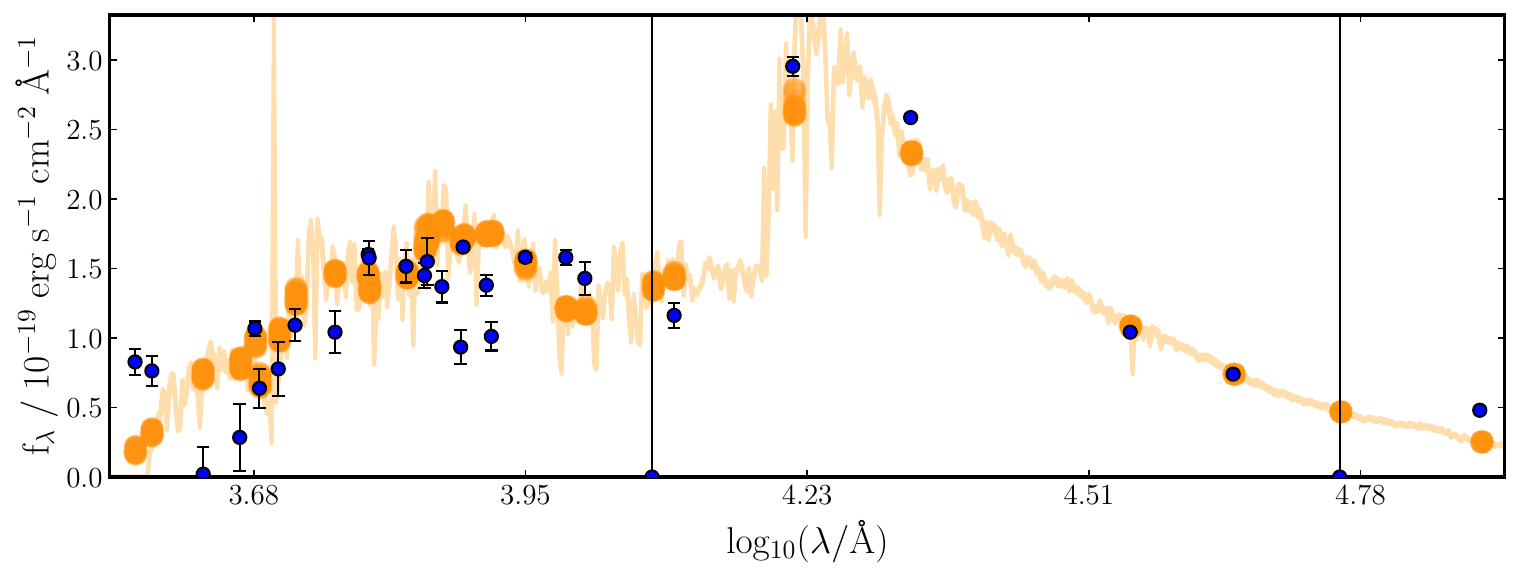}
    \includegraphics[width=0.49\textwidth]{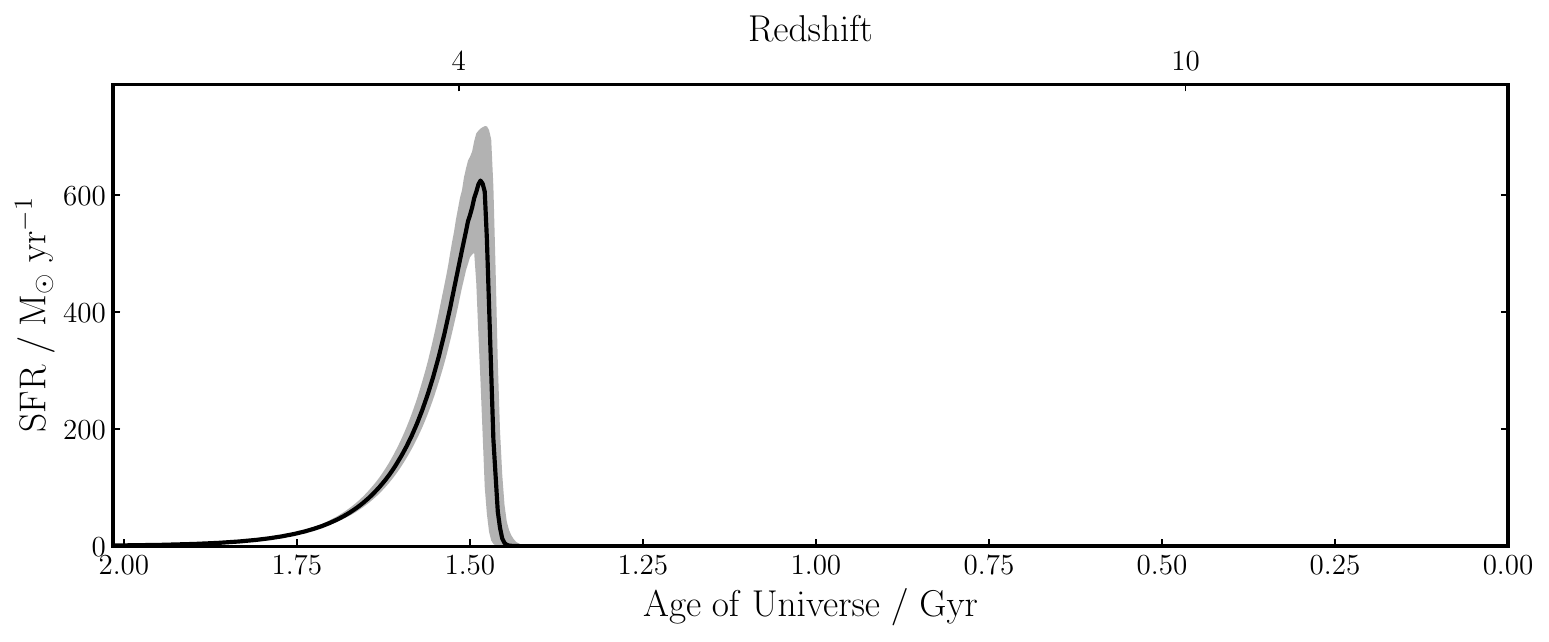}
    \caption{SED and SFH for COSMOS ID 881980}
    \label{SED_881980}
\end{figure*}

\begin{figure*}[!h]
    \centering
    \includegraphics[width=0.49\textwidth]{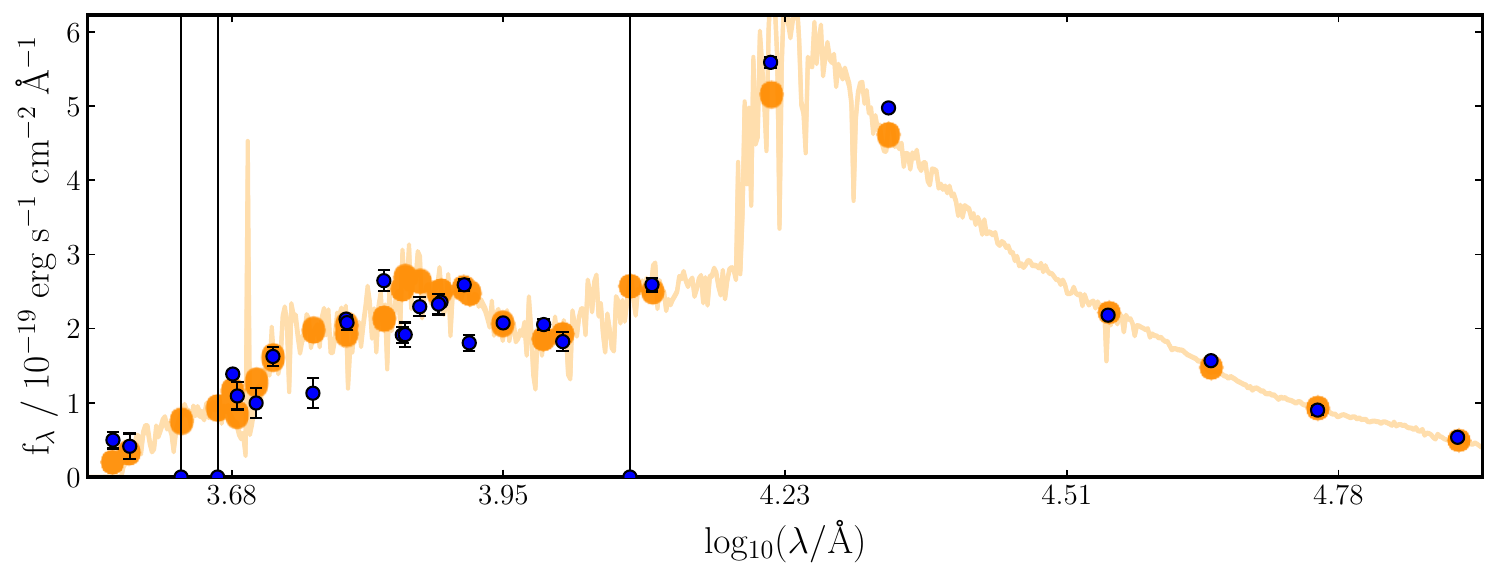}
    \includegraphics[width=0.49\textwidth]{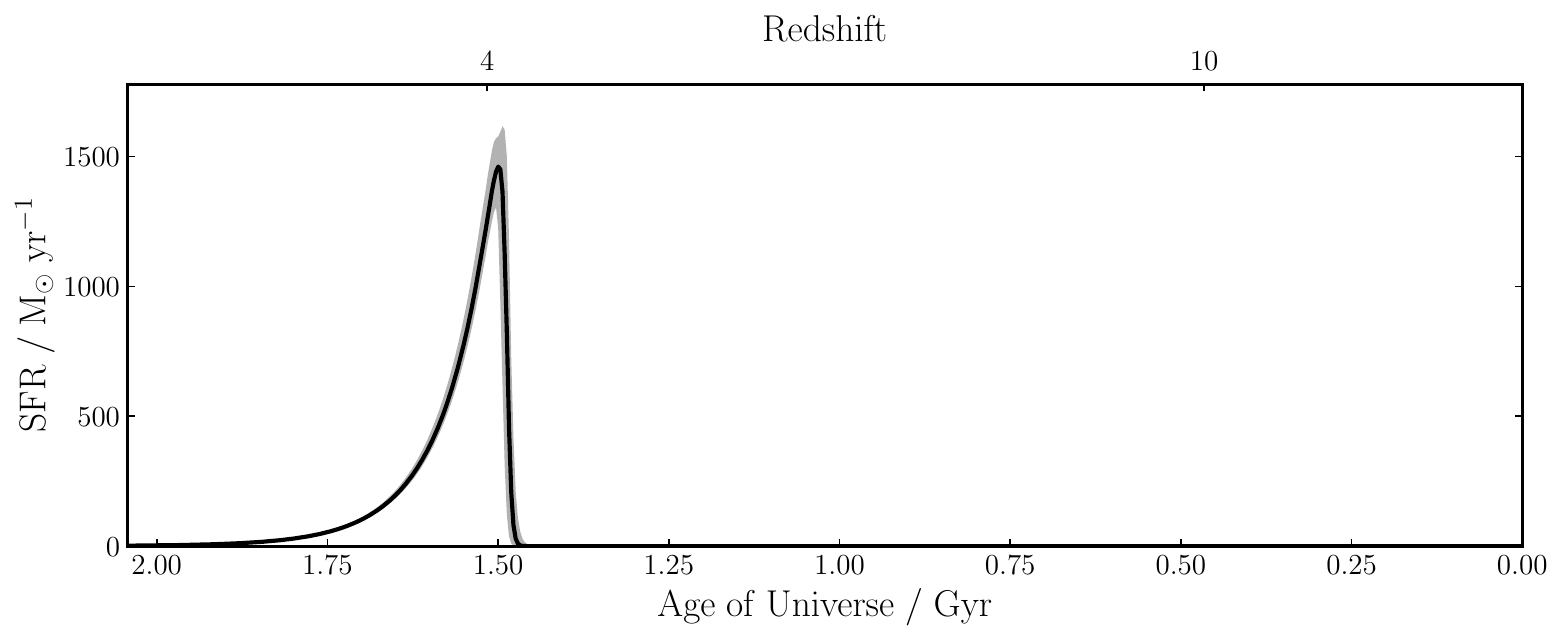}
    \caption{SED and SFH for COSMOS ID 911001}
    \label{SED_911001}
\end{figure*}

\begin{figure*}[!h]
    \centering
    \includegraphics[width=0.49\textwidth]{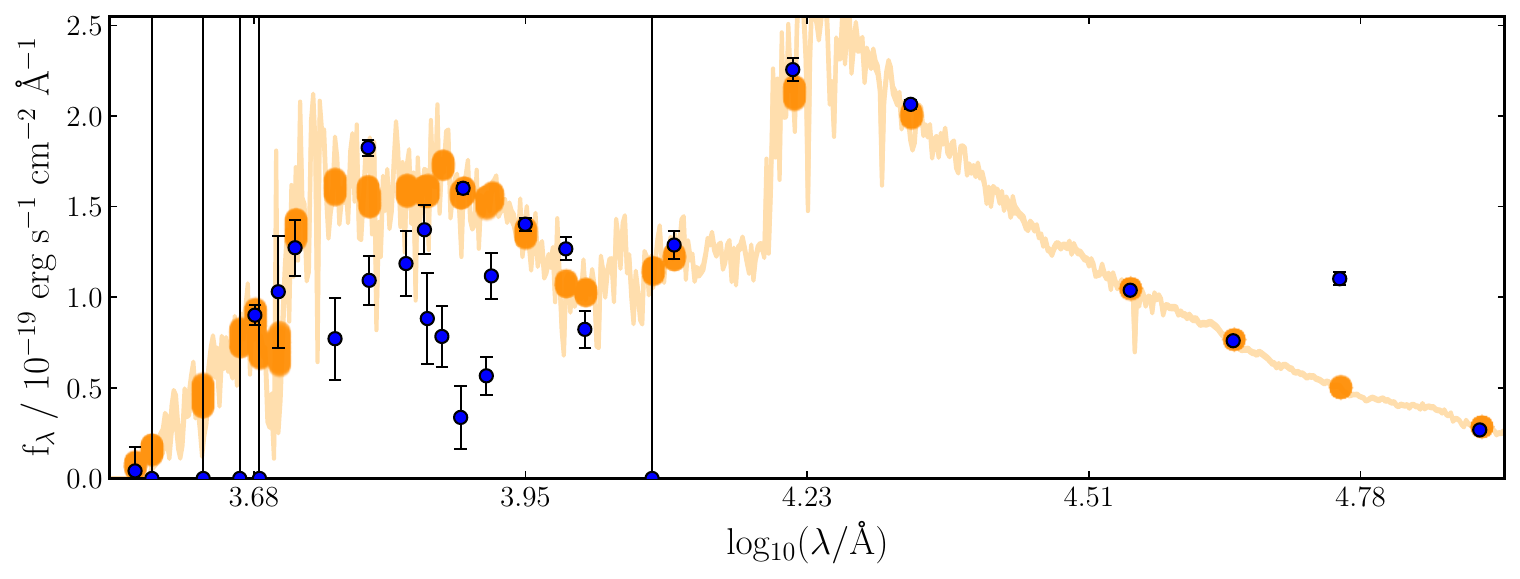}
    \includegraphics[width=0.49\textwidth]{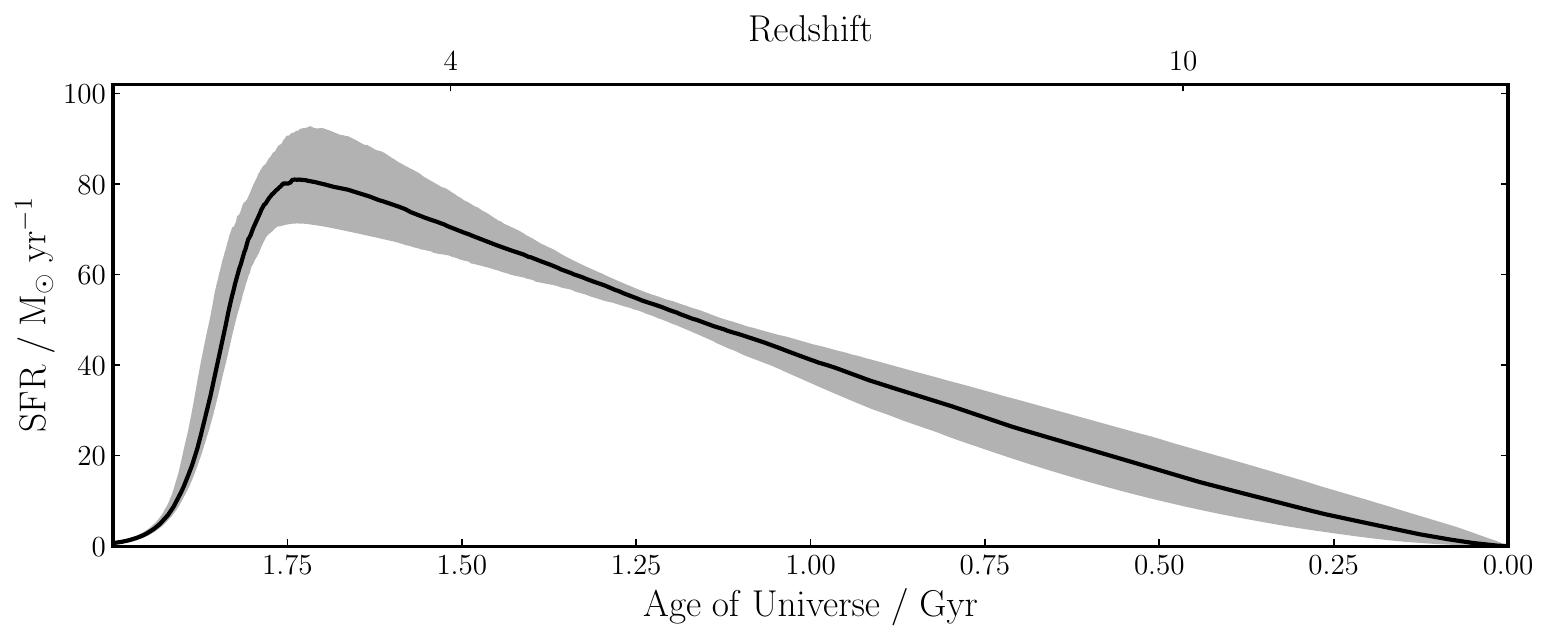}
    \caption{SED and SFH for COSMOS ID 961549}
    \label{SED_961549}
\end{figure*}

\begin{figure*}[!h]
    \centering
    \includegraphics[width=0.49\textwidth]{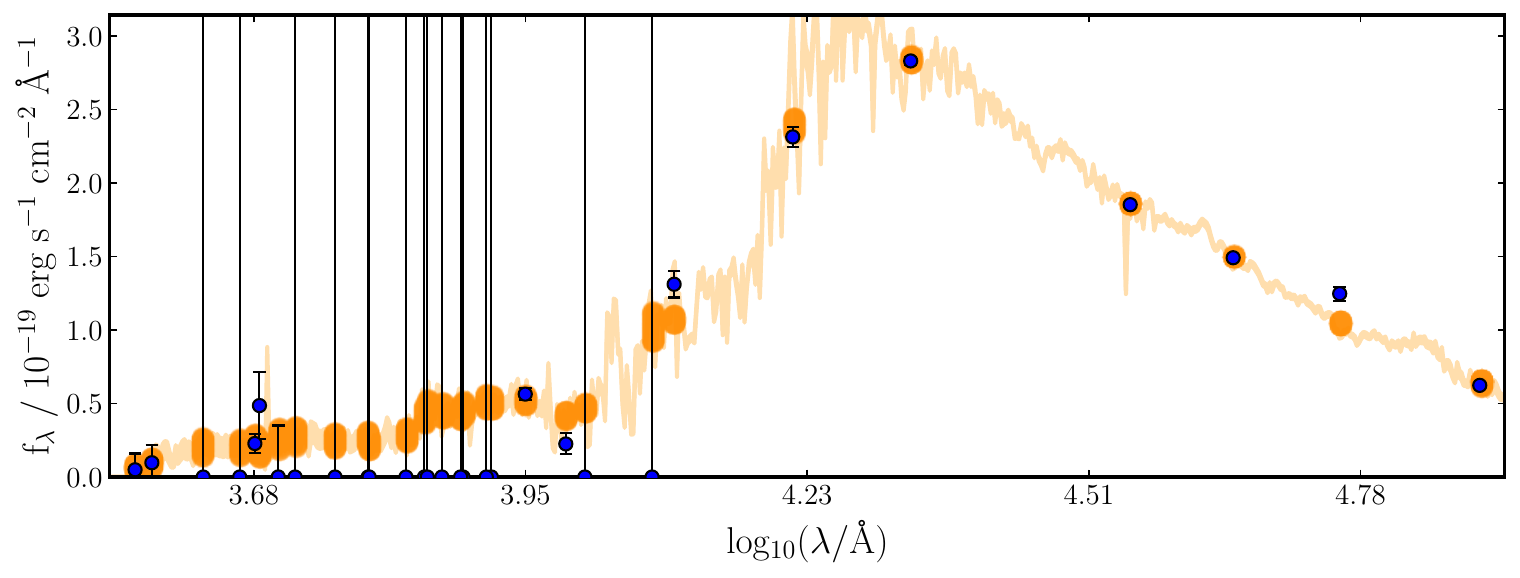}
    \includegraphics[width=0.49\textwidth]{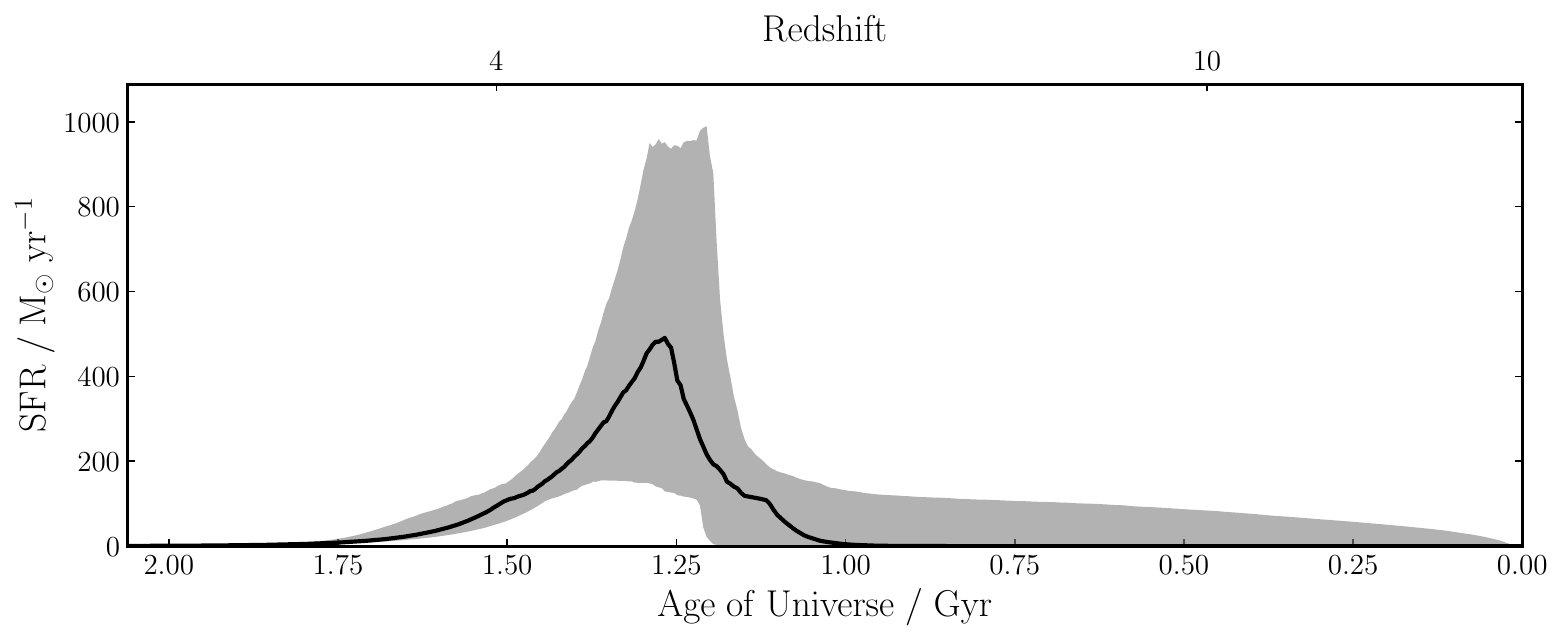}
    \caption{SED and SFH for COSMOS ID 962569}
    \label{SED_962569}
\end{figure*}

\begin{figure*}[!h]
    \centering
    \includegraphics[width=0.49\textwidth]{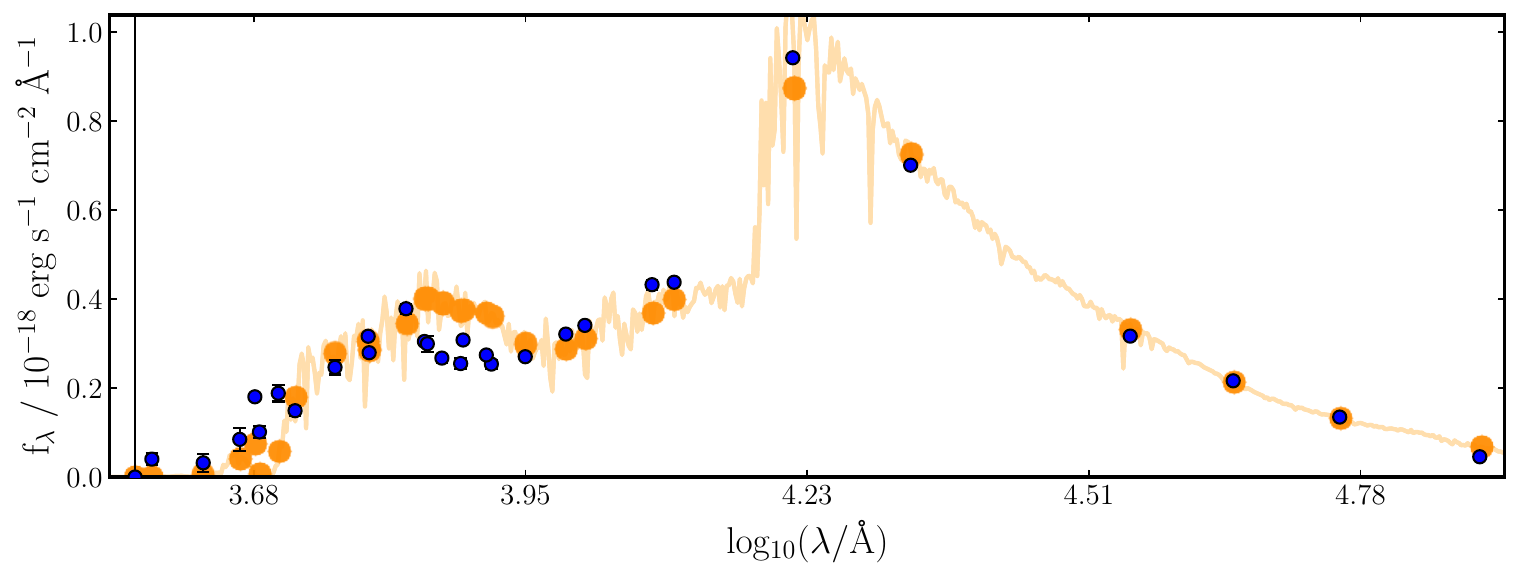}
    \includegraphics[width=0.49\textwidth]{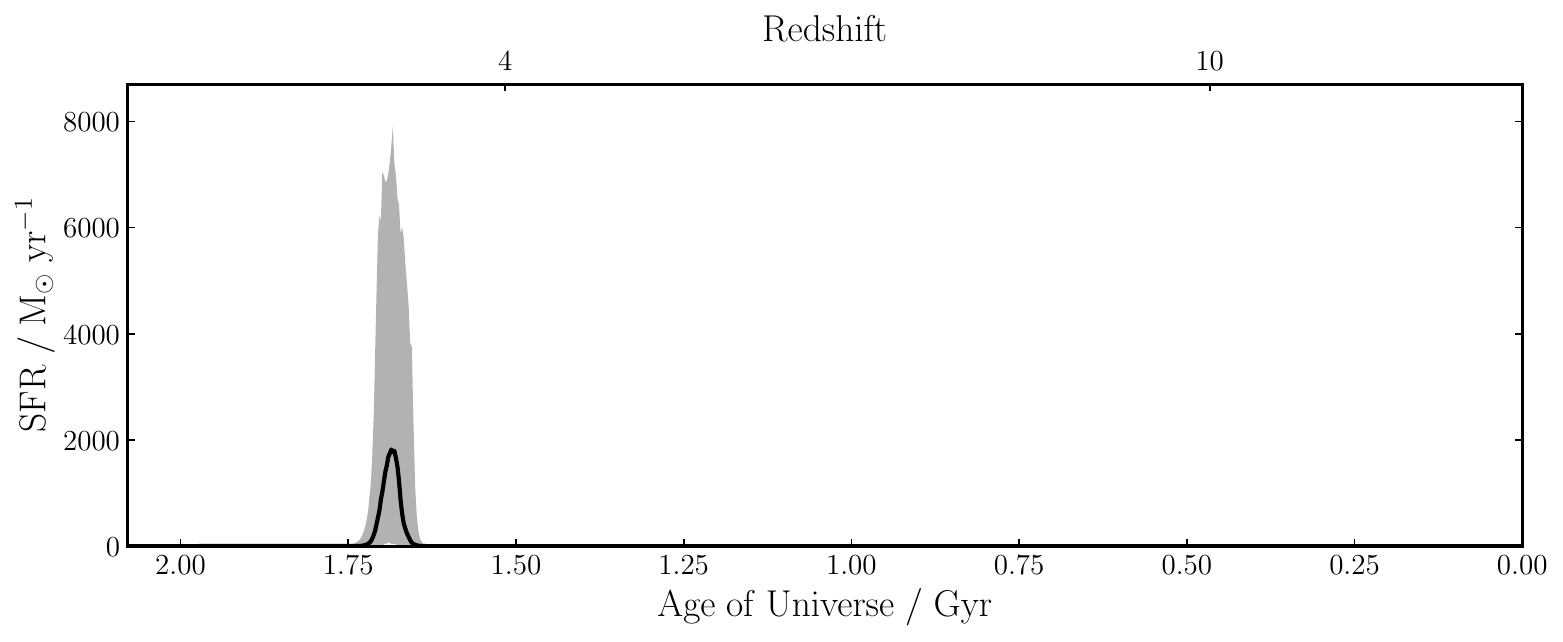}
    \caption{SED and SFH for COSMOS ID 338392}
    \label{SED_338392}
\end{figure*}

\begin{figure*}[!h]
    \centering
    \includegraphics[width=0.49\textwidth]{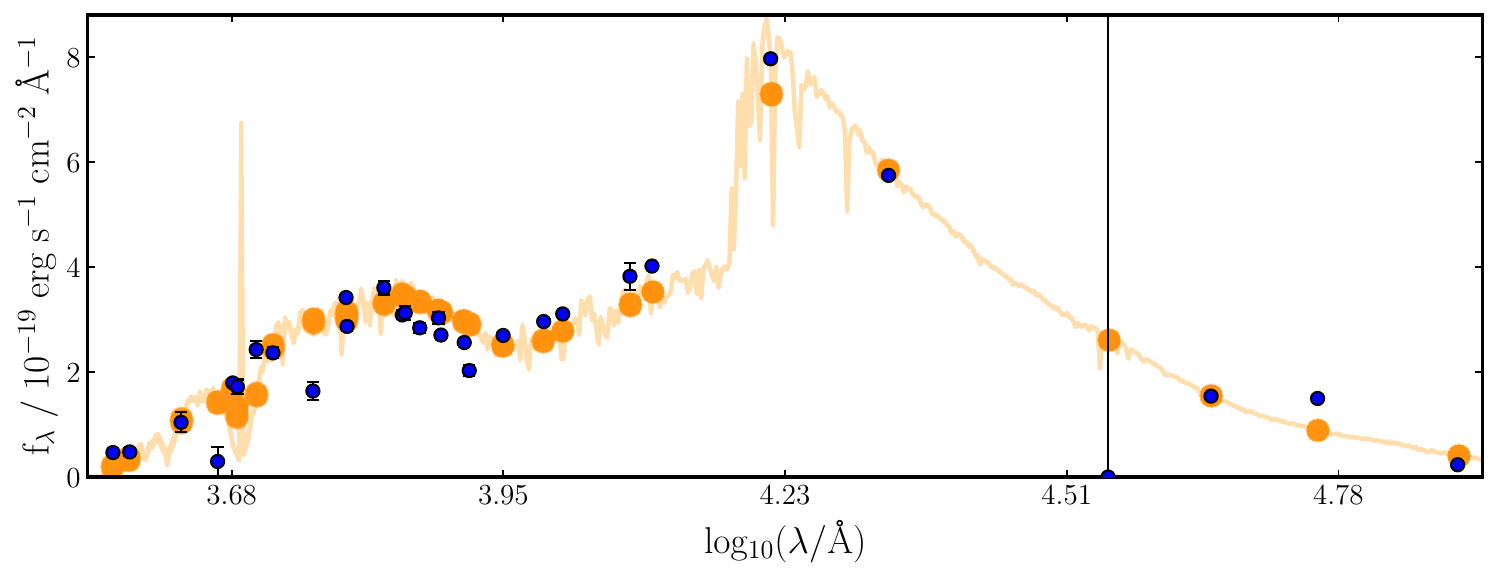}
    \includegraphics[width=0.49\textwidth]{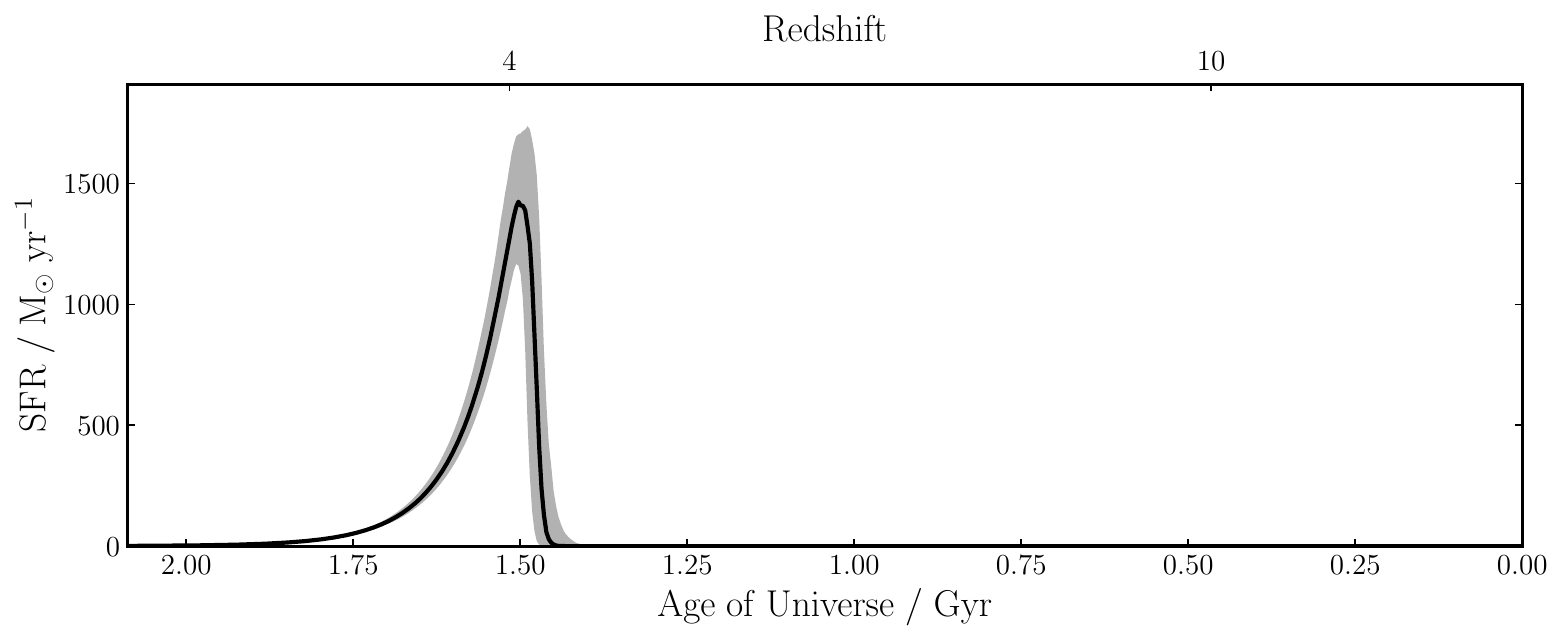}
    \caption{SED and SFH for COSMOS ID 681407}
    \label{SED_681407}
\end{figure*}

\section{Corner Plots for SED fitting}
\begin{figure*}[!h]
    \centering
    \includegraphics[width=0.49\textwidth]{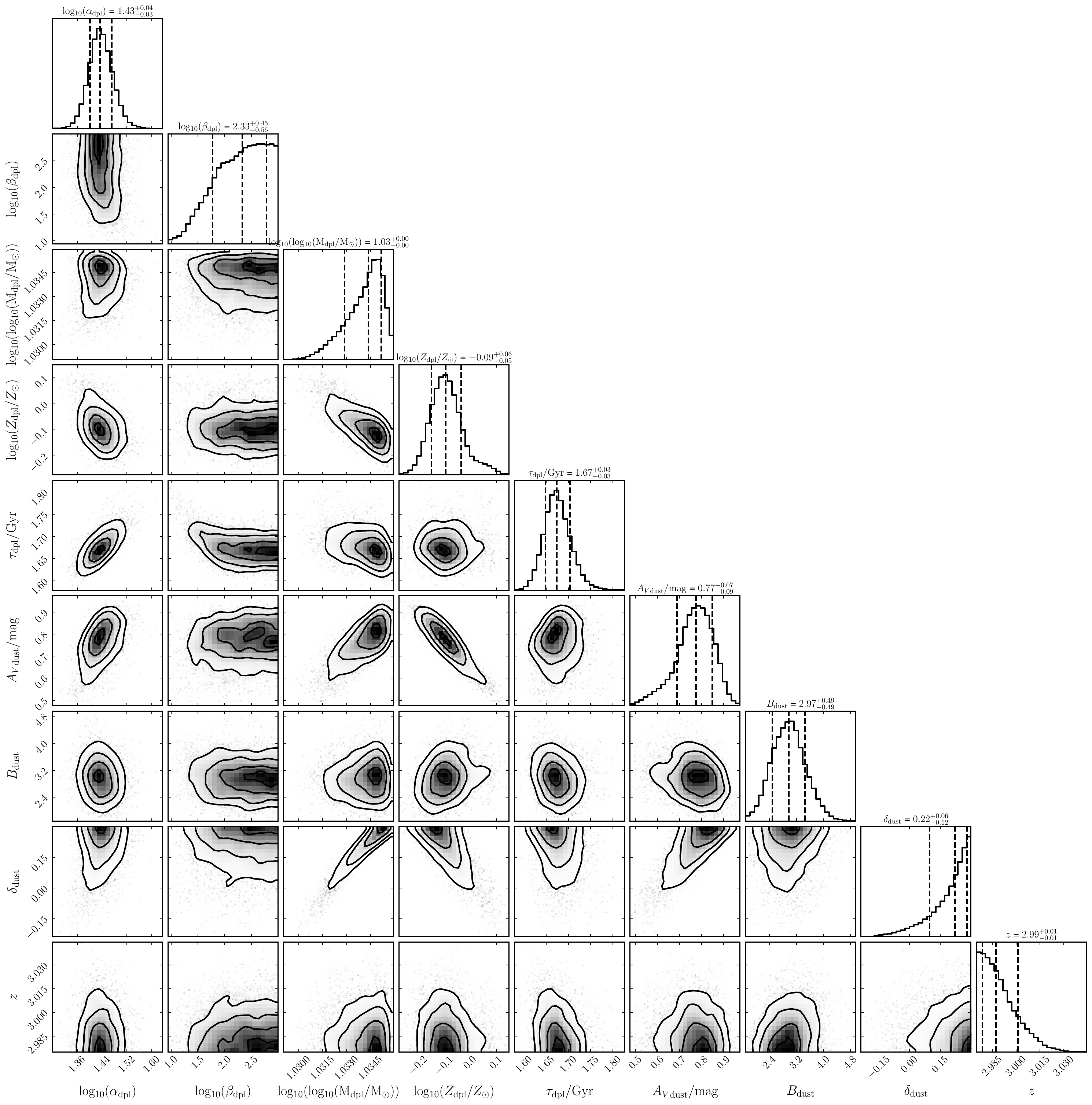}
    \includegraphics[width=0.49\textwidth]{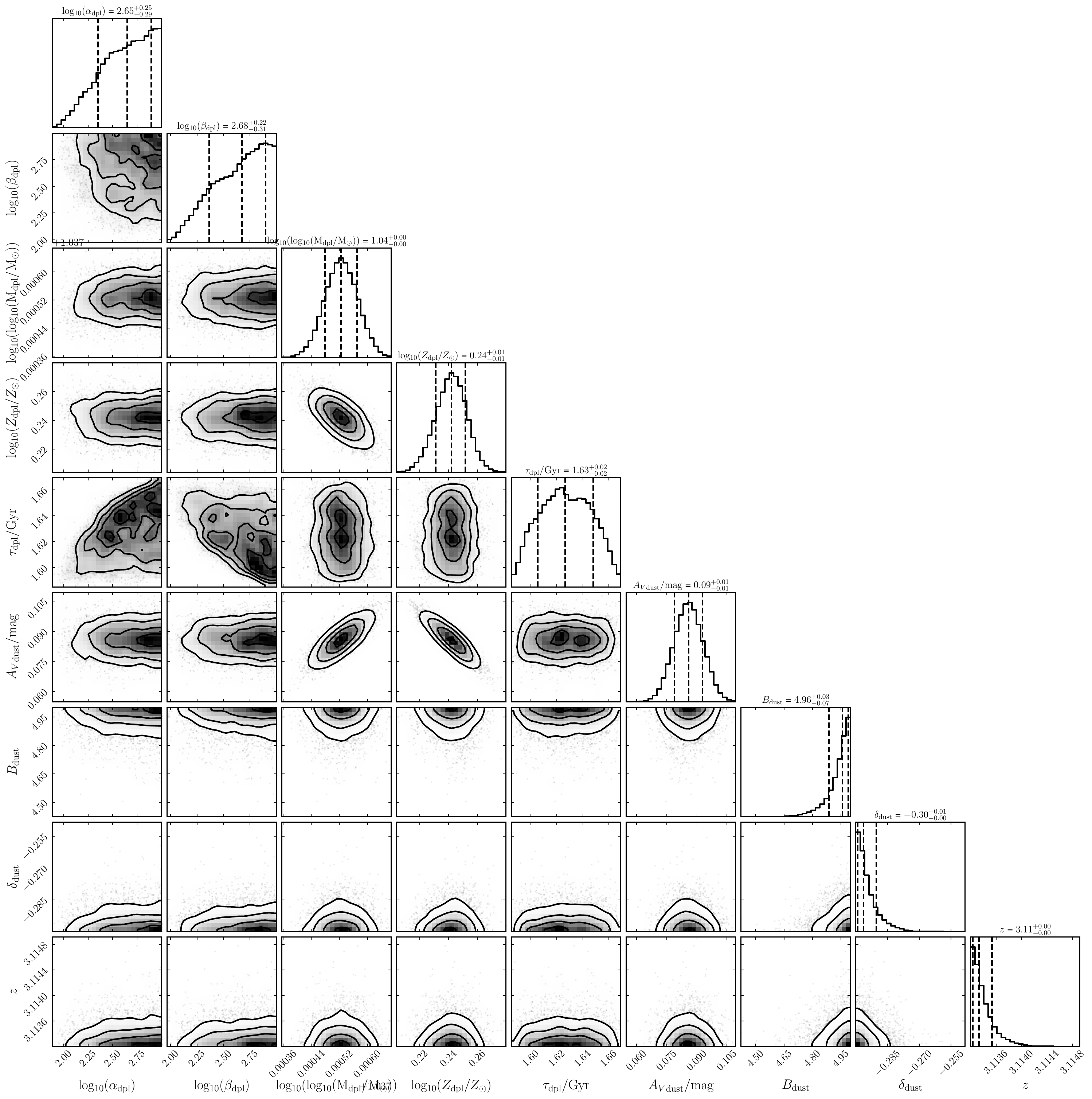}
    \caption{Corner plot for COSMOS ID 13648 and 20998}
    \label{SED_ID13648_20998}
\end{figure*}

\begin{figure*}[!h]
    \centering
    \includegraphics[width=0.49\textwidth]{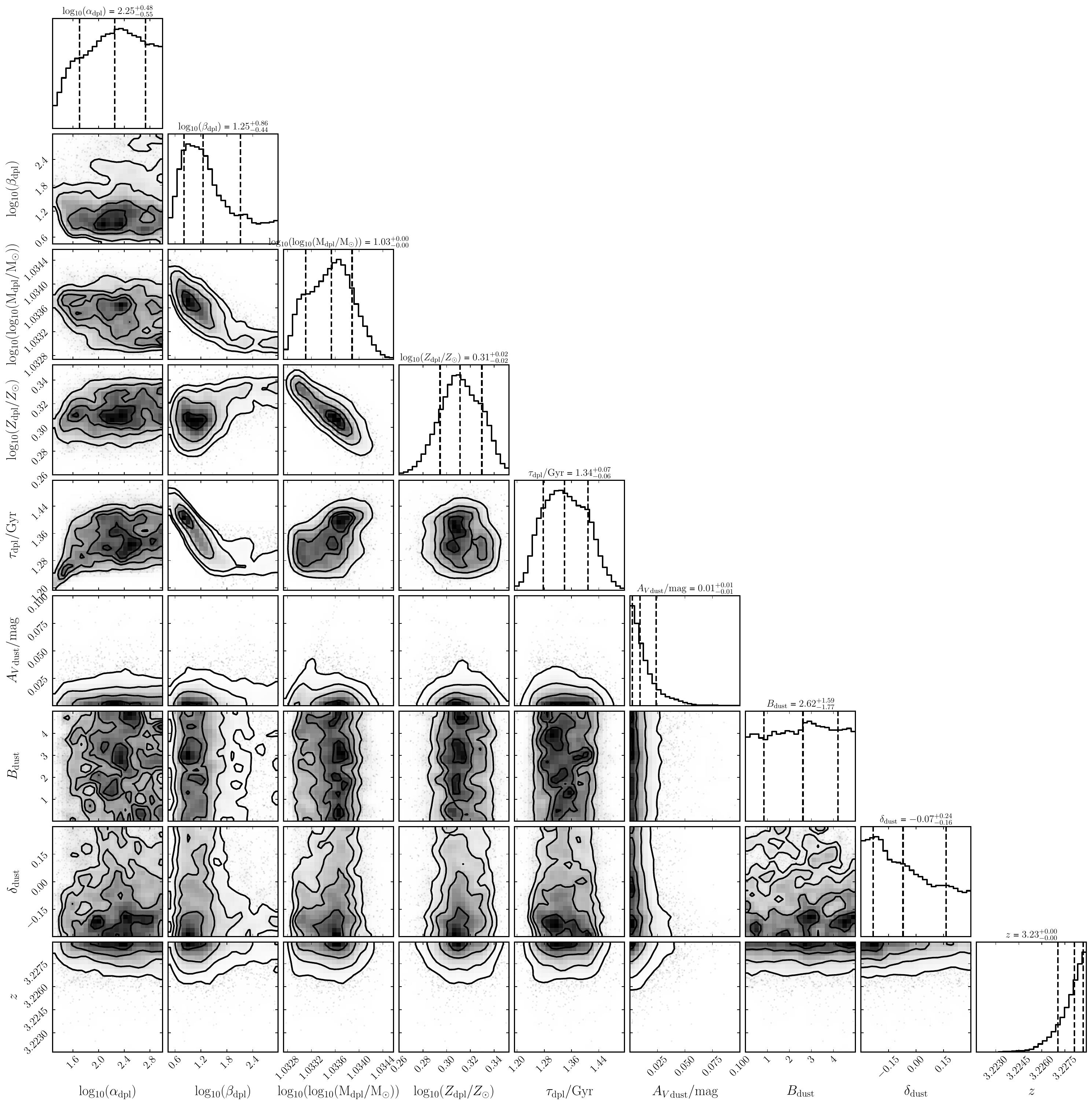}
    \includegraphics[width=0.49\textwidth]{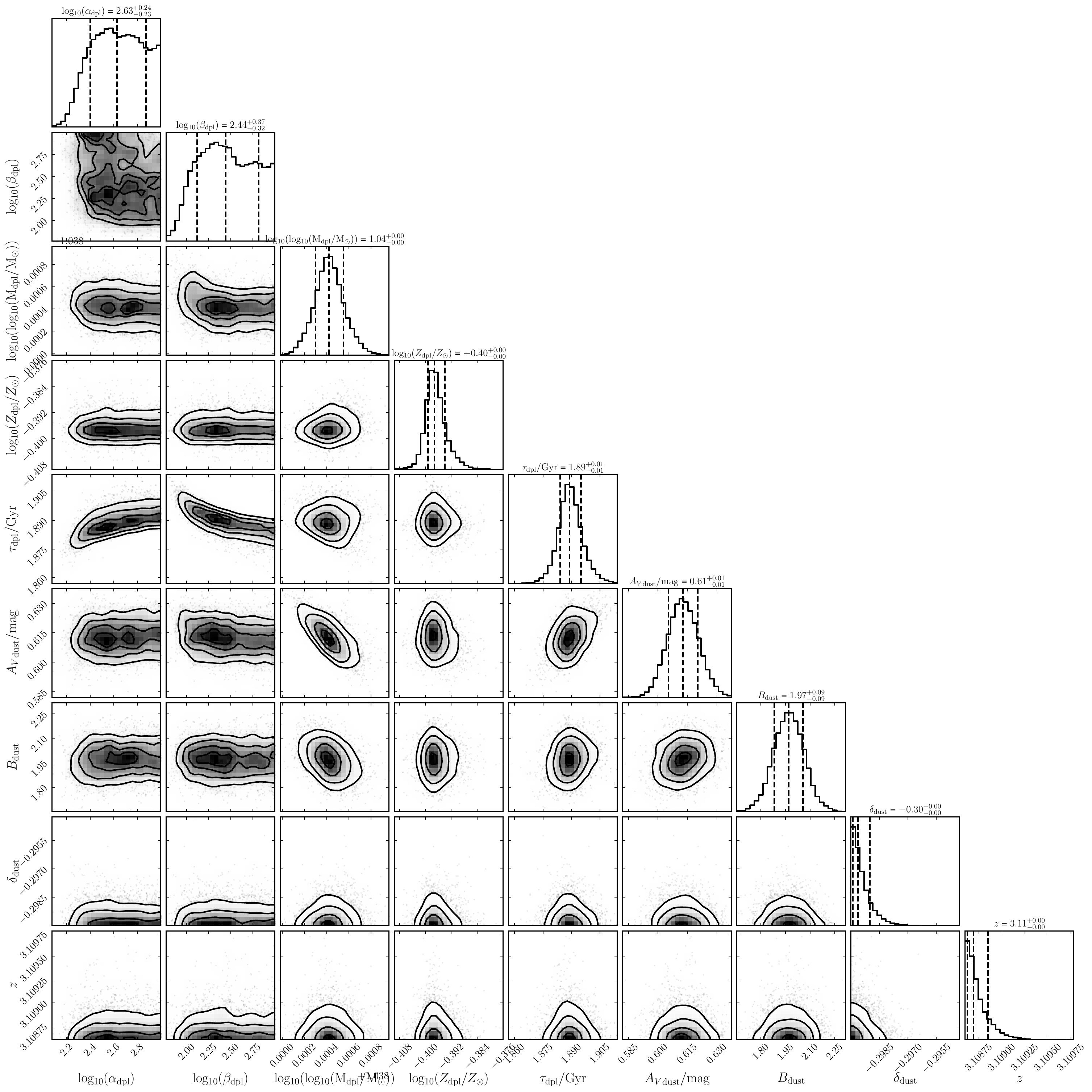}
    \caption{Corner plot for COSMOS ID 50248 and 110859}
    \label{SED_ID50248_110859}
\end{figure*}

\begin{figure*}[!h]
    \centering
    \includegraphics[width=0.49\textwidth]{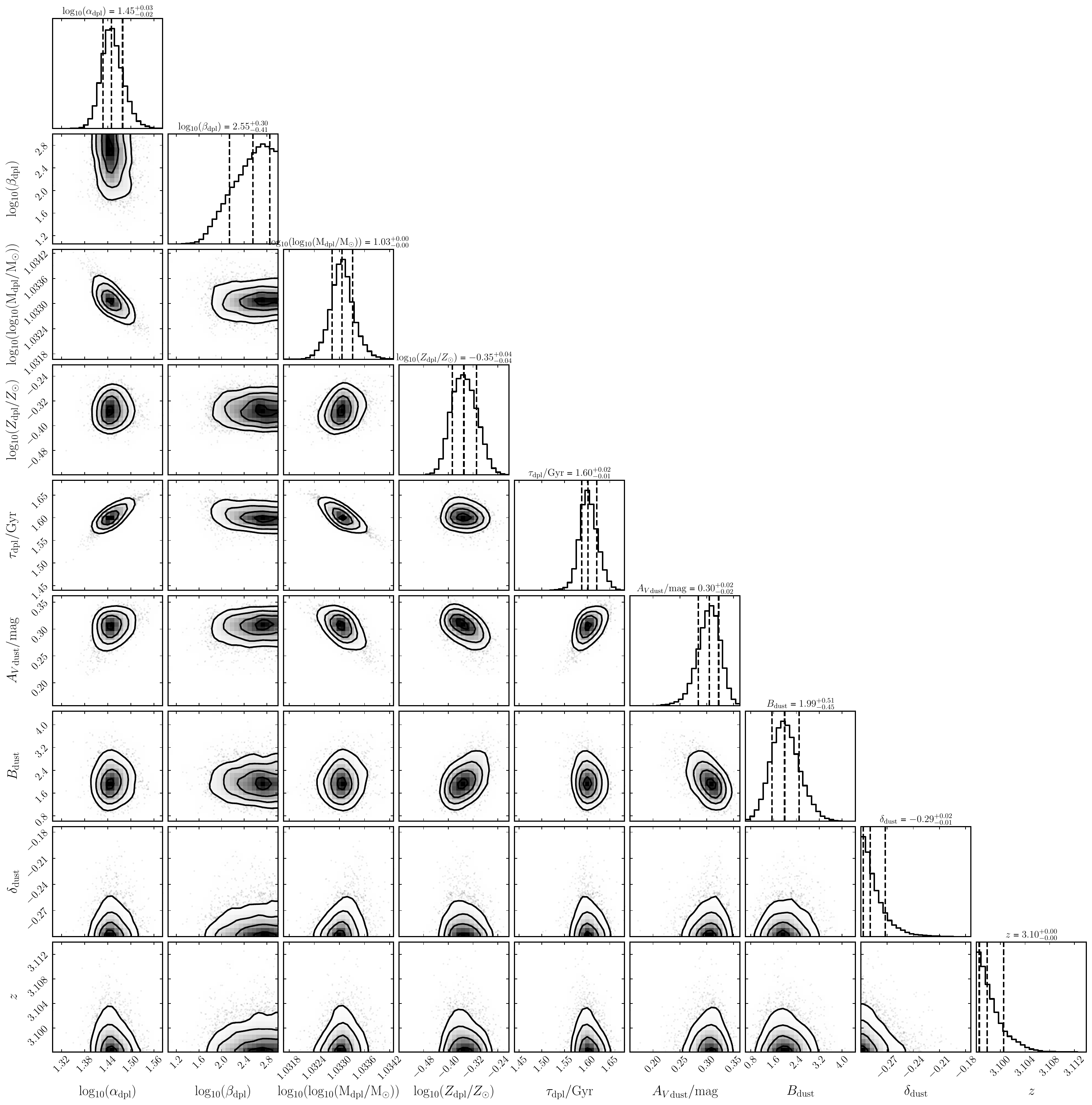}
    \includegraphics[width=0.49\textwidth]{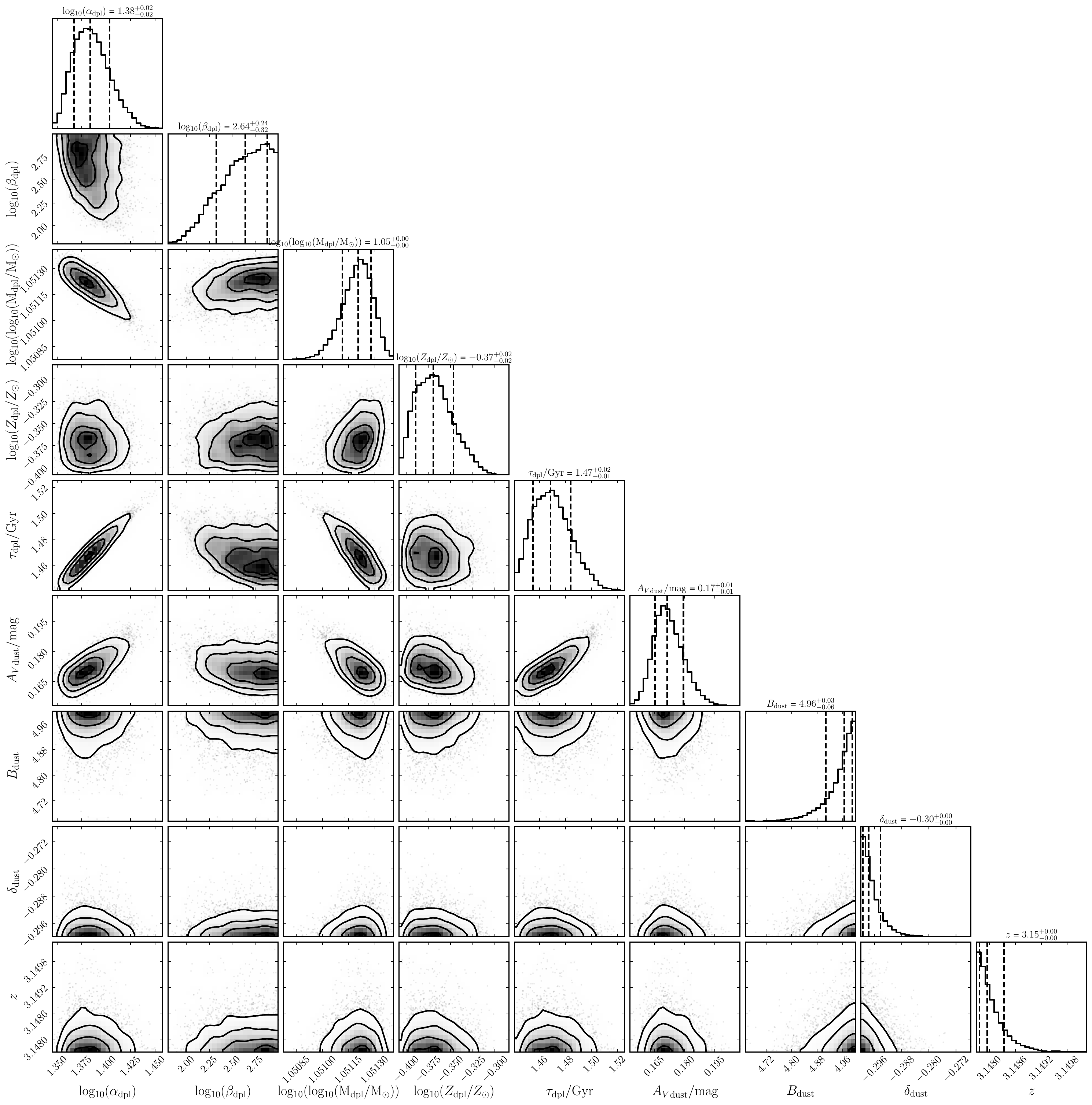}
    \caption{Corner plot for COSMOS ID 278297 and 282964}
    \label{SED_ID278297_282964}
\end{figure*}

\begin{figure*}[!h]
    \centering
    \includegraphics[width=0.49\textwidth]{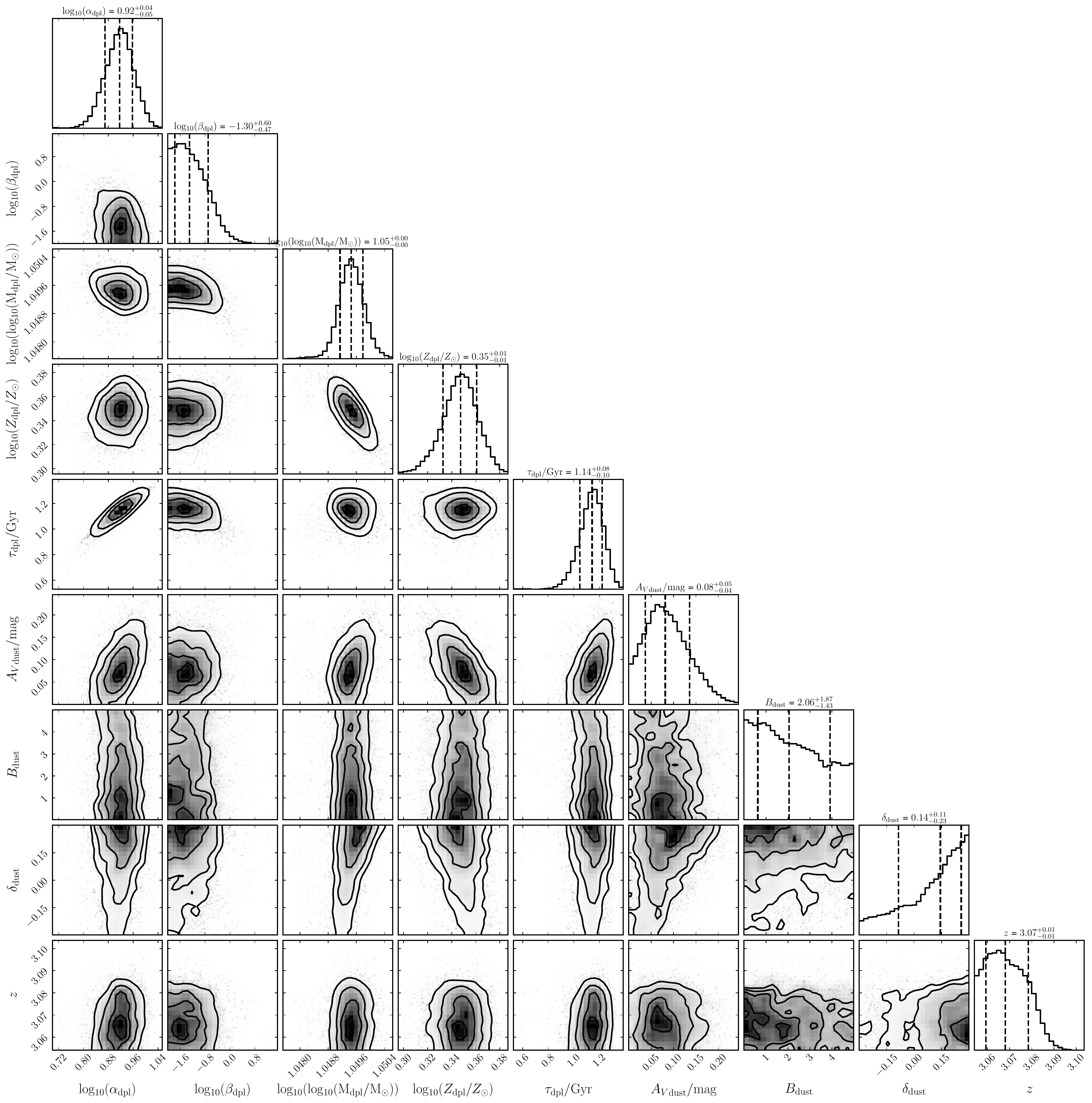}
    \includegraphics[width=0.49\textwidth]{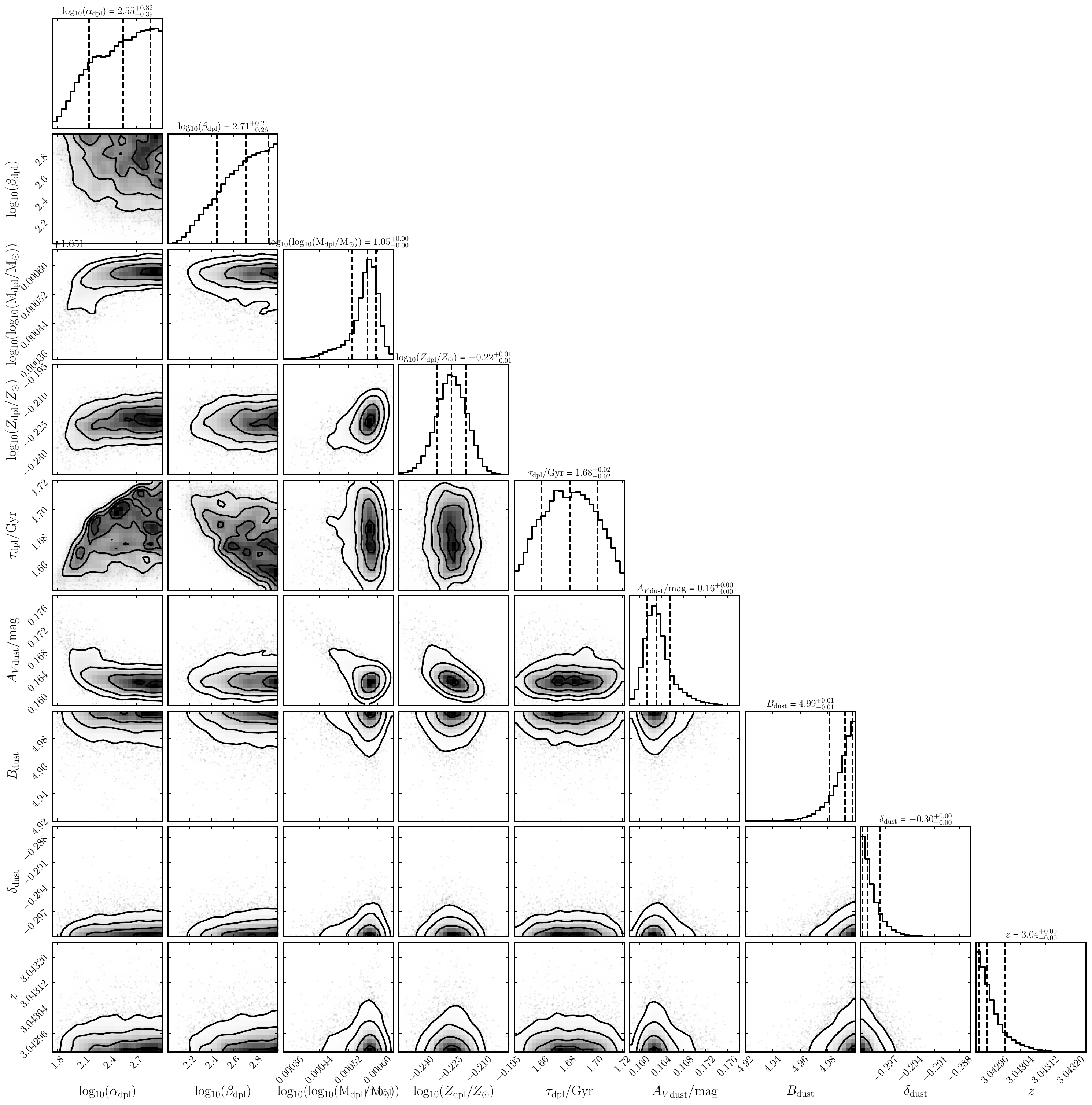}
    \caption{Corner plot for COSMOS ID 310229 and 338392}
    \label{SED_ID310229_338392}
\end{figure*}

\begin{figure*}[!h]
    \centering
    \includegraphics[width=0.49\textwidth]{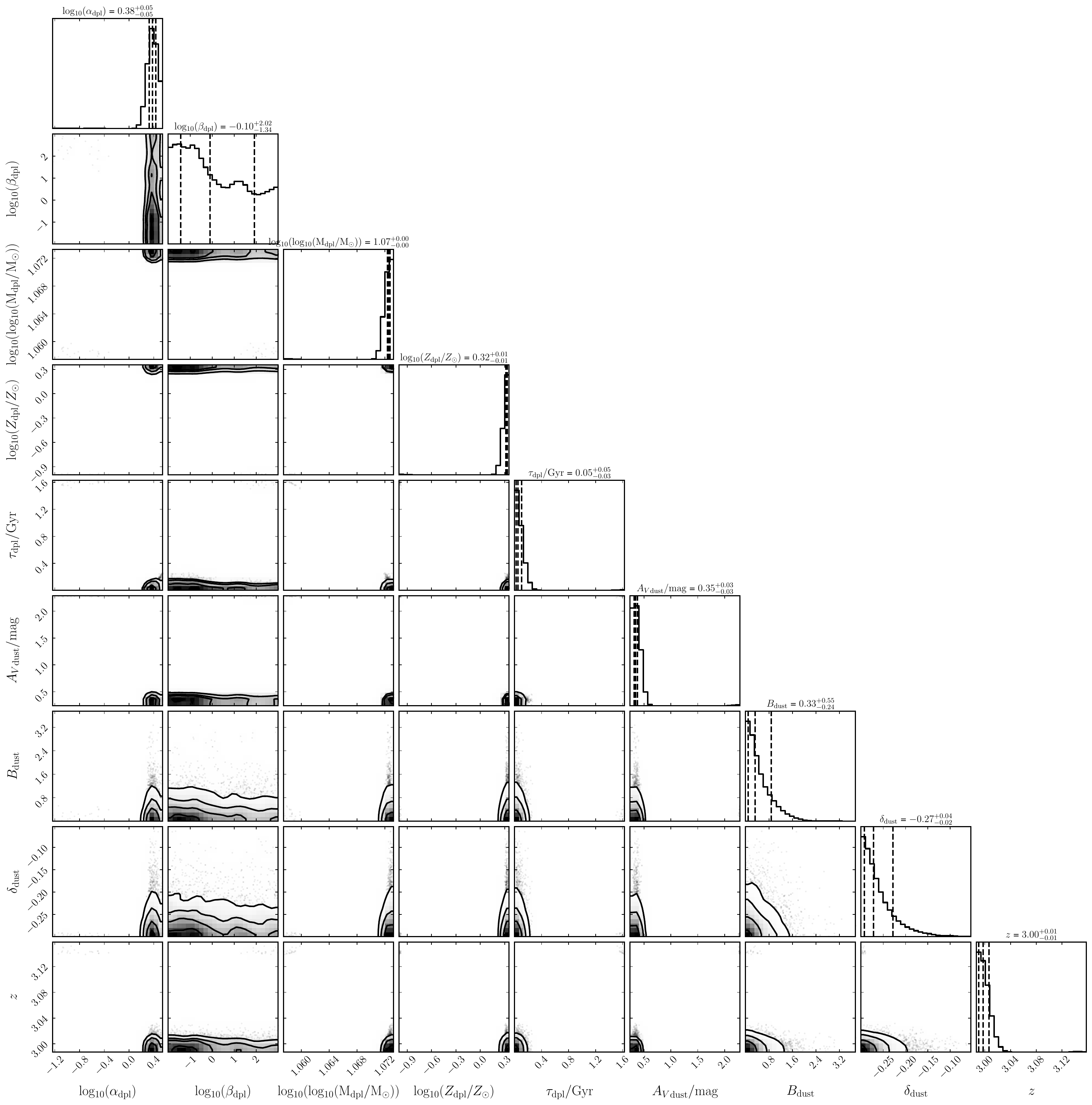}
    \includegraphics[width=0.49\textwidth]{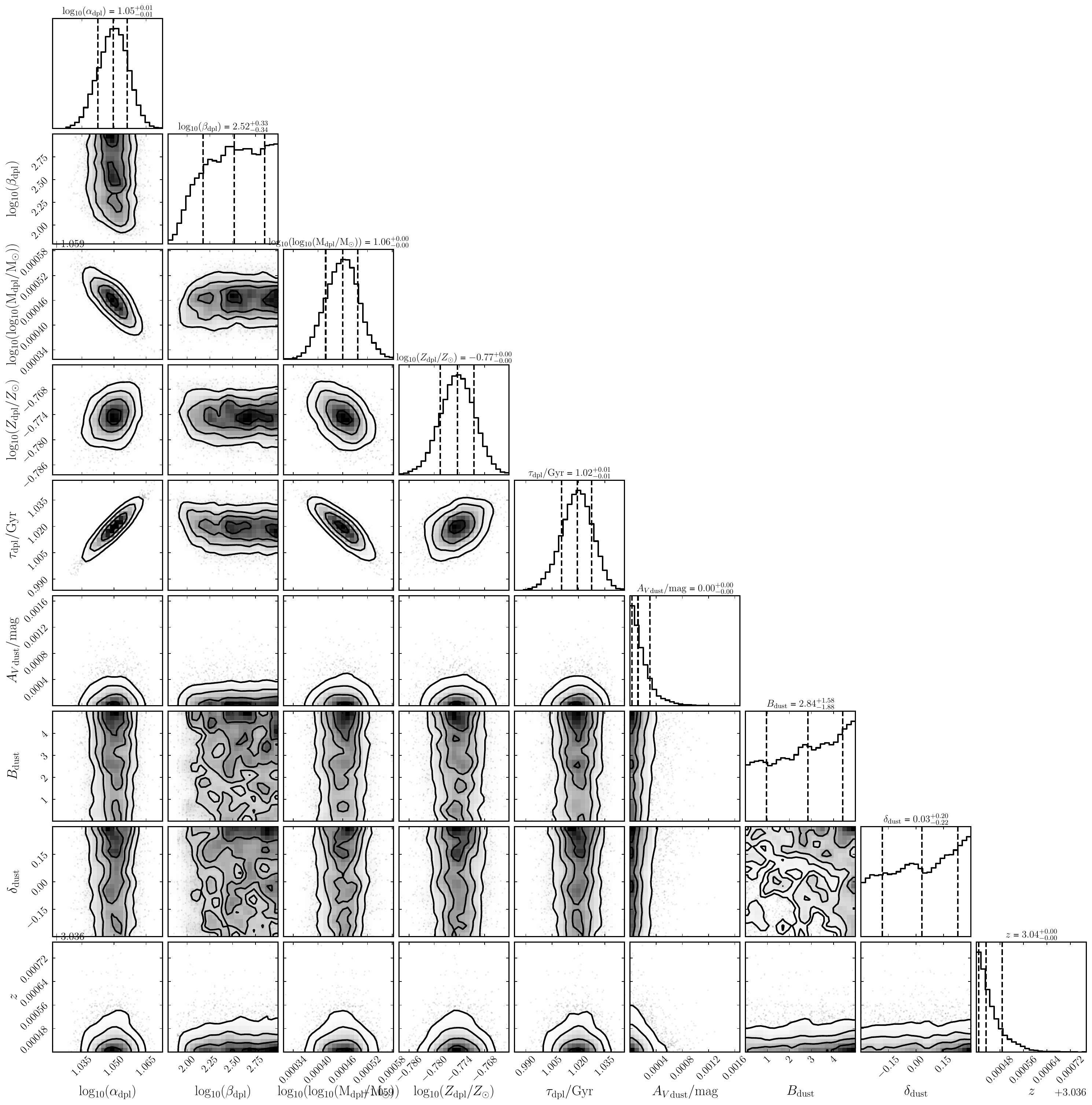}
    \caption{Corner plot for COSMOS ID 341682 and 343373}
    \label{SED_ID341682_343373}
\end{figure*}

\begin{figure*}[!h]
    \centering
    \includegraphics[width=0.49\textwidth]{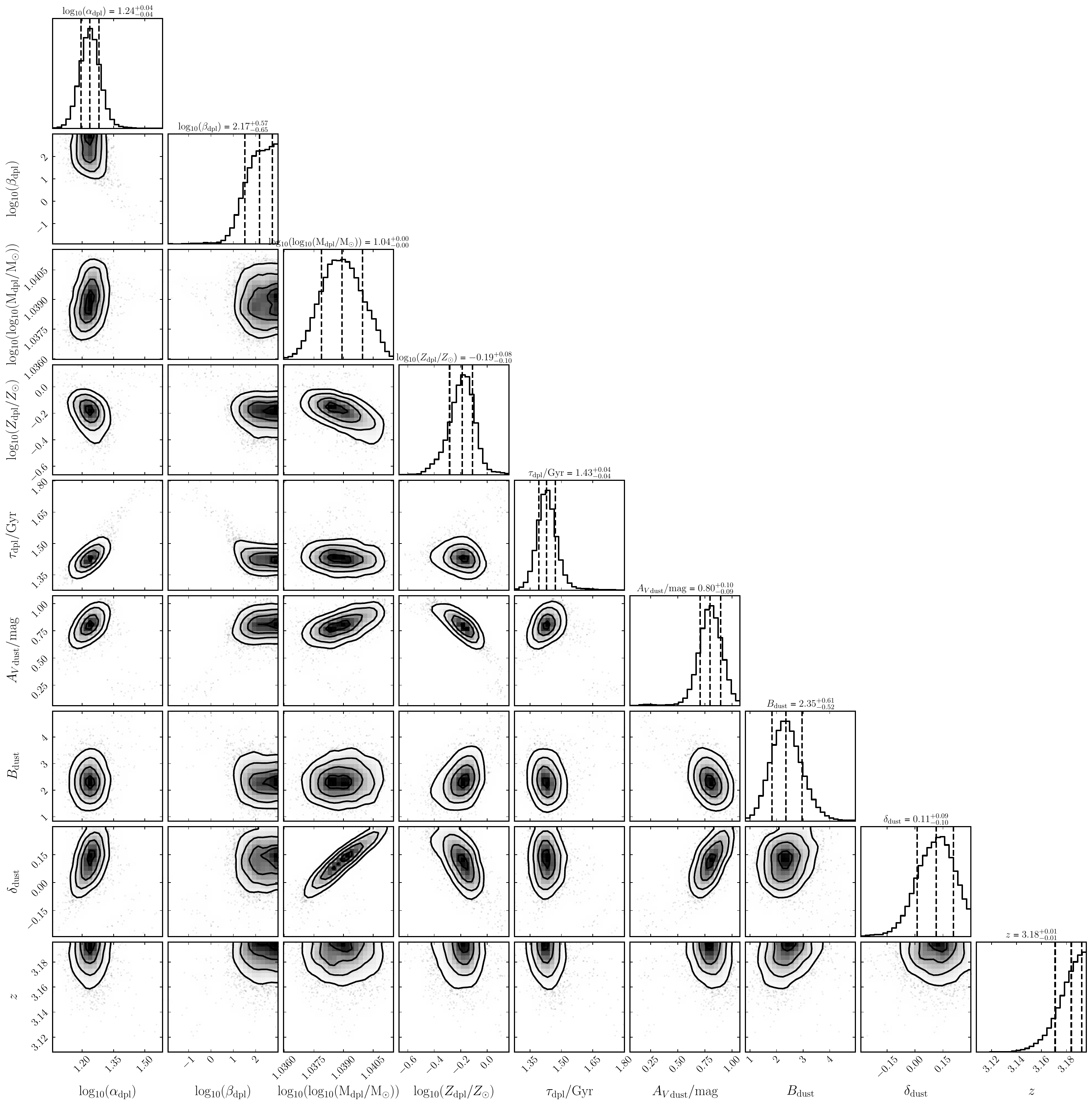}
    \includegraphics[width=0.49\textwidth]{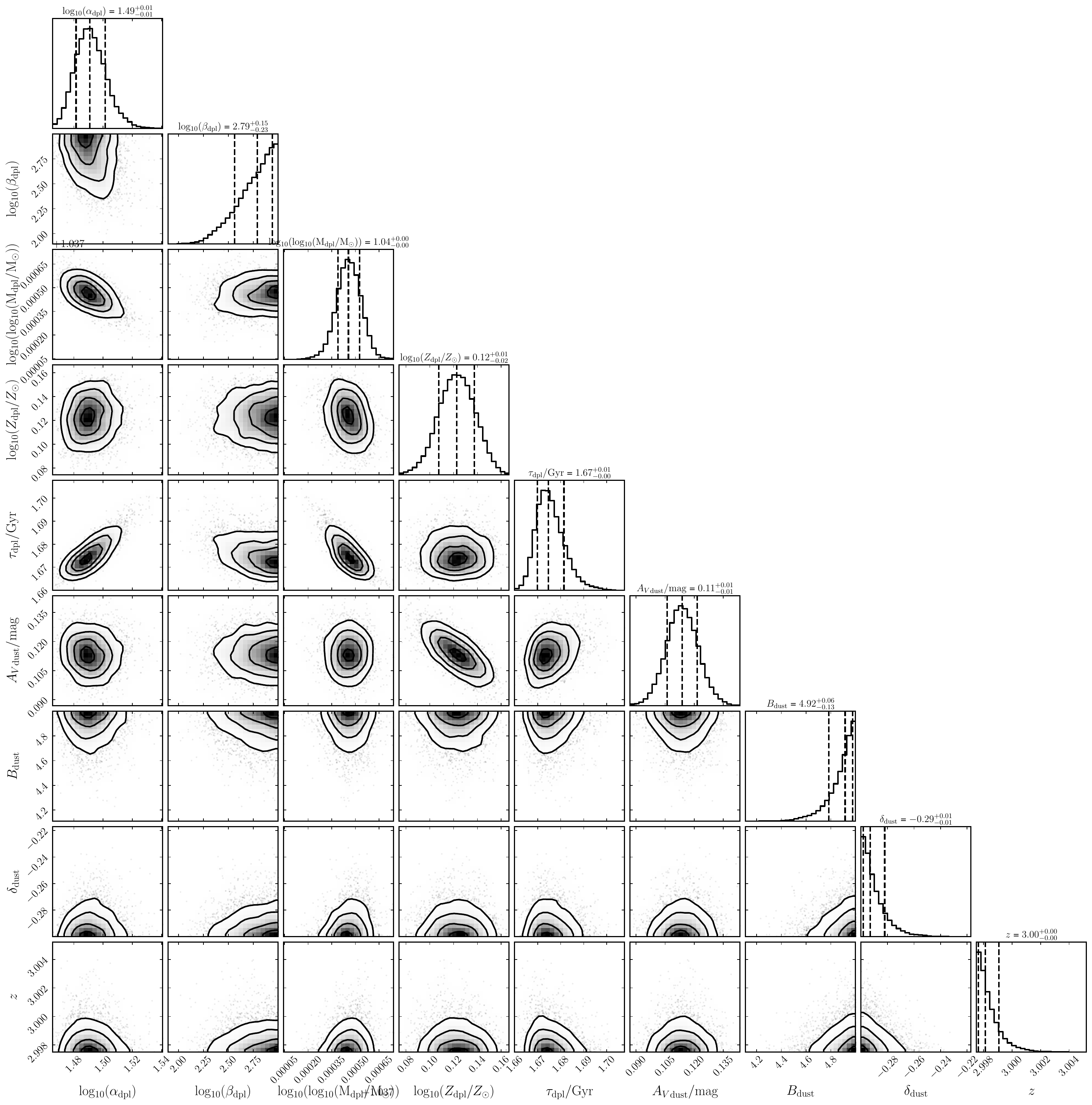}
    \caption{Corner plot for COSMOS ID 383298 and 405872}
    \label{SED_ID383298_405872}
\end{figure*}

\begin{figure*}[!h]
    \centering
    \includegraphics[width=0.49\textwidth]{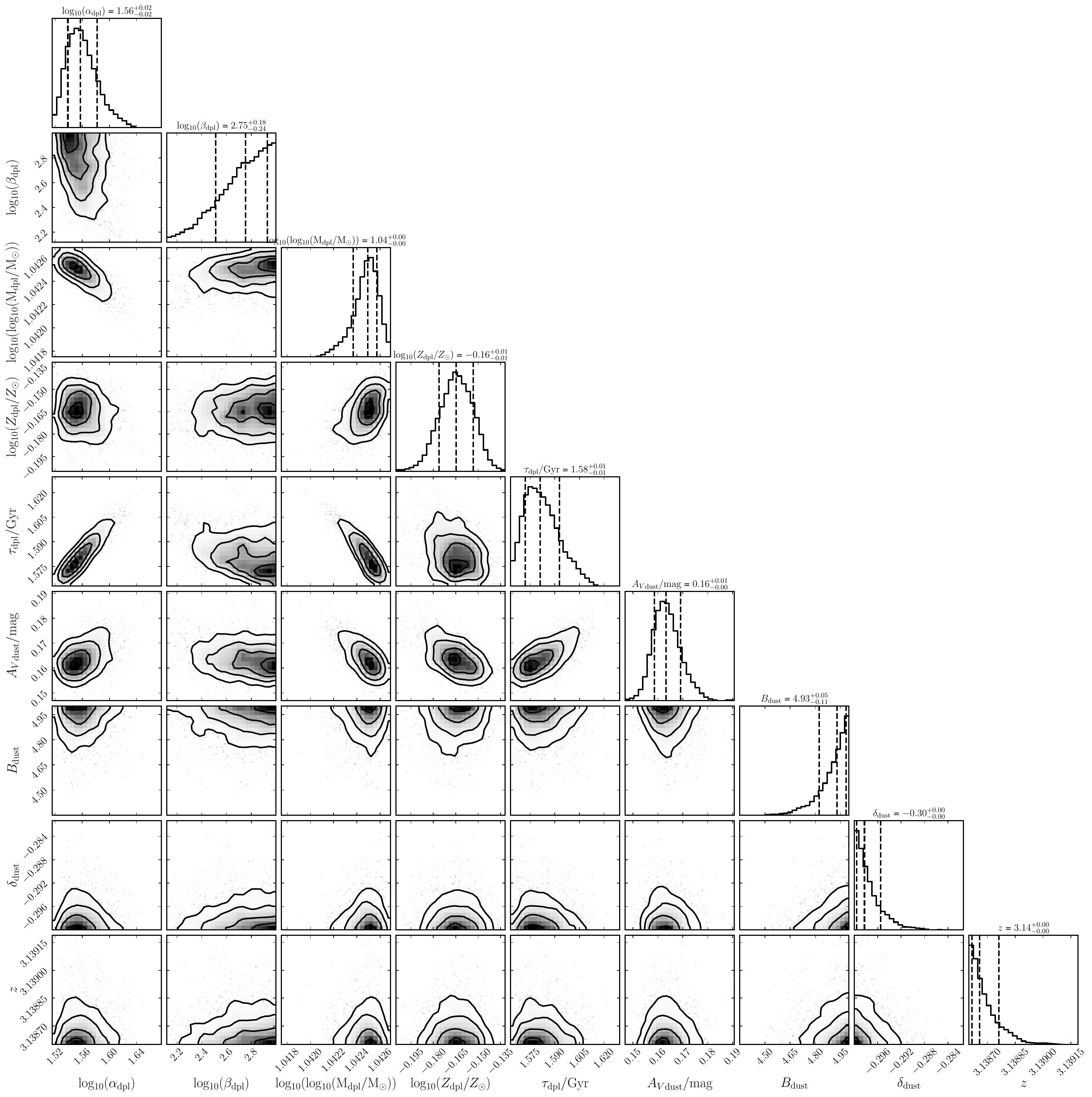}
    \includegraphics[width=0.49\textwidth]{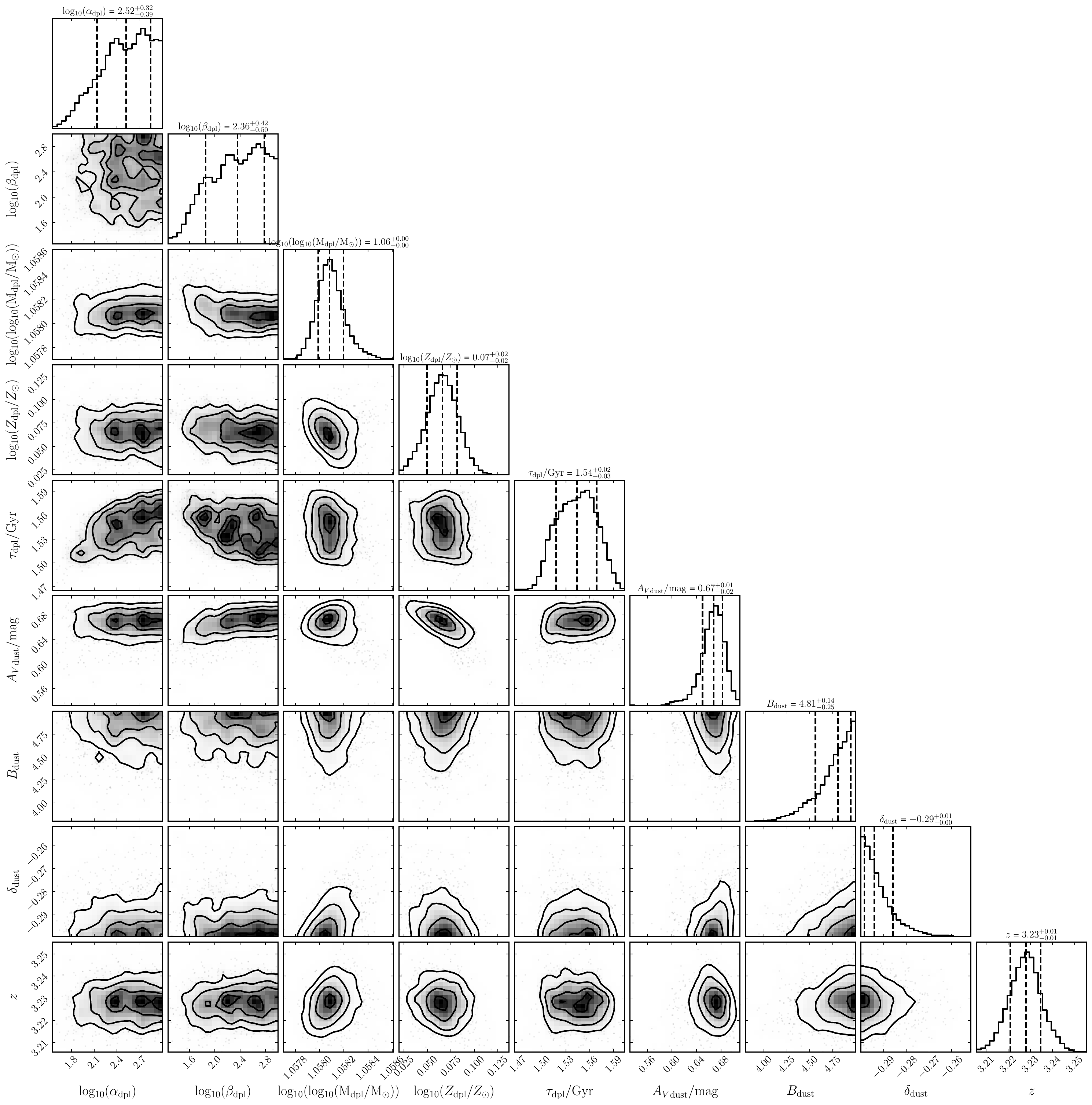}
    \caption{Corner plot for COSMOS ID 412439 and 489177}
    \label{SED_ID412439_489177}
\end{figure*}

\begin{figure*}[!h]
    \centering
    \includegraphics[width=0.49\textwidth]{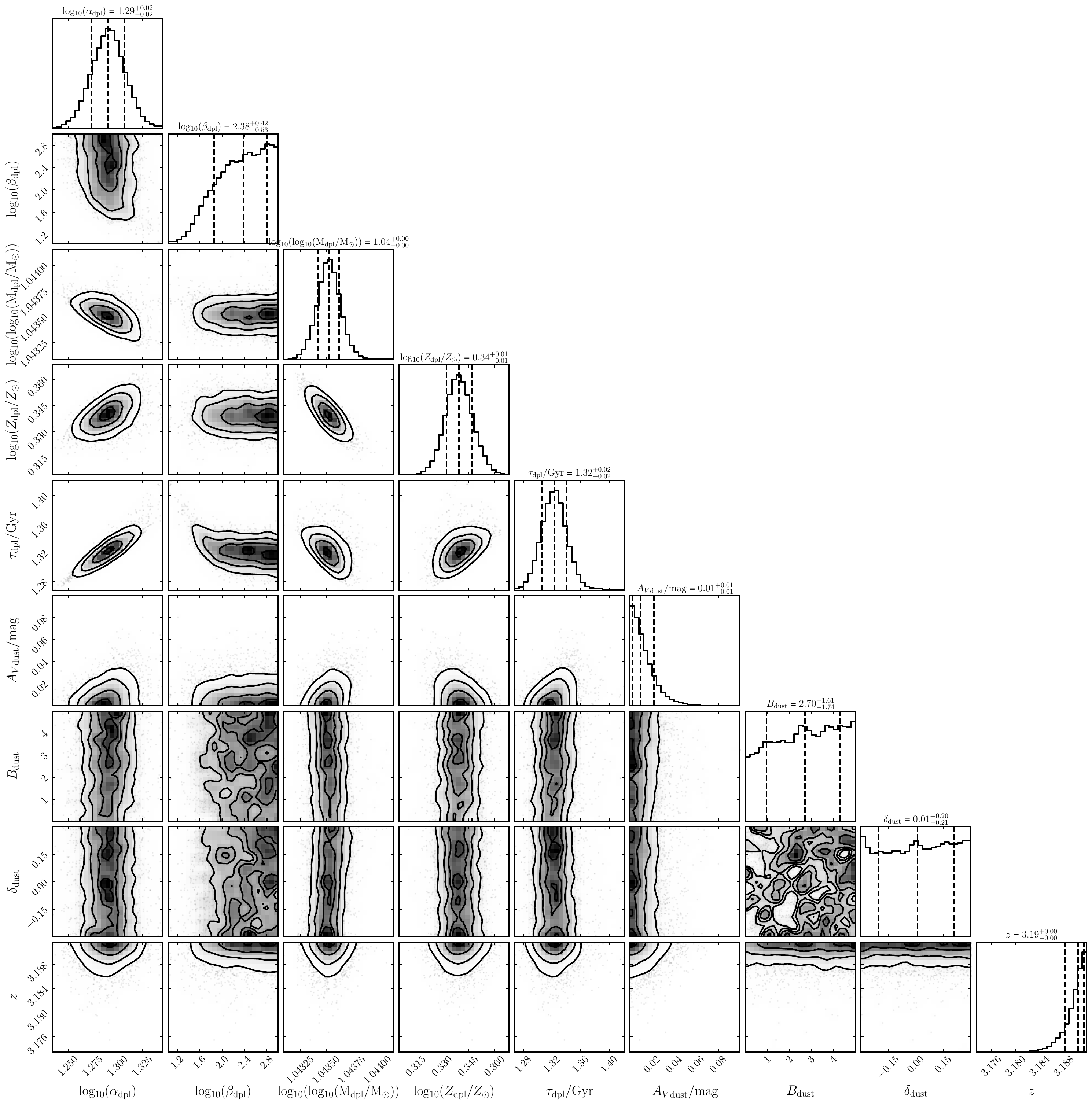}
    \includegraphics[width=0.49\textwidth]{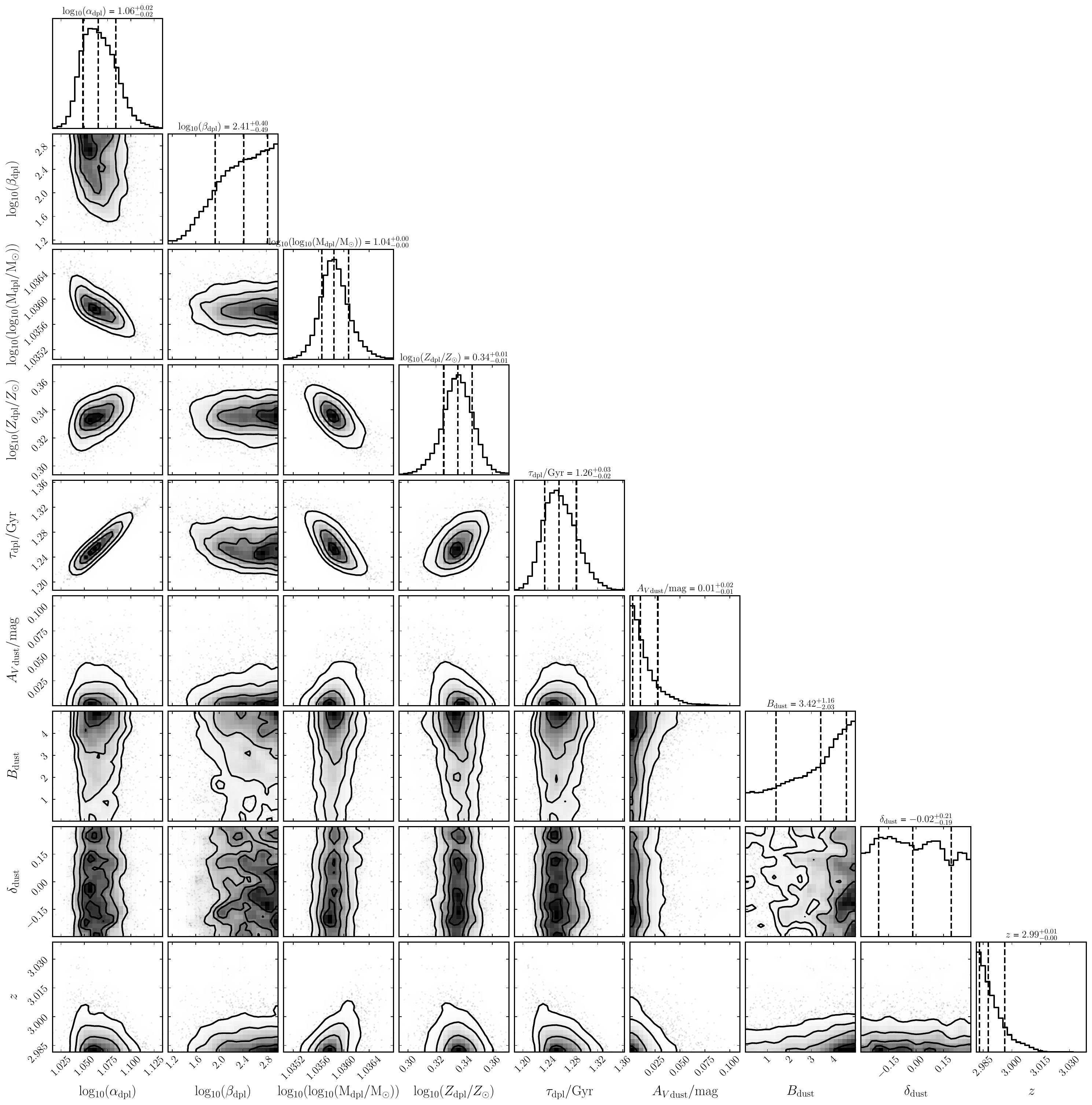}
    \caption{Corner plot for COSMOS ID 642338 and 658452}
    \label{SED_ID642338_658452}
\end{figure*}

\begin{figure*}[!h]
    \centering
    \includegraphics[width=0.49\textwidth]{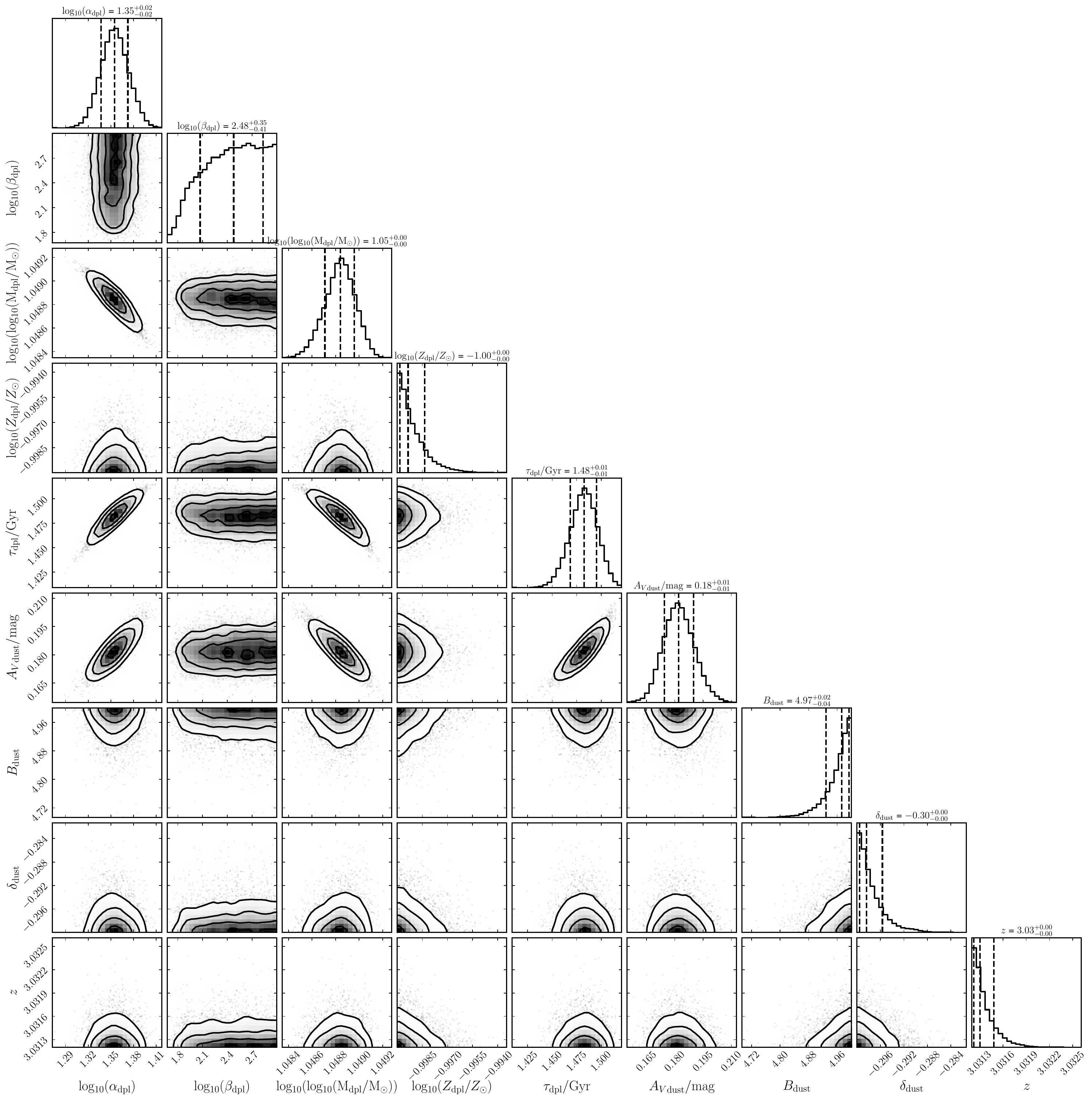}
    \includegraphics[width=0.49\textwidth]{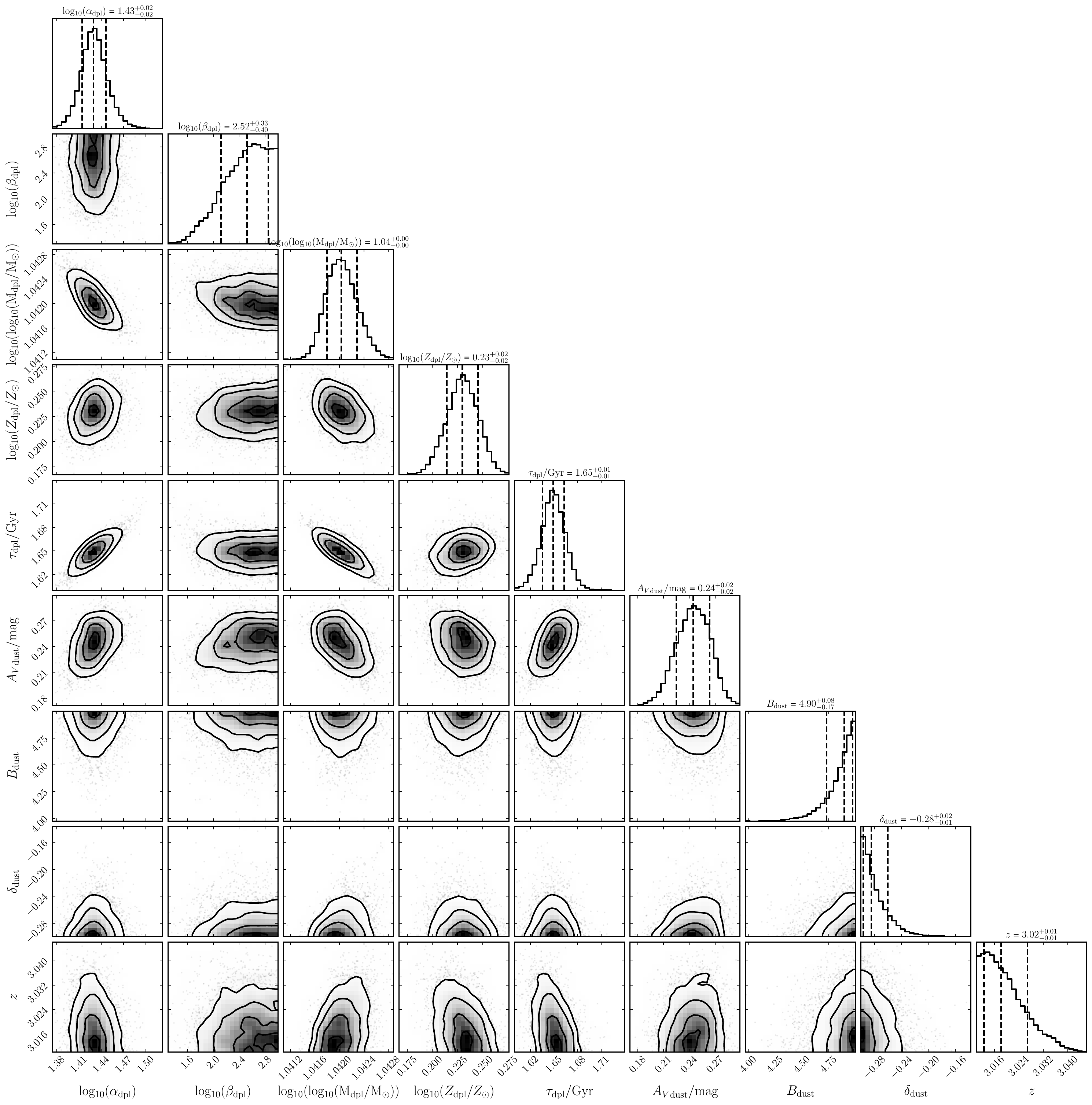}
    \caption{Corner plot for COSMOS ID 681407 and 706985}
    \label{SED_ID681407_706985}
\end{figure*}

\begin{figure*}[!h]
    \centering
    \includegraphics[width=0.49\textwidth]{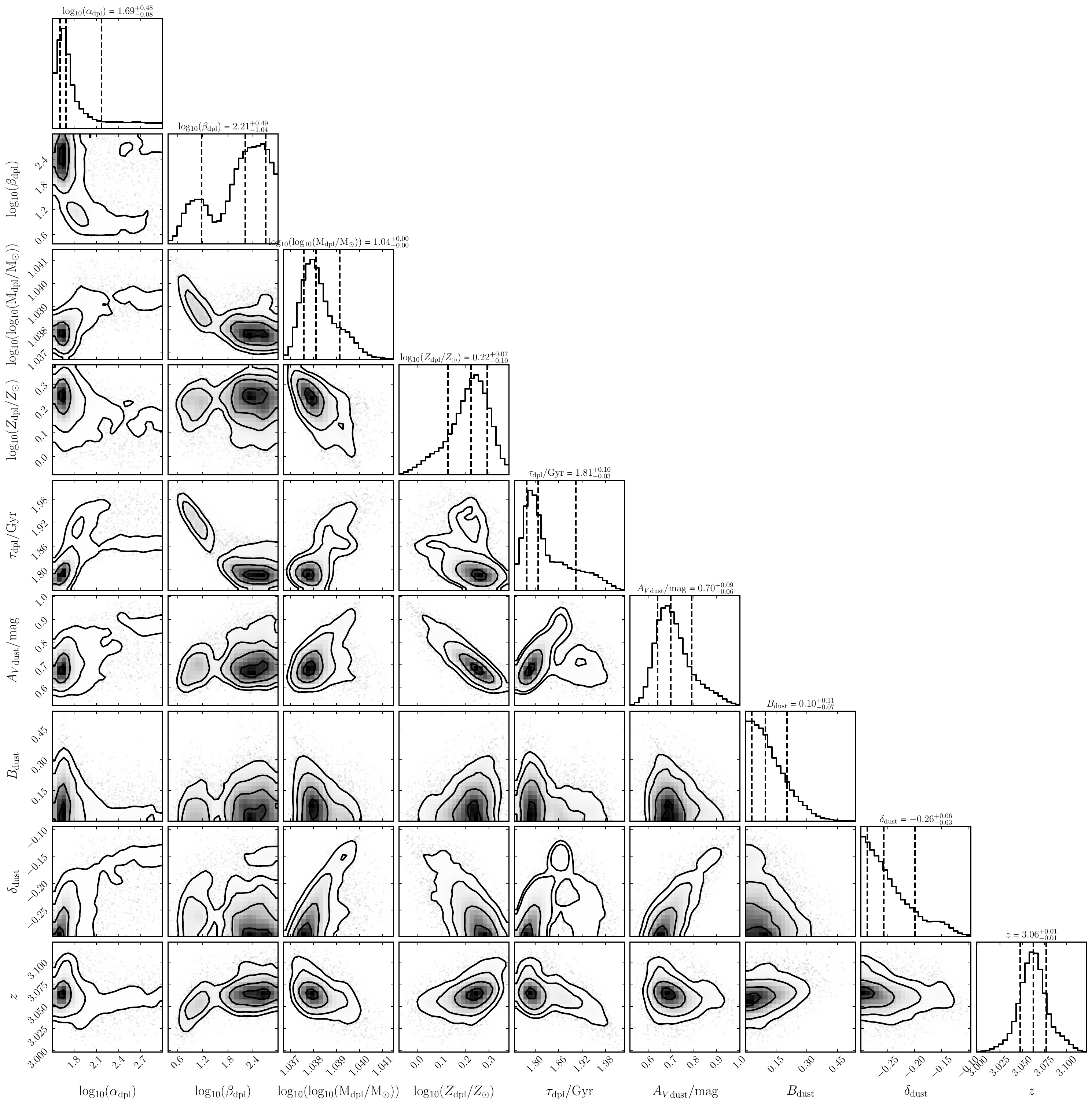}
    \includegraphics[width=0.49\textwidth]{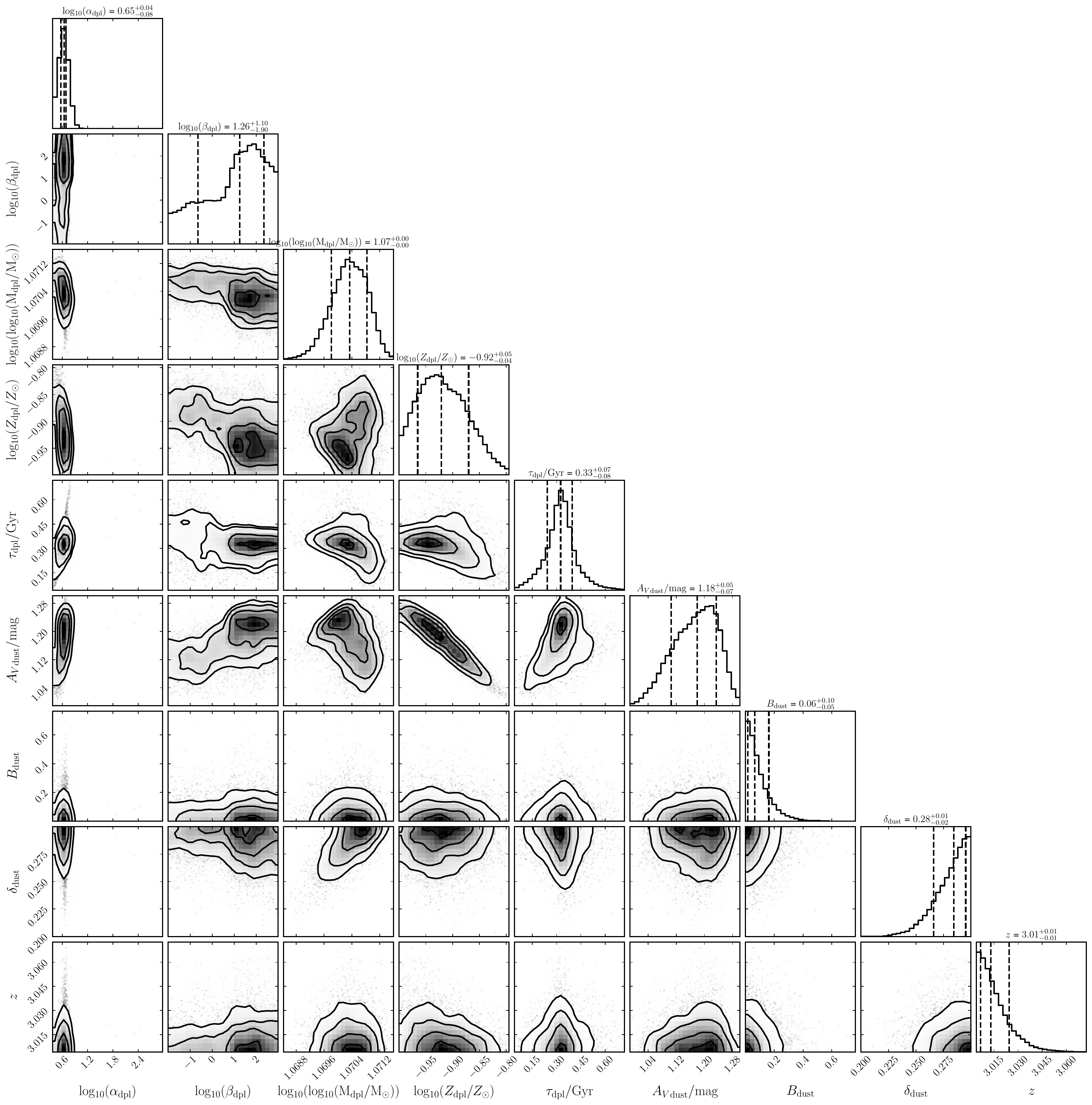}
    \caption{Corner plot for COSMOS ID 779869 and 803389}
    \label{SED_ID779869_803389}
\end{figure*}

\begin{figure*}[!h]
    \centering
    \includegraphics[width=0.49\textwidth]{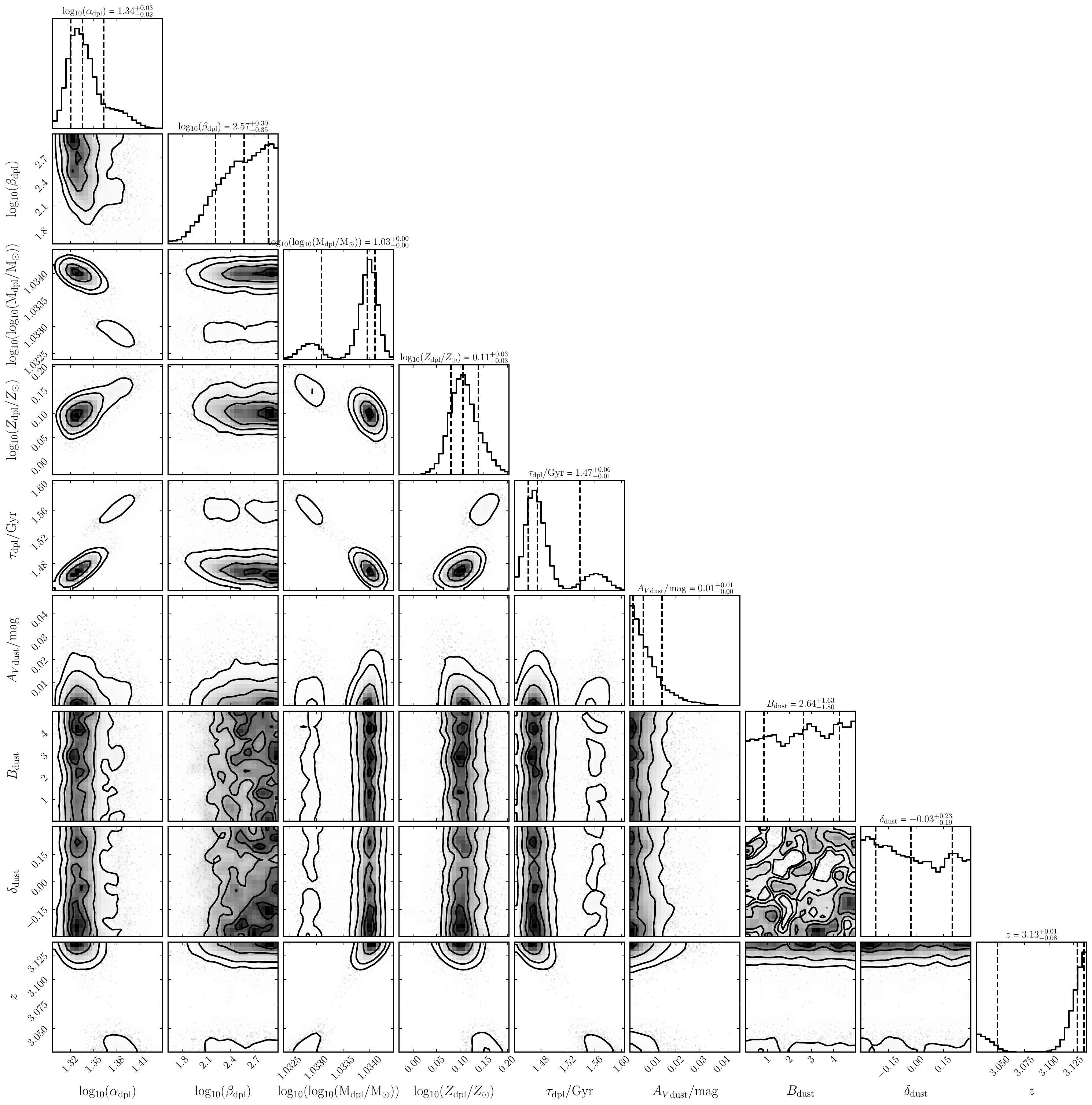}
    \includegraphics[width=0.49\textwidth]{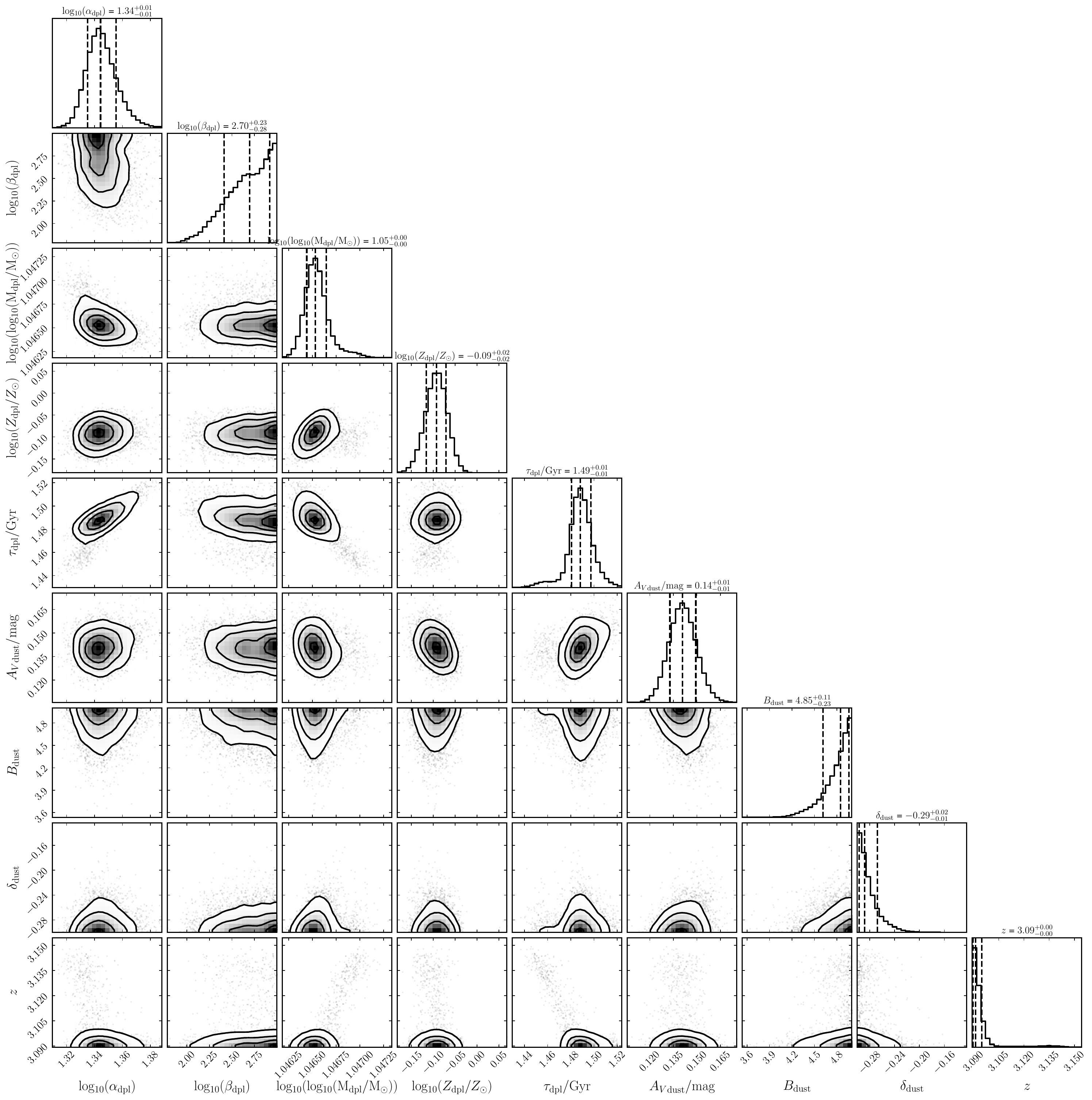}
    \caption{Corner plot for COSMOS ID 881980 and 911001}
    \label{SED_ID881980_911001}
\end{figure*}

\begin{figure*}[!h]
    \centering
    \includegraphics[width=0.49\textwidth]{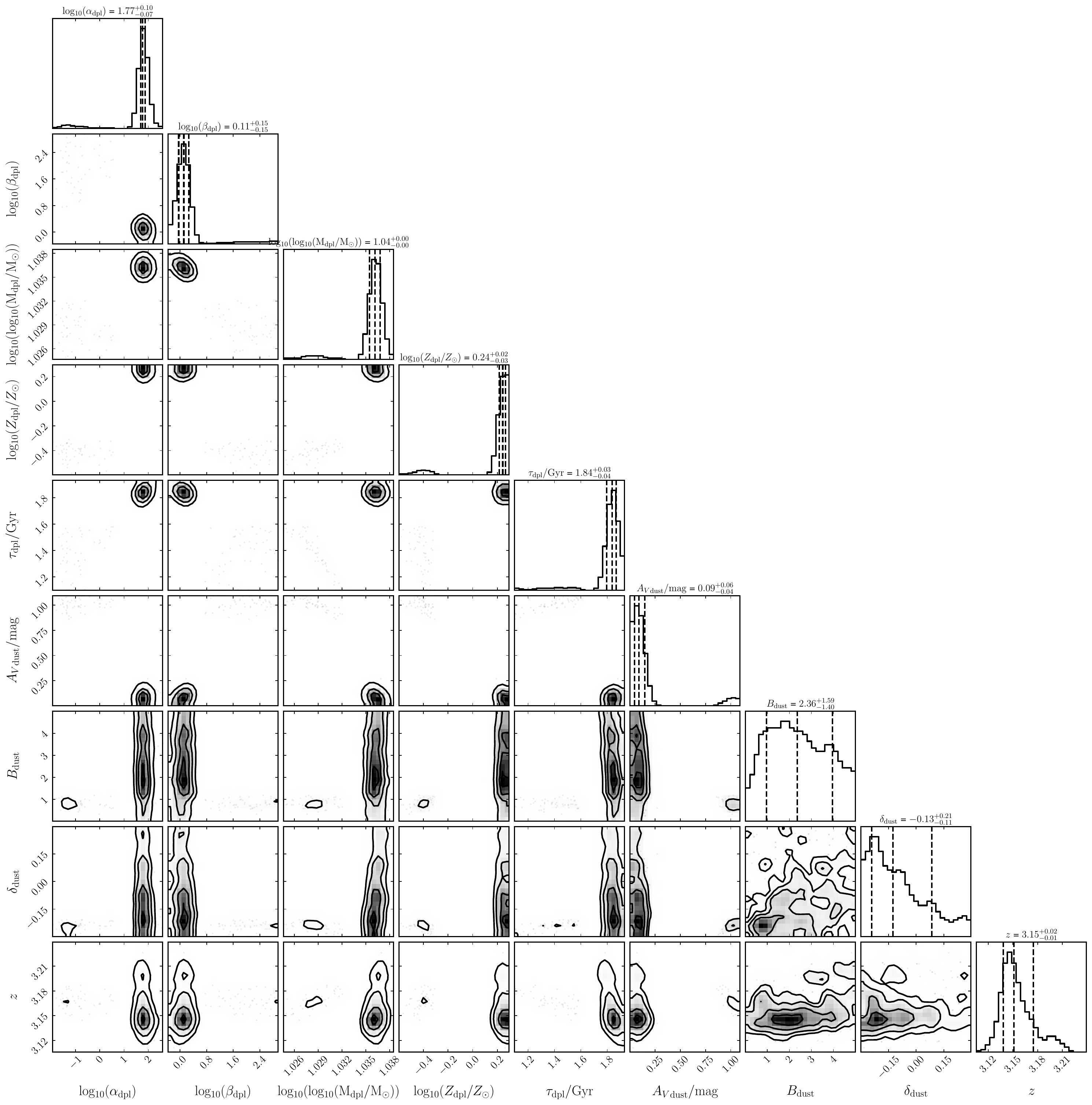}
    \includegraphics[width=0.49\textwidth]{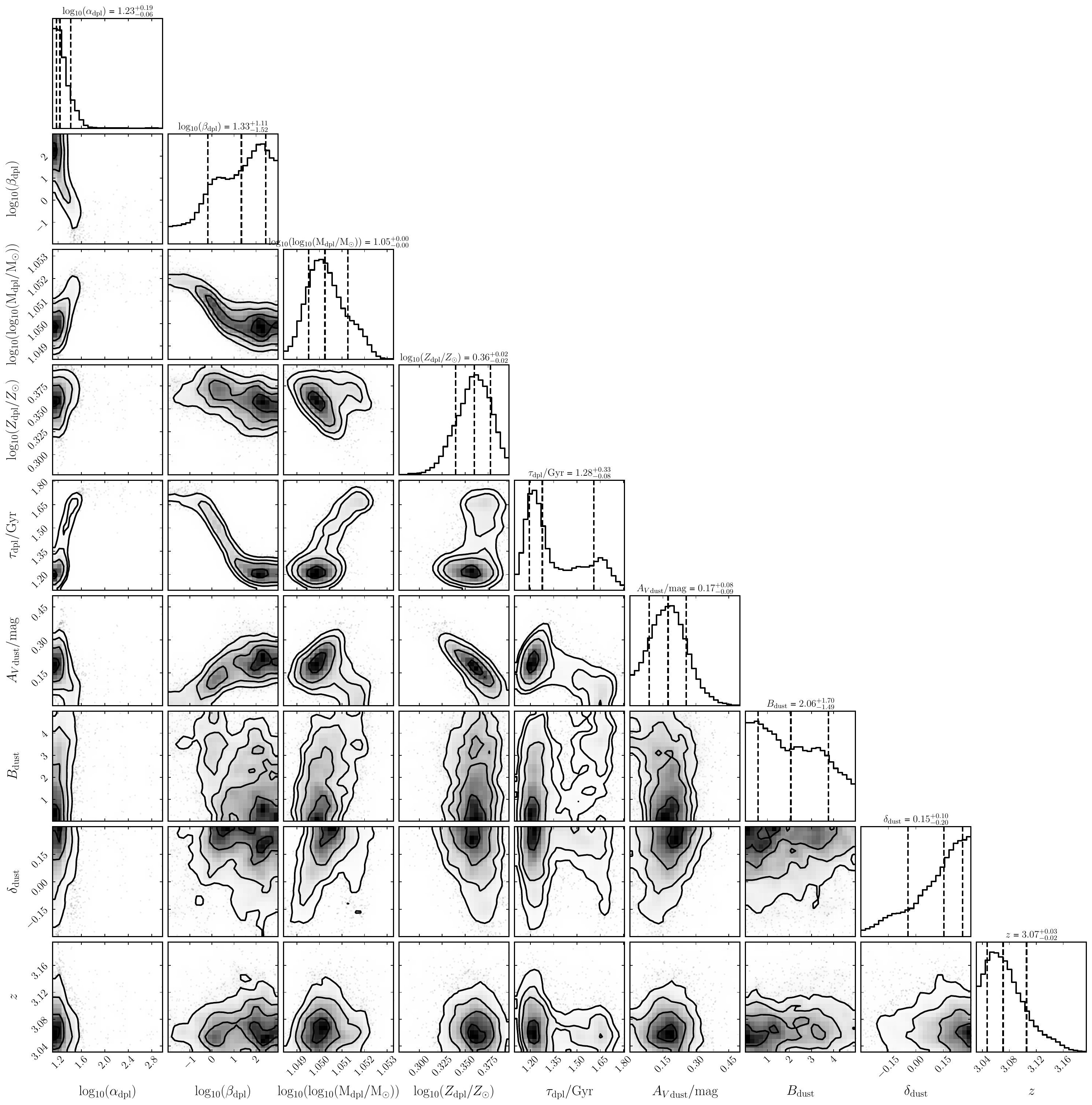}
    \caption{Corner plot for COSMOS ID 961549 and 962569}
    \label{SED_ID961549_962569}
\end{figure*}

\section{Visual inspection of galaxies}\label{reject}
\begin{figure*}[]
    \centering
    \includegraphics[width=\textwidth]{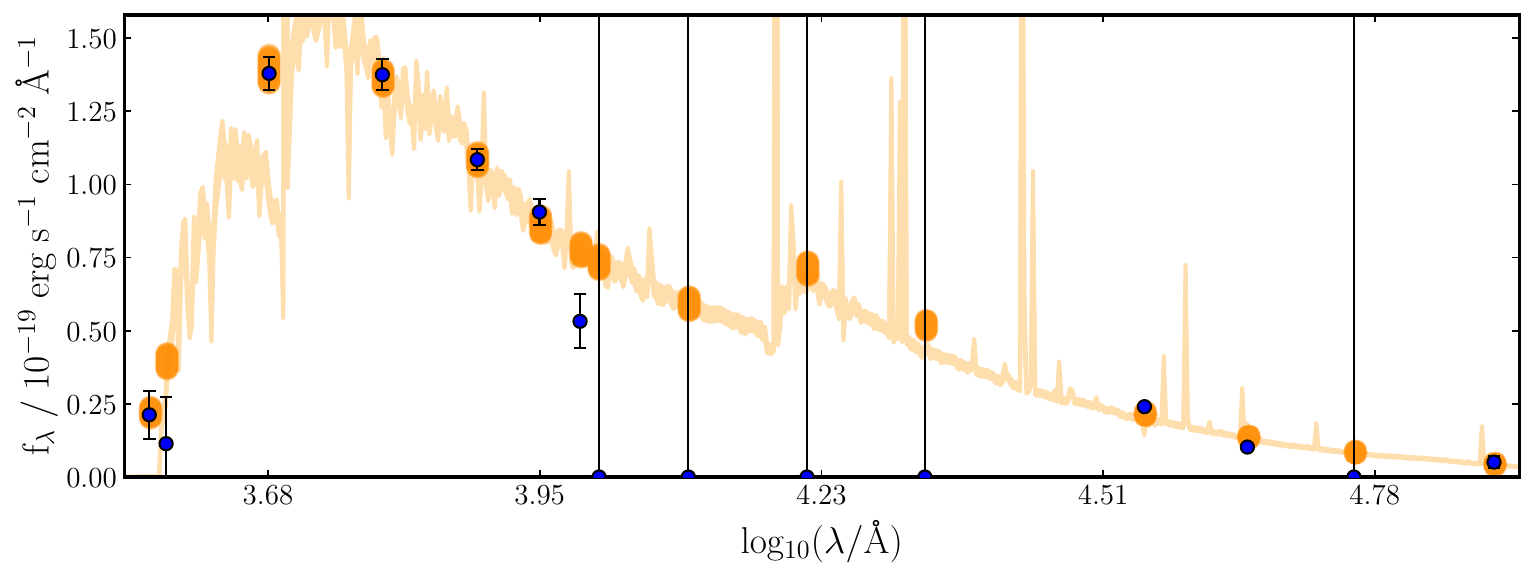}
    \includegraphics[width=\textwidth]{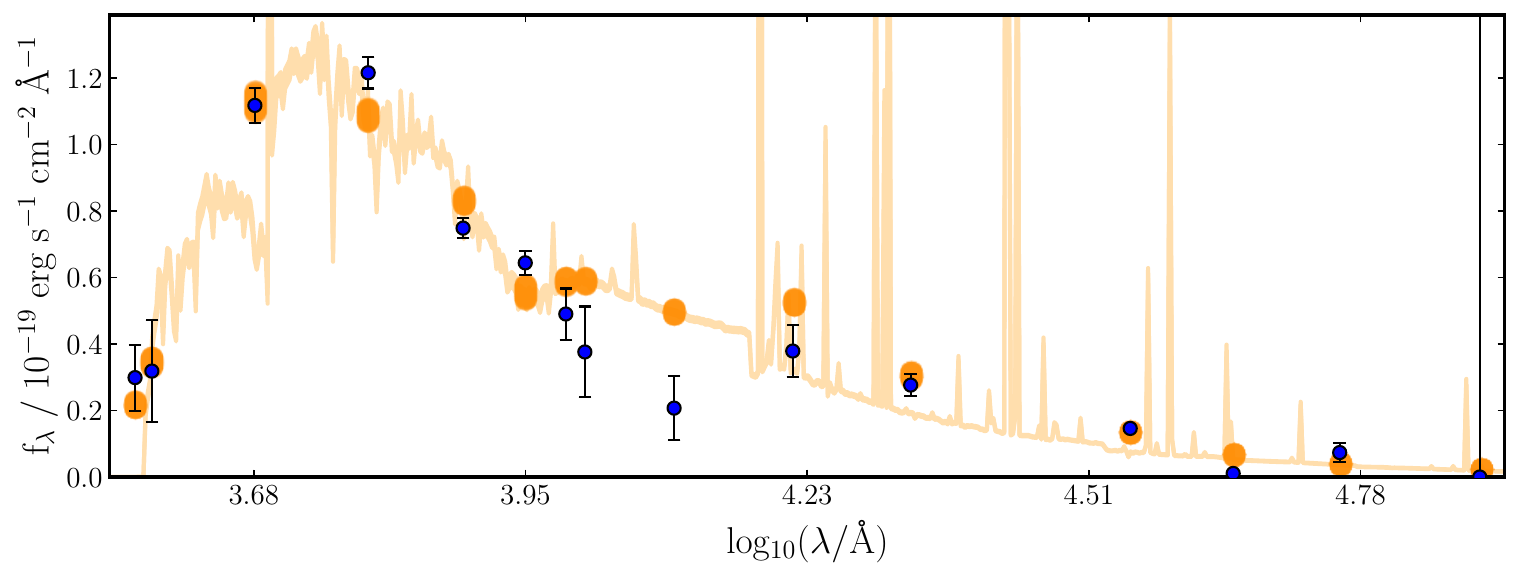}
    \includegraphics[width=\textwidth]{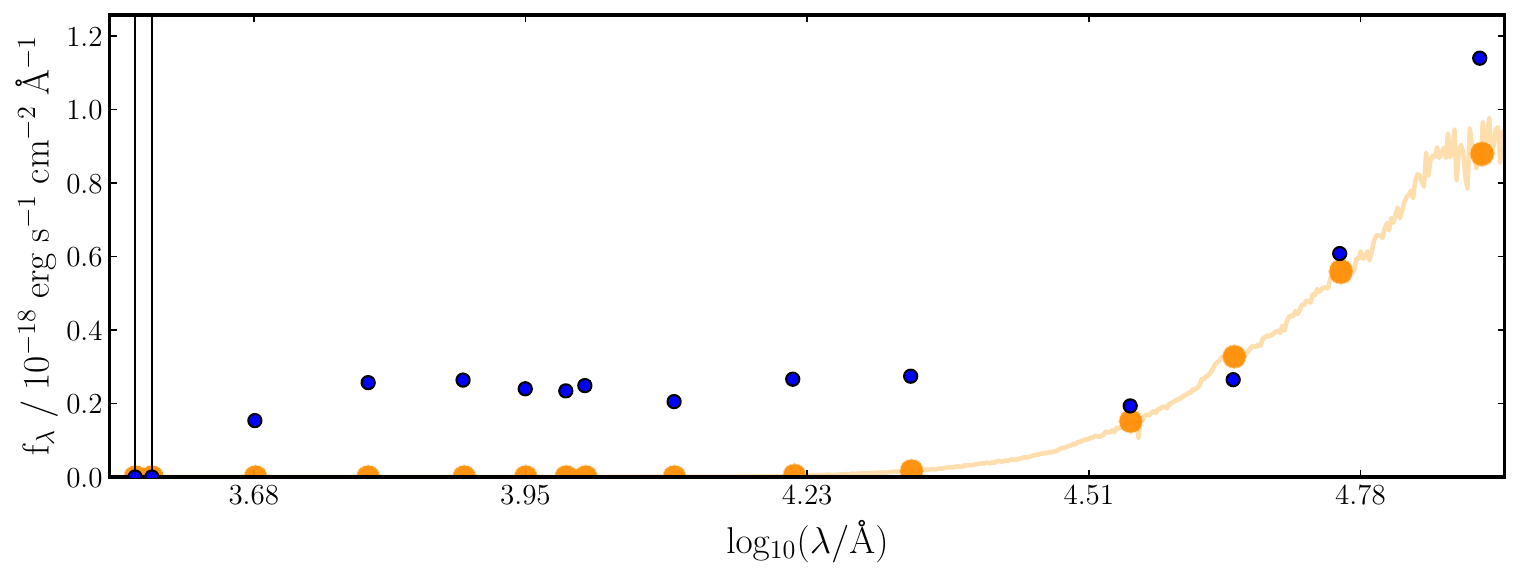}
    \caption{Examples of galaxies rejected from the MQG sample after visual inspection.}
     \label{fig:wide_image}
 \end{figure*}

\section{Nearest neighbours}\label{nn}

\begin{figure*}[]
    \centering
    \begin{subfigure}[b]{0.49\textwidth}
        \includegraphics[width=\textwidth]{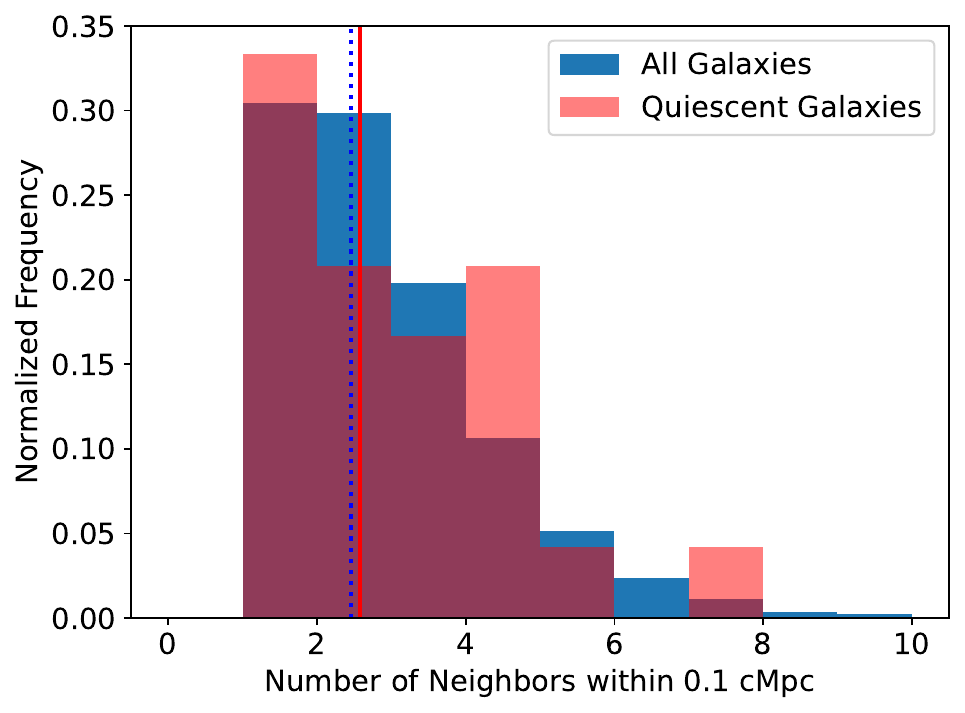}
        
    \end{subfigure}
    \begin{subfigure}[b]{0.49\textwidth}
        \includegraphics[width=\textwidth]{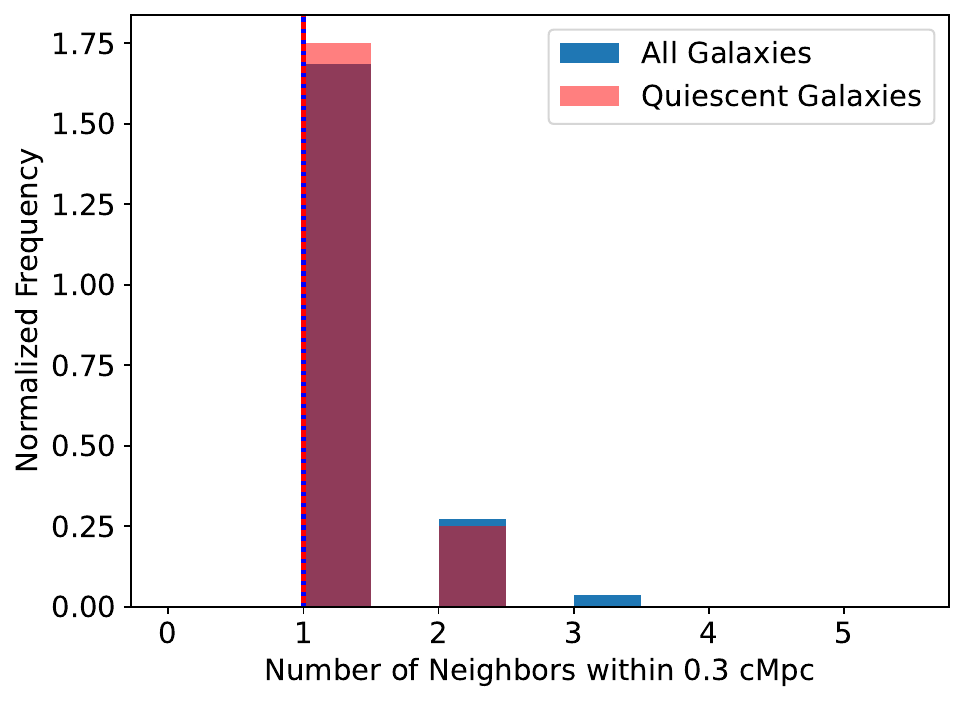}
       
    \end{subfigure}

    \begin{subfigure}[b]{0.49\textwidth}
        \includegraphics[width=\textwidth]{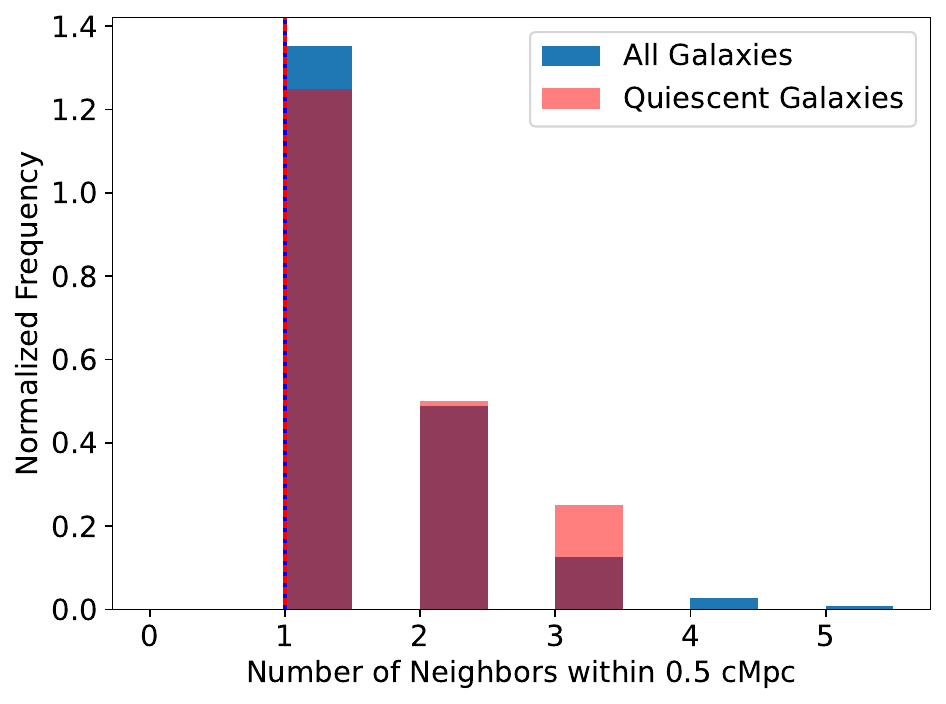}
        
    \end{subfigure}
    \begin{subfigure}[b]{0.49\textwidth}
        \includegraphics[width=\textwidth]{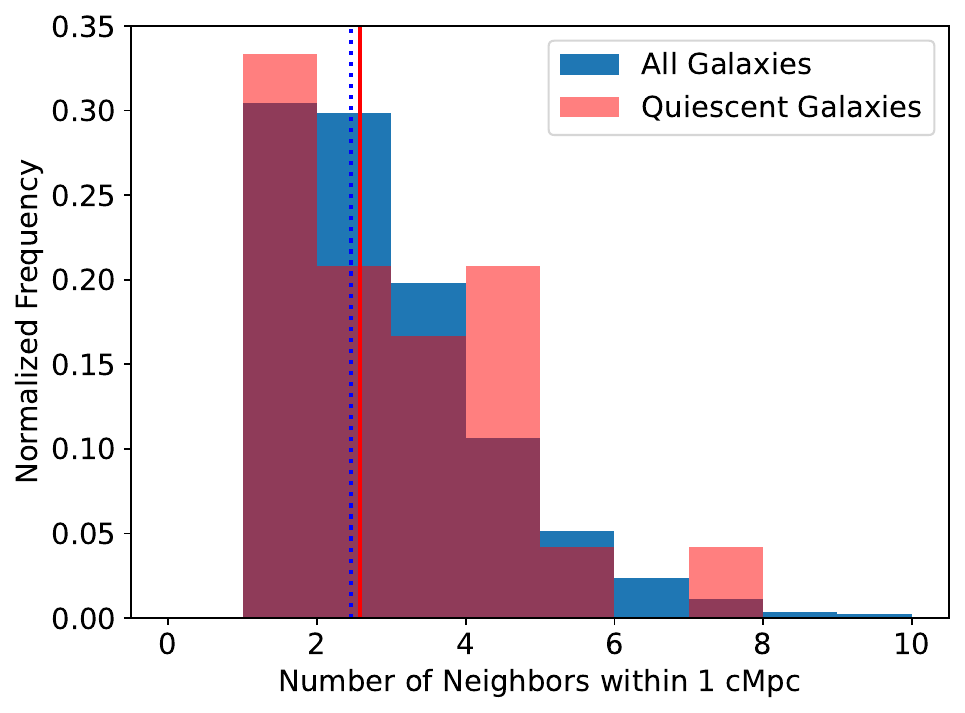}
        
    \end{subfigure}

    \begin{subfigure}[b]{0.49\textwidth}
        \includegraphics[width=\textwidth]{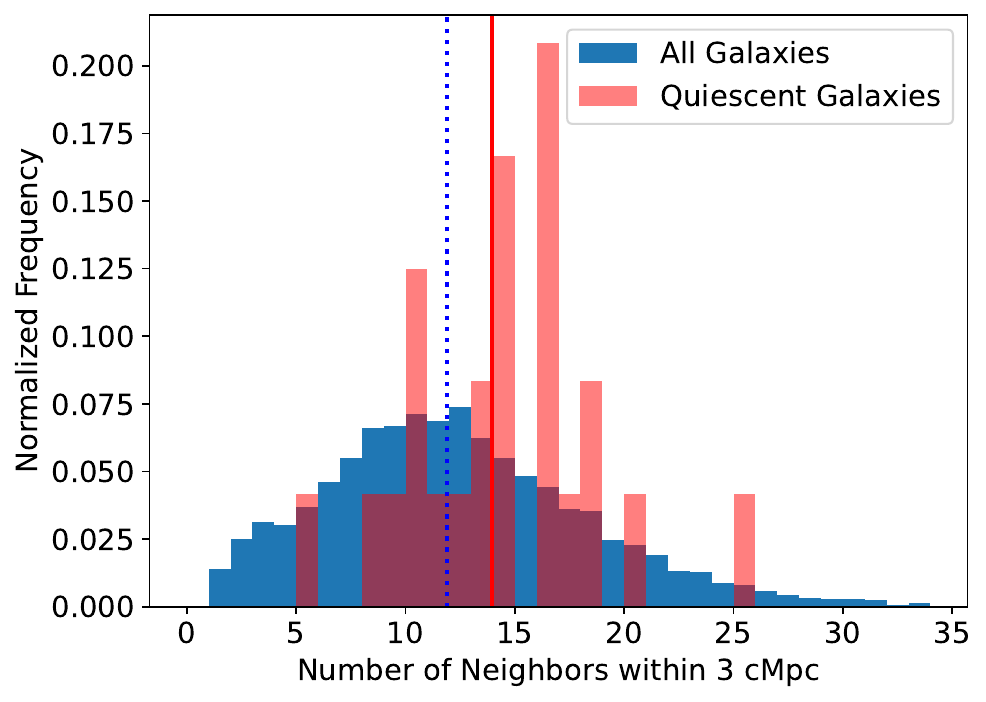}
       
    \end{subfigure}
    \begin{subfigure}[b]{0.49\textwidth}
        \includegraphics[width=\textwidth]{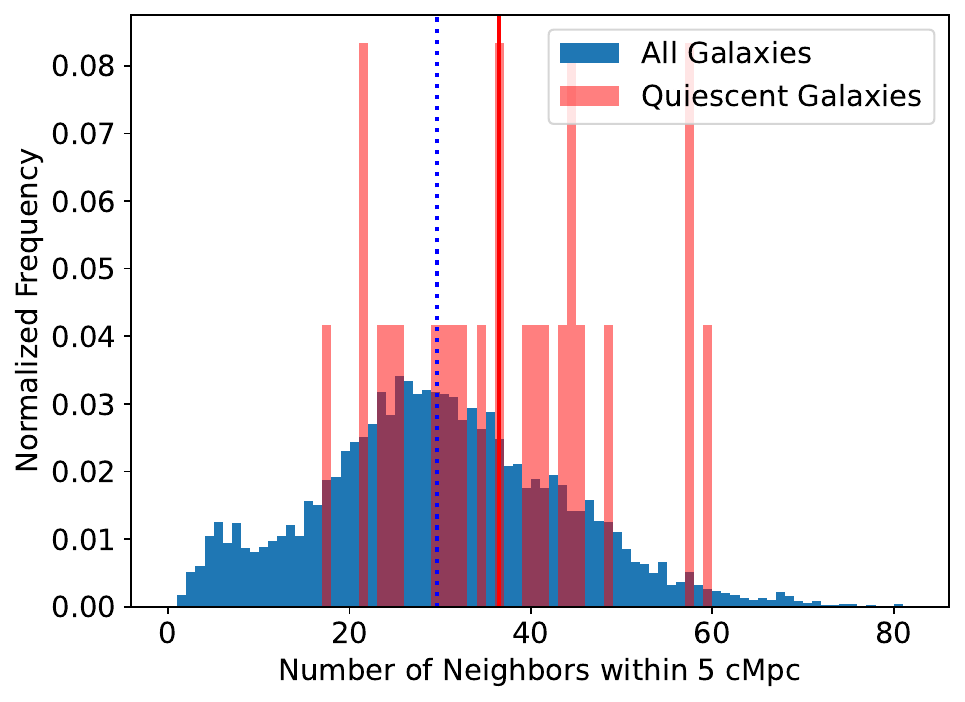}
       
    \end{subfigure}
    \caption{Number of neighbours within radii of 0.1-6 Mpc for all galaxies and MQGs. The vertical line represents the mean of the distributions. }
    \label{6nnall}
\end{figure*}
\end{appendix}
\end{document}